\pdfoutput=1

\documentclass[12pt,reqno]{article}
\usepackage{jheppub}
\usepackage{amsmath,amssymb,amsfonts}

\graphicspath{ {./Graphs/} }

\usepackage{float}

\usepackage{mathtools}

\usepackage{blkarray}

\usepackage[usenames,dvipsnames]{xcolor}
\usepackage{epsfig}
\usepackage{epstopdf}
\usepackage{dcolumn}
\usepackage{tikz}
\usetikzlibrary{shapes.geometric, arrows}
\usepackage{upgreek}
\usepackage{setspace}
\usepackage{enumitem}
\usepackage{array,multirow,bigdelim}
\usepackage{cleveref}
\usepackage{bm}

\def\be{\begin{equation}}
\def\ee{\end{equation}}
\def\ba{\begin{eqnarray}}
\def\ea{\end{eqnarray}}

\def\a{\alpha}
\def\b{\beta}

\def\b#1{\overline{#1}}

\def\CP1{\mathbb{CP}^1}
\def\SL2C{\mathrm{SL}(2,\mathbb{C})}

\def\Z2{\mathbb{Z}_2}

\def\su2{{SU(2)}}
\def\eps{{\epsilon}}

\def\a{{\alpha}}

\def\[{\left[}
\def\]{\right]}

\def\L{\Lambda}

\def\s{\sigma}
\def\a{\alpha}
\def\b{\beta}

\def\({\left(}
\def\){\right)}
\def\[{\left[}
\def\]{\right]}

\def\<{\langle}
\def\>{\rangle}

\def\i2{\frac{i}{2}}

\def\2F1{\,_2{\rm F}_1}

\def\A{\mathsf{A}}

\newcommand{\beq}{\begin{equation}}
\newcommand{\eeq}{\end{equation}}
\newcommand{\beqq}{\begin{equation*}}
\newcommand{\eeqq}{\end{equation*}}
\newcommand\beqa{\begin{eqnarray}}
\newcommand\eeqa{\end{eqnarray}}
\newcommand\beqaa{\begin{eqnarray*}}
\newcommand\eeqaa{\end{eqnarray*}}
\newcommand\bea{\begin{array}}
\newcommand\eea{\end{array}}

\newcommand{\ie}{{\it i.e.}}

\title{Scattering Equations and Factorization of Amplitudes II: Effective Field Theories}

\author{Humberto Gomez$^{1,2}$, 	Andreas Helset$^1$}
\affiliation{$^1$Niels Bohr International Academy and Discovery Center, Niels Bohr Institute, University of Copenhagen,
Blegdamsvej 17, DK-2100 Copenhagen , Denmark.\\
\\
$^{2}$Facultad de Ciencias Basicas,  Universidad Santiago de Cali,\\
Calle 5 $N^\circ$  62-00 Barrio Pampalinda, Cali, Valle, Colombia.}

\emailAdd{humgomzu@gmail.com, ahelset@nbi.ku.dk}

\abstract{
We continue the program of extending the scattering equation framework by Cachazo, He and Yuan 
to a double-cover prescription.
We discuss how to apply the double-cover formalism to effective field theories, with a 
special focus on the non-linear sigma model.
A defining characteristic of the double-cover formulation is the emergence of new factorization
relations. We present several factorization relations, along with a novel recursion relation.
Using the recursion relation and a new prescription for the integrand, 
any non-linear sigma model amplitude can be expressed in terms
of off-shell three-point amplitudes. The resulting expression is purely algebraic, and we do not
have to solve any scattering equation.
We also discuss soft limits, boundary terms in BCFW recursion, and application of the 
double-cover prescription to other effective field theories, like the special Galileon theory.
}

\begin{document}
{\setstretch{1}
\maketitle
}
\onehalfspacing
\section{Introduction}

The S-matrix elements of gravity, gauge theories and various scalar theories can be 
calculated using the novel scattering equation framework by Cachazo, He and Yuan (CHY) 
\cite{Cachazo:2013gna,Cachazo:2013hca,Cachazo:2013iea}.
The $n$-point scattering amplitude in the CHY-formalism is expressed as contour integrals localized
to the solutions of the scattering equations
\begin{align}
	\label{eq:scatteringeq}
	S_a = 0, \qquad {\rm where} \quad S_a = \sum_{b\neq a} \frac{s_{ab}}{z_{ab}},
\end{align}
with $z_{ab}=z_a - z_b$ and $z_a$ are auxiliary variables on the Riemann sphere.
Unless otherwise specified, we let $a,b \in \{1,\dots,n\}$.
The momentum of the $a^{\rm th}$ external particle is $k_a^\mu$ and $s_{ab} = 2k_a\cdot k_b$ are 
the usual Mandelstam variables.
The scattering equations are invariant under ${\rm PSL}(2,\mathbb C)$ transformations of the variables,
\begin{align}
	z_a \rightarrow z_a^\prime = \frac{A z_a + B}{C z_a + D}, \qquad \textrm{where} \quad  AD-BC = 1,
\end{align}
using overall momentum conservation, $\sum k_a = 0$, and the massless condition, $k_a^2 = 0$.
This means that if $z_a$ is a solution to \cref{eq:scatteringeq}, then so is $z_a^\prime$.
Thus, only $(n-3)$ of the scattering equations are independent, which can be seen
from the fact that 
\begin{align}
	\sum_a S_a = \sum_az_a S_a = \sum_a z_a^2 S_a = 0.
\end{align}
There is a redundancy in the integration variables which needs to be fixed, 
similar to how gauge redundancy is fixed. We choose three of the
integration variables to be fixed, leaving $(n-3)$ unfixed variables, which are integrated over.
Thus, the number of integration variables and the number of constraints from 
the scattering equations are equal, which fully
localizes the integral to the solutions of the scattering equations.
However, the number of independent solutions to the scattering equations is $(n-3)!$, and it becomes
impractical to deal with them when $n$ is not small.
The computational cost becomes huge when the number of external particles increases.
Integration rules have been developed to circumvent this problem, both at tree 
\cite{Baadsgaard:2015voa,Baadsgaard:2015ifa,Bjerrum-Bohr:2016juj,Bjerrum-Bohr:2016axv,Cachazo:2015nwa,Cardona:2016gon} and 
loop level \cite{Baadsgaard:2015hia}, where no scattering equation has 
to be explicitly solved. A formal proof of the CHY-formalism was provided in Ref.~\cite{Dolan:2013isa}.
See also Ref.~\cite{Gomez:2013wza}.

Recently, one of us extended the scattering equation formalism to a double cover of the Riemann 
sphere (called the $\L$-algorithm in Refs.~\cite{Gomez:2016bmv,Cardona:2016bpi,Cardona:2016wcr,Gomez:2016cqb}).
The auxiliary double-cover variables live in $\mathbb{CP}^2$, contrasted with the original auxiliary variables $z_a$,
which live in $\mathbb{CP}^1$ in the standard CHY formulation. 
More precisely, we consider curves in $\mathbb{CP}^2$ defined by
\begin{align}
\label{eq:curveLambda}
	C_a \equiv y_a^2 - \s_a^2 + \L^2 = 0,
\end{align}
where $\Lambda$ is a non-zero constant. This curve is invariant under a simultaneous scaling of
the parameters $y,\sigma,\Lambda$. In the new double-cover formulation, the punctures on the
Riemann sphere are given by the pair $(\sigma_a, y_a)$. As \cref{eq:curveLambda} is 
a quadratic equation, two branches develop. The value of $y_a$ specifies which branch the 
solution is on. To make sure we pick up the puncture on the correct branch, the scattering
equations have to be modified
\begin{align}
	\tilde S_a^{\tau}(\s,y) = \sum_{b\neq a} \frac{1}{2}\left( \frac{y_b}{y_a}+1\right)
	\frac{s_{ab}}{\s_{ab}},
\end{align}
where $\s_{ab} = \s_a - \s_b$. The factor $\frac{1}{2}\left( \frac{y_b}{y_a} + 1\right)$
projects out the solution where $y_b$ approaches $-y_a$, and gives $1$ when 
$y_b$ approaches $y_a$.
Another (equivalent) way of defining the double cover scattering equations is to 
postulate the map
\begin{align}
	S_a(z) = \sum_{a\neq b}\frac{s_{ab}}{z_{ab}} \rightarrow
	S_a^\tau(\s,y) = \sum_{a\neq b} s_{ab} \tau_{(a,b)},
	\qquad 
	{\rm where} \quad
	\tau_{(a,b)} = \frac{1}{2\s_{ab}}\left(\frac{y_a + y_b + \s_{ab}}{y_a}\right).
\end{align}
It is easy to check that the two prescriptions for the double cover scattering equations are
equivalent by using overall momentum conservation and the on-shell condition.
The map $z_{ij} \rightarrow \tau_{(i,j)}^{-1}$ will be useful later when we define the 
double cover integrand.
For a full formulation of the double-cover prescription, see Ref.~\cite{Gomez:2016bmv}.

In the double cover prescription, three variables need to be fixed due to M\"obius invariance. 
In addition, the integrand is invariant under a scale transformation. This gives an additional 
redundancy which needs to be fixed (as the integrand is ${\rm PSL}(2,\mathbb C)$ and scale invariant, 
{\it i.e.} ${\rm GL}(2,\mathbb C)$ invariant).
Using the scale symmetry, we fix an extra puncture, and promote $\L$ to a variable and include
a scale invariant measure $\frac{d \Lambda}{\Lambda}$.
Using the global residue theorem, we can deform the 
integration contour to go around $\L=0$ instead of the solution to the scattering equation
for the puncture fixed by the scale symmetry. 
This scattering equation is left free.
Thus, in the double-cover prescription we gauge fix four points, three from the 
usual gauge fixing procedure, and one from the scale transformation.

The two sheets of the Riemann sphere are separated by a branch cut, and by integrating over $\L$,
lead to the factorization into two regular lower-point CHY amplitudes. This is the origin of the
new factorization relations which we will discuss in the main part of this paper.
By iteratively promoting the scattering amplitudes to the double-cover formulation, and using certain
matrix identities, any $n$-point scattering amplitude for the non-linear sigma model can be fully
factorized into off-shell three-point amplitudes.

This paper is organized as follows. In \Cref{sec:CHYFormalism} we formulate the non-linear
sigma model amplitudes in the usual CHY formalism. In \Cref{sec:eftDoubleCover} we introduce the double-cover prescription for effective field theories.
In \Cref{graphR} we describe the graphical representations for the scattering amplitudes in the
double-cover formalism. In \Cref{sec:integrationRules} we list the double-cover integration rules.
In \Cref{three-point} we define the three-point functions which will serve as the building
blocks for higher-point amplitudes.
In \Cref{sec-Ex-NLSM,sec:factorizationGeneral} we present the new factorization formulas for the non-linear
sigma model.
In \Cref{sec:recursion} we present a novel recursion relation, which fully factorizes the 
non-linear sigma model amplitudes in terms of off-shell three-point amplitudes.
This is one of the main results of the paper.
\Cref{sec:softLimit} takes the soft limit of the non-linear sigma model amplitudes, and presents a new relation for
${\rm NLSM}\oplus \phi^3$ amplitudes.
In \Cref{sec:sGalileon} we apply the double-cover prescription to the special Galileon theory.
We end with conclusions and outlook in \Cref{sec:conclusion}.
The \Cref{apx:matrixProof,6-pts-comp} contain matrix identities and details of the six-point calculation.

\section{CHY Formalism}
\label{sec:CHYFormalism}
We briefly review the construction of non-linear sigma model (NLSM) scattering amplitudes in the 
CHY formalism to fix notation.
The flavor-ordered partial ${\rm U}(N)$ amplitude for the non-linear sigma model in the scattering 
equation framework is defined by the integral
\begin{align}
	\mathcal{A}_n(\a) &= \int {\rm d} \mu_n^{\rm CHY} (z_{pq}z_{qr}z_{rp})^2H_n(\alpha), \\
	{\rm d}\mu_n^{\rm CHY} &=  \prod_{a=1,a\neq p,q,r}^n \frac{{\rm d}z_a}{S_a},
\end{align}
where a partial ordering is denoted by $(\alpha) = (\alpha_1, \dots, \alpha_n)$. 
We have fixed the punctures $\{z_p,z_q,z_r\}$.
The integrand is given by the Parke-Taylor factor ${\rm PT}(\alpha)$ and the reduced 
Pfaffian of the matrix $\mathsf{A}_n$, ${\rm Pf}'\mathsf{A}_n$,
\begin{align}
	H_n(\alpha) &= {\rm PT}(\alpha) \, \left({\rm Pf}'\mathsf{A}_n\right)^2,  \\
	{\rm PT}(\alpha) &= \frac{1}{z_{\a_1\a_2}z_{\a_2\a_3}\dots z_{\a_n\a_1}}, \\
	\label{eq:detAred}
	\left({\rm Pf}'\mathsf{A}_n\right)^2 &= \frac{(-1)^{i+j+l+m}}{z_{ij}z_{lm}}\,\, {\rm Pf} \left[ (\mathsf{A}_n)^{ij}_{ij}\right]\times {\rm Pf} \left[ (\mathsf{A}_n)^{lm}_{lm}\right].
\end{align}
The matrix $\mathsf{A}_n$ is $n\times n$ and antisymmetric,
\begin{align}
	\label{eq:defA}
	(\mathsf{A}_n)_{ab} = \begin{cases} 
		\frac{s_{ab}}{z_{ab}} \qquad &{\rm for}\quad a\neq b \\
		0 \qquad &{\rm for} \quad a=b.
	\end{cases}	
\end{align}
We will in general denote a reduced matrix by $(\mathsf{A}_n)^{i_1\dots i_p}_{j_1\dots j_p}$, where we have
removed rows $\{i_1,\dots ,i_p\}$ and columns $\{j_1,\dots ,j_p\}$ from the matrix $\mathsf{A}_n$.
As an example, we can remove rows $\{i,j\}$ and columns $\{j,k\}$ from $\mathsf{A}_n$ in \cref{eq:defA},
denoted by $(\mathsf{A}_n)^{ij}_{jk}$.

With the conventional choice $\{l,m\} = \{i,j\}$, the product of Pfaffians turns into a 
determinant
\begin{align}
	\left( {\rm Pf}^\prime \mathsf{A}_n \right)^2 = - {\rm PT}(i,j) \, {\rm det}\left[(\mathsf{A}_n)^{ij}_{ij}\right].
\end{align}
We will denote the amplitude with this choice by
\begin{align}
	\label{eq:NLSMold}
	A_n(\a) = -\int {\rm d}\mu_n^{\rm CHY} (z_{pq}z_{qr}z_{rp})^2\, {\rm PT}(\a) \, {\rm PT}(i,j) \, {\rm det} \left[(\mathsf{A}_n)^{ij}_{ij}\right].
\end{align}
We can make a different choice, specifically $\{l,m\} = \{j,k\}$. We will make use of the matrix identities
\begin{align}
	\label{eq:matrixid1}
	{\rm Pf}\left[(\mathsf{A}_n)^{ij}_{ij}\right]\times {\rm Pf}\left[(\mathsf{A}_n)^{jk}_{jk}\right]
	&= {\rm det}\left[(\mathsf{A}_n)^{ij}_{jk}\right], \\
	\label{eq:matrixid2}
	{\rm det} \left[(\mathsf{A}_n)^{ij}_{jk}\right] &= 0 \qquad \textrm{if } n \textrm{ is odd.}
\end{align}
\Cref{eq:matrixid2} depends on momentum conservation and the massless condition.
A proof of the matrix identities in \cref{eq:matrixid1,eq:matrixid2}
is found in \cref{apx:matrixProof}.
The amplitude with this new choice is denoted by
\begin{align}
	\label{eq:NLSMnew}
	A_n^{\prime}(\a) = \int {\rm d}\mu_n^{\rm CHY} (z_{pq}z_{qr}z_{rp})^2\, {\rm PT}(\a) \, \frac{(-1)^{i+k}}{z_{ij}z_{jk}}
	\, {\rm det}\left[(\mathsf{A}_n)^{ij}_{jk}\right].
\end{align}
This definition differs from the conventional one, and will be of great practical use in the 
following \cite{Bjerrum-Bohr:2018lpz}. 
It will often be useful to remove columns and rows from the set of fixed punctures.
For the objects in \cref{eq:NLSMold,eq:NLSMnew}, we will encode the information of which rows
and columns are removed in the labeling of the partial ordering $\a$. When removing columns and 
rows $(i,j)$, we bold the corresponding elements in the partial ordering, {\it i.e.} 
$A_n(\dots,{\bm i},\dots,{\bm j},\dots)$. 
For the new prescription, the choice $(ijk)$
is labeled by ${A}_n^{\prime}(\dots,{\bm i},\dots,{\bm j},\dots,{\bm k},\dots)$, where the 
set is chosen to be ordered as $i<j<k$.
Unless otherwise specified, we assume the set of removed rows and columns are in the two or three
first positions, {\it i.e.} $A_n = A_n({\bm i},{\bm j},\dots)$ and $A_n^\prime = A_n^{\prime}({\bm i},{\bm j},{\bm k},\dots)$.
In this case, we will suppress the bold notation.
For an odd number of external particles $n$, ${\rm det}\left[(\mathsf{A}_n)^{ij}_{ij}\right] = {\rm det}\left[(\mathsf{A}_n)^{ij}_{jk}\right] = 0$, 
and the amplitudes vanish.

When evaluating the double cover amplitudes, it will be necessary to relax the requirement of 
masslessness, as the full amplitude is splits into off-shell lower-point amplitudes.
The off-shell punctures are part of the set of fixed punctures.
We will also use the object
\begin{align}
	\label{eq:Aij}
	A_n^{(ij)}(\a) = \int {\rm d}\mu_n^{\rm CHY} (z_{pq}z_{qr}z_{rp})^2\, {\rm PT}(\a) \, \frac{(-1)^{i+j}}{z_{ij}}  \,
	{\rm det}\left[(\mathsf{A}_n)^i_j\right].
\end{align}
As the matrix $\mathsf{A}_n$ has co-rank 2 on the support of the massless condition and the scattering 
equations, $\{k_a^2=0, S_a=0\}$, $A_n^{(ij)}(\a)$ vanishes trivially. However,
when some of the particles are off-shell, $A_n^{(ij)}(\a)$ is non-zero in general.
Similarly, the object $A_n^\prime(\a)$ is non-zero for odd number of particles, if and only
if some of the particles are off-shell.

\section{Effective Field Theories in the Double-Cover Prescription}
\label{sec:eftDoubleCover}

In Ref.~\cite{Bjerrum-Bohr:2018lpz}, it was argued that the $n$-point NLSM scattering amplitude
in the double-cover language is given by the integral
\begin{align}
	\label{eq:nlsmfullamp}
	\mathcal{A}_n^{\rm NLSM} (\alpha) &= \int_\Gamma {\rm d}\mu_n^\L \, \frac{(-1)\Delta(pqr)\Delta(pqr|m)}{S_m^\tau} \,
	\mathcal{I}_n^{\rm NLSM}(\alpha), \\
	{\rm d}\mu_n^\L &= \frac{1}{2^2} \frac{{\rm d}\L}{\L} \prod_{a=1}^n
	\frac{y_a {\rm d}y_a}{C_a} \prod_{d=1,d\neq p,q,r,m}^n \frac{{\rm d}\sigma_d}{S_d^\tau}, \\
	\Delta(pqr) &= \frac{1}{\tau_{(p,q)}\tau_{(q,r)}\tau_{(r,p)}}, \\
	\Delta(pqr|m) &= \s_p\Delta(qrm) - \s_q\Delta(rmp) + \s_r\Delta(mpq) 
	- \s_m\Delta(pqr).
\end{align}
In this section we will include a superscript to denote the amplitudes. In the rest of the 
paper we keep this superscript implicit. When not otherwise specified, an amplitude without
a superscript refers to an NLSM amplitude.
The integration contour $\Gamma$ is constrained by the $(2n-3)$ equations
\begin{align}
	\L = 0, \qquad S_d^\tau(\sigma,y)=0, \qquad C_a = 0, 
\end{align}
for $d\neq\{p,q,r,m\}$ and $a=1,\dots,n$.

In a similar fashion, one can obtain the expressions for the ${\rm NLSM}\oplus \phi^3$ and 
special Galileon amplitudes, {\it i.e.} for  
$ A^{\rm NLSM\oplus \phi^3}_n(\a || \b) $ and $ A^{\rm sGal}_n $, by specifying the integrand.
The integrands in the double-cover scattering equation framework for the NLSM, 
${\rm NLSM}\oplus \phi^3$ and special Galileon theory are given by the expressions
\begin{align}\label{InteNLSM}
{\cal I}^{\rm NLSM}_n(\a) &=  {\rm PT}^\tau(\a) 
\times
{\bf det}^{\prime} \mathsf{A}^\L_n 
\,, \\
\label{InteNLSMphi}
{\cal I}^{\rm NLSM\oplus \phi^3}_n(\a || \b) &=  {\rm PT}^\tau(\a) 
\left(\left[ \prod_{a=1}^n   \frac{(y\s)_a}{y_a}    \right] {\rm PT}^T(\b)  \,
{\rm det} \left[ \mathsf{A}^\L_n \right]^{\b_1...\b_p}_{\b_1...\b_p} \right)  \,,\\
	\label{sGal}
{\cal I}^{\rm \, sGal}_n &= 
{\bf det}^{\prime} \mathsf{A}^\L_n
  \times 
{\bf det}^{\prime} \mathsf{A}^\L_n 
\,,
\end{align}
where $(y\s)_a \equiv y_a + \s_a$. The bold reduced determinant is defined as
\begin{align}
 {\bf det}^{\prime} \mathsf{A}_{n}^\L &=  
\left[ \prod_{a=1}^n   \frac{(y\s)_a}{y_a}    \right] (-1)\,
{\rm PT}^T(i,j)
 {\rm det} \left[\mathsf{A}^\L_n \right]^{ij}_{ij} \\
 &=   
\left[ \prod_{a=1}^n   \frac{(y\s)_a}{y_a}    \right] 
(-1)^{i+k}T_{ij}T_{jk}
 {\rm det} \left[\mathsf{A}^\L_n \right]^{ij}_{jk},
\end{align}
where the second equality is used to define the $A^\prime$ amplitude in the double cover
language,
similar to \cref{eq:NLSMnew}.
The Parke-Taylor factors and the kinematic matrix are defined by the following 
replacement
\begin{align}
	\label{eq:ALambda}
	\mathsf{A}_n \rightarrow \mathsf{A}_n^\L \qquad &{\rm for} \quad z_{ab} \rightarrow T_{ab}^{-1}, \\
	\label{eq:PTT}
	{\rm PT} \rightarrow {\rm PT}^T \qquad &{\rm for} \quad z_{ab} \rightarrow T_{ab}^{-1}, \\
	\label{eq:PTtau}
	{\rm PT} \rightarrow {\rm PT}^\tau \qquad &{\rm for} \quad z_{ab} \rightarrow \tau_{(a,b)}^{-1}, 
\end{align}
where $T_{ab}^{-1} = (y\s)_a - (y\s)_b$.

Notice that the generalization to theories such as ${\rm sGal}\oplus {\rm NLSM}^2\oplus \phi^3$ or Born-Infeld theory, 
among others, is straightforward \cite{Cachazo:2014xea,Cachazo:2016njl,Mizera:2018jbh}.

\subsection{The $\Pi$ Matrix}

Most integrands in the CHY approach depend on the auxiliary variable $z_i$ through the combination 
$z_{ij} = z_i - z_j$.
As shown in \cref{eq:ALambda,eq:PTT,eq:PTtau}, we can construct the double cover integrand 
by replacing
$z_{ij}$ with $T_{ij}^{-1}$ or $\tau_{(i,j)}^{-1}$.\footnote{Of course, the measure is also 
redefined in the double cover prescription.}
This makes for an easy map between the traditional CHY approach and the new double cover method for
most integrands.

However, the $\Pi$ matrix, defined in Refs.~\cite{Cachazo:2014xea,Cachazo:2016njl,Mizera:2018jbh},  
has elements such as, $\frac{z_a\, k_a\cdot k_b}{z_{ab}}$, 
which so far have not been studied in the double cover framework.  
Explicitly, the $\Pi_{\b_1,\b_2,...,\b_m}$ matrix, defined 
in Ref.~\cite{Mizera:2018jbh}, is
{\small
\begin{eqnarray}
\Pi_{\b_1,...,\b_m} \, = \, 
\begin{blockarray}{ccccc}
 j \in \overline{\b}  &   b \in \{\b_1,...,\b_m\}  &  j \in \overline{\b}  &   b^\prime \in \{\b_1,...,\b_m\}  \\
\begin{block}{(c|c|c|c)c}
{\mathsf A}_{ij} & \Pi_{ib} & {\mathsf A}_{ij} & \Pi_{ib^\prime} &  i \in \overline{\b}  \\
----& --------& ---- & --------\\
\Pi_{aj} & {\Pi}_{ab}  &{\Pi}_{aj} & {\Pi}_{ab^\prime} &  a \in \{\b_1,...,\b_m\}   \\
---- & --------& ----  & --------\\
{\mathsf A}_{ij} & {\Pi}_{ib} & 0 & {\Pi}_{ib^\prime}   & i \in \overline{\b}  \\
----& --------& ---- & --------\\
{\Pi}_{a^\prime j} & {\Pi}_{a^\prime b}  & {\Pi}_{a^\prime j} &{\Pi}_{a^\prime b^\prime} &  a^\prime \in \{\b_1,...,\b_m\}   \\
\end{block}
\end{blockarray} ~~~  .
\nonumber
\end{eqnarray}
}
\vskip-0.2cm\noindent
Here, the $\b_a$'s sets are such that  $\b_a\cap \b_b=\emptyset$, $a\neq b$,  and $\overline{\b}$ is the complement, namely, $\overline{\b}=\{1,2,...,n \} \setminus \b_1\cup\b_2\cup\cdots \cup \b_m$, where $n$ is the total number of particles. The $\Pi$ submatrices are given by the expressions
{\small
\begin{eqnarray}
&&\Pi_{ib} = \sum_{c\in \b_b}\frac{k_i\cdot k_c}{z_{ic}}, \quad
\Pi_{ib^\prime} = \sum_{c\in \b_b}\frac{z_c\, k_i\cdot k_c}{z_{ic}}, \quad
\Pi_{ab} = \sum_{c\in \b_a \atop d\in \b_b}\frac{k_c\cdot k_d}{z_{cd}}, \quad  \nonumber\\
&&\Pi_{ab^\prime} = \sum_{c\in \b_a \atop d\in \b_b}\frac{z_d\, k_c\cdot k_d}{z_{cd}}, \quad
\Pi_{a^\prime b^\prime} = \sum_{c\in \b_a \atop d\in \b_b}\frac{z_c\, z_d \, k_c\cdot k_d}{z_{cd}}. \quad
\end{eqnarray}
}
\vskip-0.2cm\noindent
As shown in Refs.~\cite{Bjerrum-Bohr:2018lpz,Gomez:2018cqg}, 
to obtain the usual CHY matrices in the double-cover prescription we use the identification
$\frac{1}{z_{ab}} \rightarrow T_{ab}=\frac{1}{(y_a+\s_a)-(y_b+\s_b)}$  
(see the above section), which gives us the naive identification 
$z_a\rightarrow (y_a+\s_a)$. 
However, we need all elements of $\Pi_{\b_1,...,\b_m}$ to transform in the same way
under a global scaling $(y_1,\s_1,...,y_n,\s_n,\L)\rightarrow \rho \,(y_1,\s_1,...,y_n,\s_n,\L), \, \rho\in \mathbb{C}^\ast$.
We use the map\footnote{This is in agreement with the single and double-cover equivalence given in
Ref.~\cite{Gomez:2016bmv}.} $z_a \rightarrow \frac{(y_a+\s_a)}{\L }$.  
Thus, the $\Pi$ matrix in the double-cover representation is given by the replacement, 
\begin{equation}
	\label{eq:pimatrix}
\Pi_{\b_1,\b_2,...,\b_m}^\L \equiv \Pi_{\b_1,\b_2,...,\b_m}
\qquad {\rm for}\quad  \frac{1}{z_{ab}}  \rightarrow T_{ab},\quad z_a  \rightarrow  \frac{(y\s)_a}{\L} .
\end{equation}
The multi-trace amplitude for interactions among NLSM pions and bi-adjoint scalars is given by 
the integrand \cite{Mizera:2018jbh} 
\vspace{-0.2cm}
{\small
\begin{eqnarray}\label{InteNLSMBA}
{\cal I}^{\rm NLSM\oplus BA}_n(\a || \b_1|\cdots | \b_m) =  {\rm PT}^\tau(\a) \times 
\left(\left[ \prod_{a=1}^n   \frac{(y\s)_a}{y_a}    \right]\times  {\rm PT}^T(\b_1)\dots {\rm PT}^T(\b_m)  \times  {\rm Pf}^\prime \left[ {\Pi}^\L_{\b_1...\b_p}\right] \right)  \, .
\nonumber
\end{eqnarray}
}
\vskip-0.45cm\noindent
The integrand is the defined using \cref{eq:pimatrix,eq:PTT,eq:PTtau}.
The reduced Pfaffian is defined as 
\begin{align}
	 {\rm Pf}^\prime \left[ {\Pi}^\L_{\b_1...\b_p}\right]
	 =
	 {\rm Pf}\left[ ({\Pi}^\L_{\b_1...\b_p})^{ab^\prime}_{ab^\prime} \right].
\end{align}

\section{Graphical Representation}\label{graphR}

The graphical representation for effective field theory amplitudes in the double-cover prescription 
is analogous to one presented in Ref.~\cite{Gomez:2018cqg}.
The only difference is that we are going to work with determinants instead of Pfaffians. 
We will briefly review the graphical notation used in this paper.

First, the Parke-Taylor factor is drawn by a sequence of arrows joining vertices. The orientation
of the arrow represents the ordering, 
\vspace{-0.45cm}
\begin{eqnarray}\label{PTgraph}
\hspace{0.1cm}
{\rm  PT}^\tau(1,\ldots, n) 
=
\hspace{-0.32cm}
\parbox[c]{4.9em}{\includegraphics[scale=0.14]{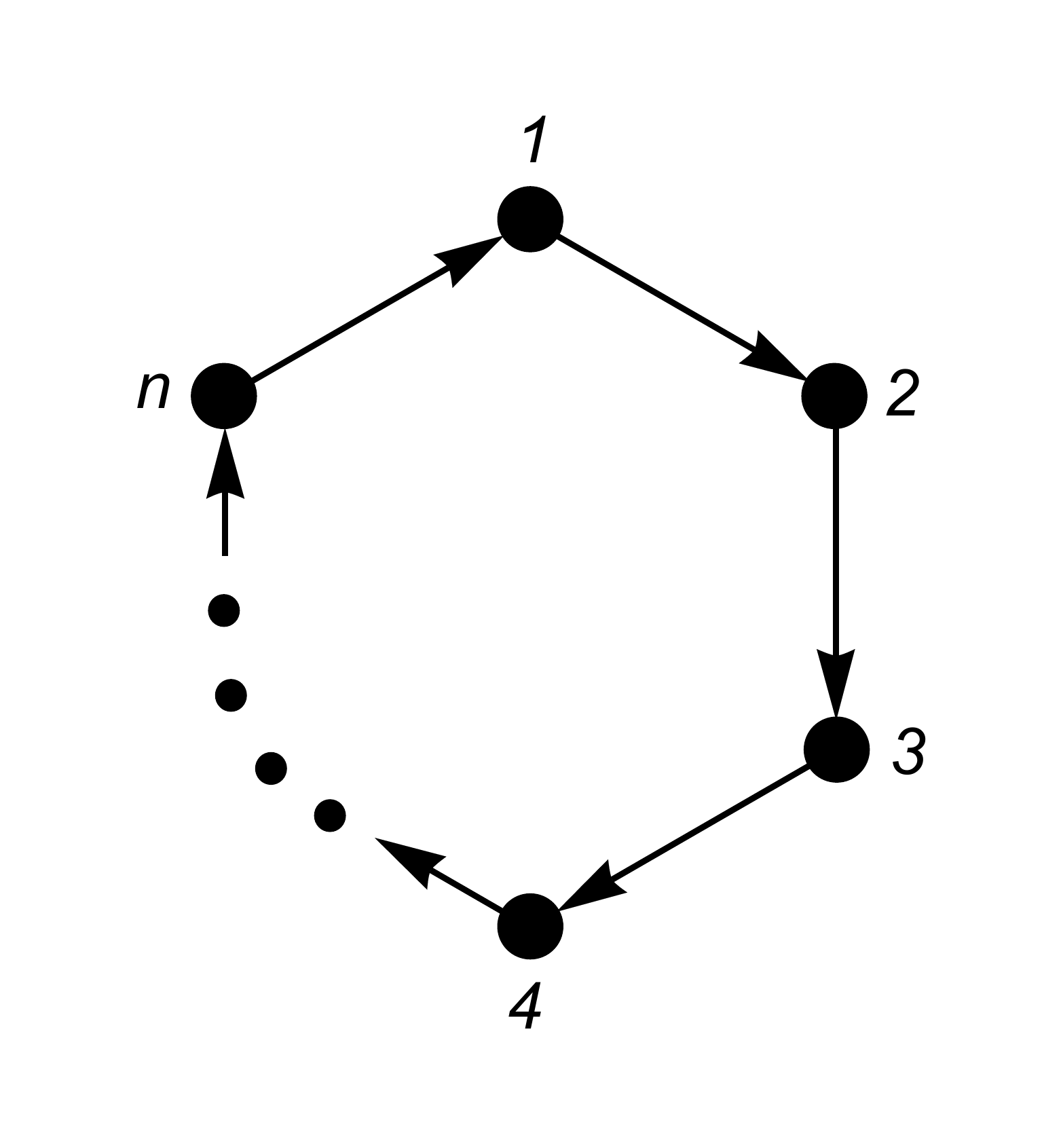}}
= (-1)^n \times
\hspace{-0.25cm}
\parbox[c]{4.9em}{\includegraphics[scale=0.14]{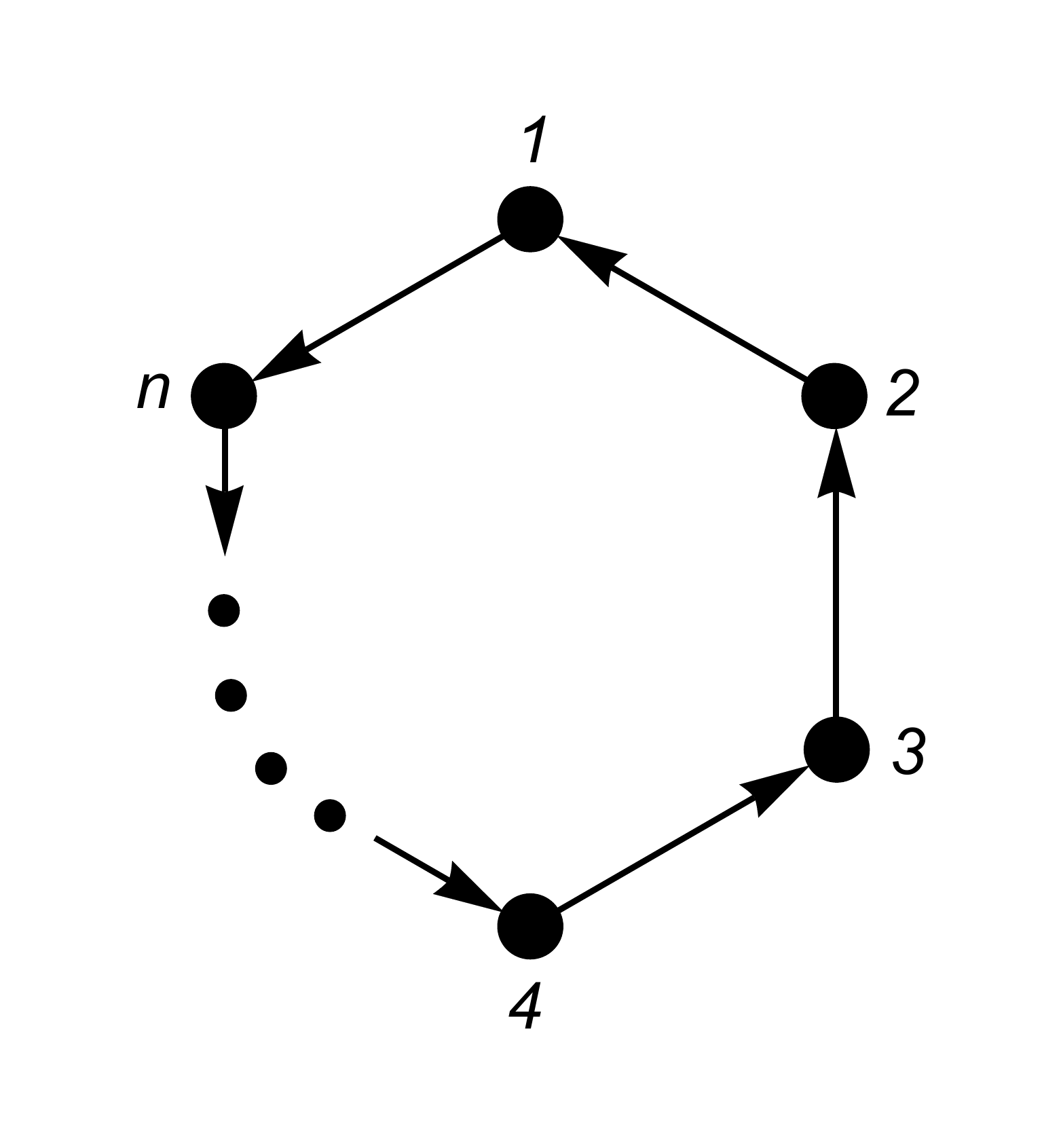}}=  (-1)^n \times {\rm  PT}^\tau(n,\ldots, 1) \, .\,\,
\end{eqnarray}
\vskip-0.45cm\noindent
To describe the half-integrand 
$(-1)
\left[ \prod_{a=1}^n   \frac{(y\s)_a}{y_a}    \right] 
(T_{ij}T_{ji})\,{\rm det}[(\mathsf{A}_n^\L)^{ij}_{ij}]$, we recall how the Pfaffian in Yang-Mills 
theory was represented \cite{Gomez:2018cqg}. In YM, the half-integrand 
\newline
$
(-1)^{i+j}\left[ \prod_{a=1}^n   \frac{(y\s)_a}{y_a}    \right] 
(T_{ij}) {\rm Pf}[(\Psi_n^\L)_{ij}^{ij}]$ was represented by a red arrow from $i \, {\color{red} \rightarrow} \, j$.
We associate this red arrow with the factor $T_{ij}$ of the reduced Pfaffian.
In the case of NLSM, we draw two red arrows, $i\, {\color{red} \rightleftarrows} \, j$,
for the factor
$T_{ij}T_{ji}$ of the reduced determinant. With the new definition of the NLSM integrand,
$
(-1)^{i+k}\left[ \prod_{a=1}^n   \frac{(y\s)_a}{y_a}    \right] 
T_{ij}T_{jk} \, {\rm det}[(\mathsf{A}_n^\L)^{ij}_{jk}]$, we draw two red arrows, 
$i \, {\color{red} \rightarrow}  \, j {\color{red} \rightarrow} \, k$.

If we choose to fix the punctures $(pqr|m)=(123|4)$ and reduce the determinant with $(i,j) = (2,p)$,
we can graphically represent the NLSM amplitude $ A_n(\a)$ by an {\it NLSM-graph},
\vspace{-0.3cm}
{\small
\begin{eqnarray}\label{YMgraph}
A_n(1,{\bm 2},3, 4, ... ,{\bm p}, ..., n)
= \int d\mu_n^\L
\hspace{-0.1cm}
\parbox[c]{5.2em}{\includegraphics[scale=0.14]{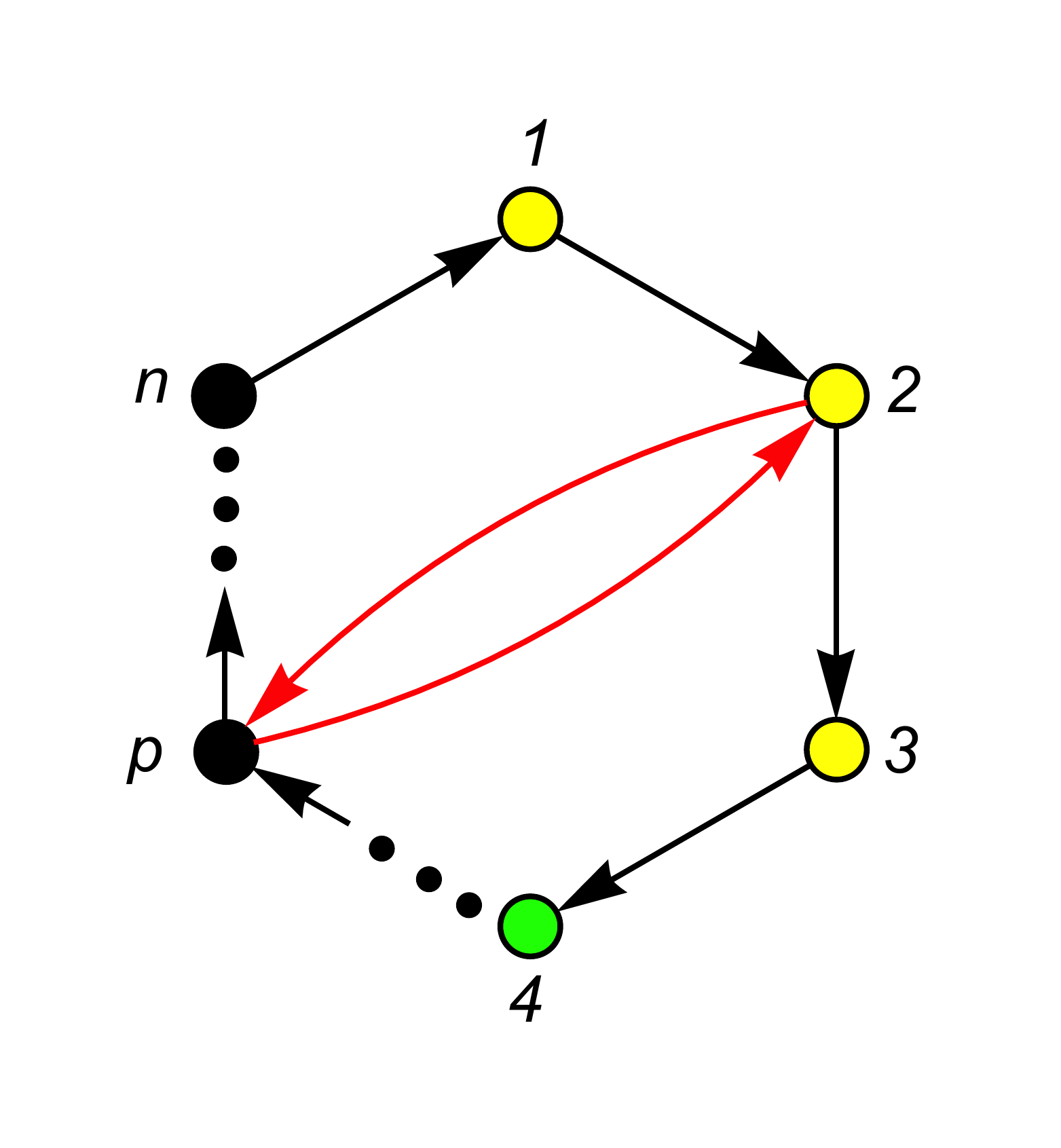}} .
\nonumber
\end{eqnarray}
}
\vskip-0.7cm\noindent
Recall that the removed columns and rows $(i,j)$ are written in bold in the partial ordering. 
The notation for the fixed punctures by yellow, green and red vertices is the same as
in Ref.~\cite{Gomez:2018cqg}.
When all particles are on-shell, the expression is independent of the choice of fixed punctures
and reduced determinant. However, as we shall see later, when we have off-shell particles, the 
expression depends on the choices.

Lastly, the following two properties
\vspace{-0.1cm}
\begin{eqnarray}
&& 
\hspace{-0.8cm}
A_n (1,{\bm 2},3, 4, ... ,{\bm p}, ..., n)=
A_n({\rm cyc}(1,{\bm 2},3, 4, ... ,{\bm p}, ..., n)), \quad
\nonumber \\
&&
\hspace{-0.8cm}
A_n(1,{\bm 2},3, 4, ... ,{\bm p}, ..., n) = 
(-1)^n\,
A_n(n,...,{\bm p},... ,4,3,{\bm 2}, 1)\,,\,\,\quad
\end{eqnarray}
are satisfied even if some of the particles are off-shell.
The graphical representation for other effective field theories are similar.
Also, the double-cover representation reduces to the usual CHY representation when
the green vertex is replaced by a black vertex.

\section{The  Double-Cover Integration Rules}
\label{sec:integrationRules}
We will formulate the double-cover integration rules, applicable for the effective field theory amplitudes
for the NLSM and special Galileon theory (sGal). 
Generalizing the integration rules to other effective field theories is straightforward.
The integration rules share a strong resemblance to the Yang-Mills integration rules
given in Ref.~\cite{Gomez:2018cqg}.

The integration of the double-cover variables $y_a$ localizes the integrand to the curves $C_a=0$, with 
the solutions $y_a = \pm\sqrt{\s_a^2 - \L^2}, \, \forall\, a$. The double cover splits into an upper and a 
lower Riemann sheet, connected by a branch-cut, defined by the branch-points $-\L$ and $\L$.
The punctures are distributed among the two sheets in all $2^n$ possible combinations.\footnote{Only $2^{n-1}$
configurations are distinct, due to a $\mathbb{Z}_2$ symmetry.}
When performing the integration of $\L$, the two sheets factorize into two single covers connected by 
an off-shell propagator (the scattering equation $S_m^\tau$ in \cref{eq:nlsmfullamp} reduces to the off-shell propagator under the
$\L$ integration). On each of the two lower-point single covers three punctures need to be fixed
due to the ${\rm PSL}(2,\mathbb{C})$ redundancy. The branch-cut closes to a point when $\L\rightarrow 0$,
which becomes an off-shell particle. The corresponding puncture is fixed. In addition, two more punctures
need to be fixed on each of the sheets. These fixed punctures must come from the fixed punctures in the original
double cover (graphically represented by colored vertices, yellow or green).
If there is not exactly two colored vertices on each of the new single covers, the configuration vanishes.
We summarize this in the first integration rule \cite{Gomez:2016bmv,Gomez:2018cqg};
\begin{itemize}
	\item {\bf Rule-I.} {\it All configurations (or cuts) with fewer (or more) than two colored 
		vertices (yellow or green) vanish trivially.}
\end{itemize}
The first integration rule, {\bf Rule-I}, is general for any theory formulated in a double-cover language.
In addition, we need to formulate supplementary integration rules specific to the NLSM and special Galileon amplitudes.

We start by determining how different parts of the integrand (and the measure) scale with $\L$. 
Without loss of generality, consider a configuration where the punctures $\{\s_{p+1},\dots,\s_n,\s_1,\s_2\}$
are located on the upper sheet, and the punctures $\{\s_3,\s_4,\dots,\s_p\}$ are located on the lower
sheet. 
This configuration (or cut) will be graphically represented by a dashed red line, which separates the two sets.
{\bf Rule-I} forces two of the fixed punctures to be on the upper sheet, and the other two to be on the
lower sheet.
By expanding around $\L=0$, the measure and the Faddeev-Popov determinants become
\vskip-0.5cm
{\small
\begin{align}
&\hspace{-0.2cm}
d\mu_n^\L \Big|^{p+1,\ldots,1,2}_{3,\, 4,\ldots ,p}  
= \frac{d\L}{\L} \times  \left[ \frac{d\s_{p+1}}{S_{p+1}}\cdots  \frac{d\s_{n}}{S_n} \right] \times  \left[\frac{d\s_{5}}{S_5}\cdots \frac{d\s_{p}}{S_p} \right]
+ {\cal O}(\L) \nonumber \\
&
\hspace{1.8cm}
=
\frac{d\L}{\L} \times
d\mu^{\rm CHY}_{n-(p-2)+1} \times d\mu^{\rm CHY}_{(p-2)+1} + {\cal O}(\L) , \\
&
\hspace{-0.4cm}
\frac{\Delta{(123)}  \Delta{(123|4)}}{ S^{\tau}_4}
\Big|^{p+1,\ldots,1,2}_{3,\, 4,\ldots ,p}  
\hspace{-0.2cm}
=  \frac{2^5}{\L^4} (\s_{12}\, \s_{2P_{3:p}} \,  \s_{P_{3:p}1}  )^2 
\left[ \frac{ 1 }{s_{34 \ldots p}} 
\right] 
(\s_{P_{p+1:2}3} \, \s_{34} \, \s_{4P_{p+1:2}})^2
 +  {\cal O}\left(\L^{-2} \right),
\label{eq:fpL0}
\end{align}
}
\vskip-0.2cm\noindent
where $P_{3:p}$ and $P_{p+1:2}$ denote the momentum of the off-shell punctures on the upper and
lower sheets, respectively. Here, $P_{3:p} = k_3 + \dots + k_p$, $P_{p+1:2}= k_{p+1} + \dots + 
k_2$
and $s_{34\dots p} = 2\sum_{i<j,i= 3}^p k_i \cdot k_j$.
For concreteness, we have fixed the punctures $(pqr|m)=(123|4)$. 
Graphically, this configuration is represented by
\vspace{-0.3cm}
{\small
\begin{eqnarray}\label{YMgraph}
A_n(1,{\bm 2},3, 4, ... ,{\bm p}, ..., n) \Big|^{p+1,\ldots,1,2}_{3,\, 4,\ldots ,p}  
= 
\hspace{0.1cm}
\parbox[c]{6.2em}{\includegraphics[scale=0.14]{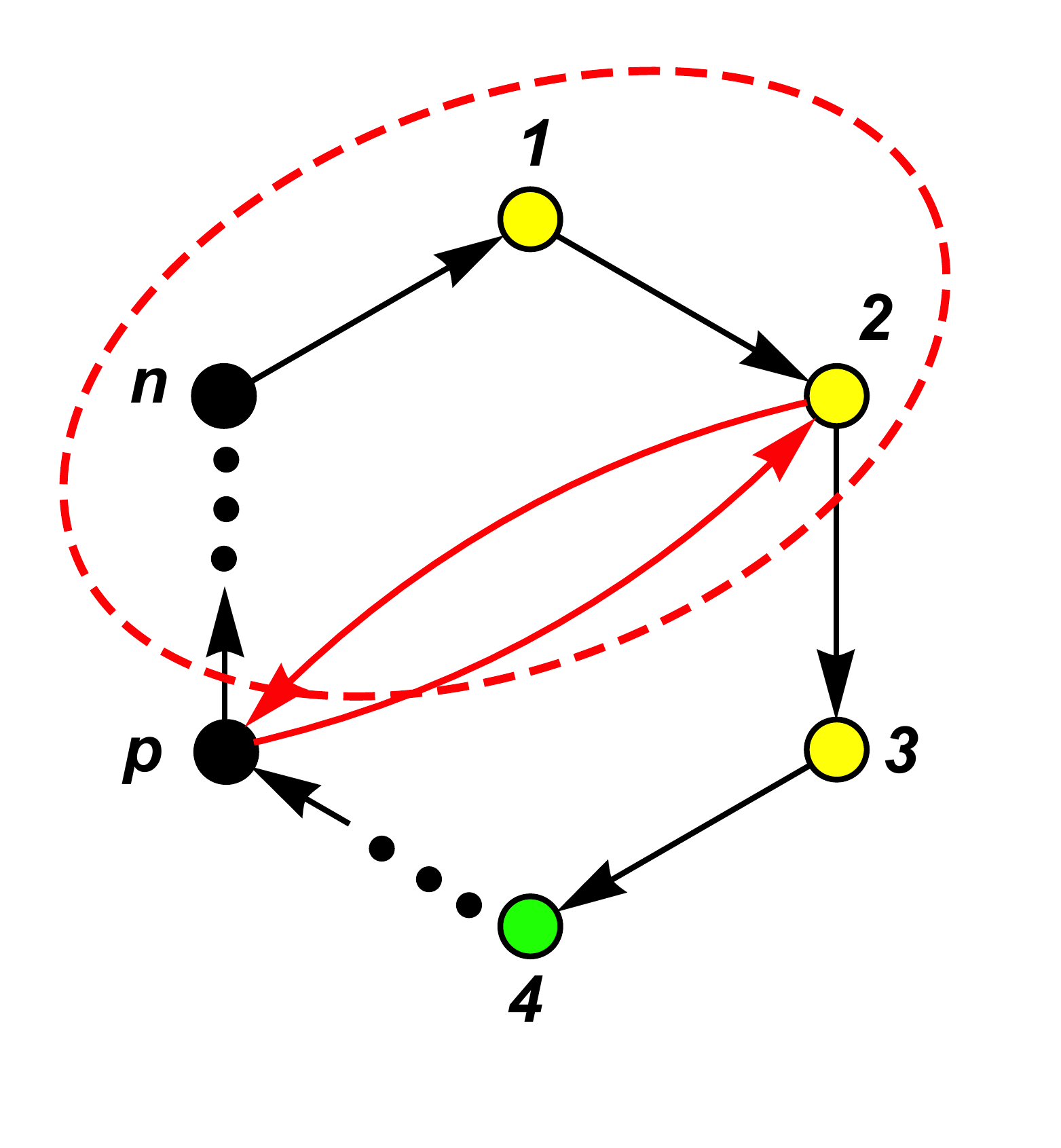}} .
\end{eqnarray}
}
\vskip-0.7cm\noindent
Notice how the measure and the Faddeev-Popov
determinants scale with $\L$ at leading order,
\begin{align}
	\label{eq:fadeevLscaling}
	d\mu_n^\L &\sim \frac{d\L}{\L}, \\
	\label{eq:measureLscaling}
	\frac{\Delta(123)\Delta(123|4)}{S_4^\tau} &\sim \frac{1}{\L^4}.
\end{align}
We also need to know how the Parke-Taylor factor and the reduced determinant scale with $\L$.
\begin{table}[ht]
	\large
	\centering
	\begin{tabular}{c|c||c|c|}
		\multicolumn{2}{c}{} & \multicolumn{2}{c}{{\small Factor}} \tabularnewline
		\cline{2-4}
		\multirow{9}*{\rotatebox{90}{\small{No. of cut arrows}}} & & 
		${\rm PT}^\tau(\a)$ & ${\bf det}^\prime (\mathsf{A}_n^\L)$ \tabularnewline[1ex]
		\cline{2-4}
		& \bfseries 0 & $\L^0$ & $\L^0$ \tabularnewline[1ex]
		& \bfseries 1 & - & $\L^2$ \tabularnewline[1ex]
		& \bfseries 2 & $\L^2$ & $\L^2$ \tabularnewline[1ex]
		& \bfseries 3 & - & - \tabularnewline[1ex]
		& \bfseries 4 & $ \L^4$ & - \tabularnewline[1ex]
		\cline{2-4}
	\end{tabular}
	\caption{
	\label{intruleTable}
		The table displays the dependence of $\L$ in the integrand factors when expanding around 
	$\L=0$. Some entries are empty, meaning that they are impossible to achieve. E.g. the Parke-Taylor 
	factor only appears when an even number of arrows are cut. This is because the PT factor forms a
	closed ring. Similarly, the reduced determinant enters with two arrows, so at most two arrows can
be cut.}
\end{table}
\Cref{intruleTable} shows how the integrand factors depend on $\L$ when expanded around $\L=0$. We
see that how the integrand scales with $\L$ is very dependent on the number of cut arrows. 
For an NLSM amplitude, for each possible non-zero cut, we find that
{\small
\begin{align}
	&{\rm PT}^\tau(1,\ldots,n) \times  {\bf det}^\prime \mathsf{A}^{\L}_n \sim  {\cal O}(\L^6),\,\quad\qquad\, \textit{The dashed red line cuts more than four arrows.}\, \nonumber \\
	&{\rm PT}^\tau(1,\ldots,n) \times  {\bf det}^\prime \mathsf{A}^{\L}_n \sim  \L^4  + {\cal O}(\L^2),\quad\, \textit{The dashed red line cuts three or four arrows.}\,\,~~~ \nonumber \\
	&{\rm PT}^\tau(1,\ldots,n) \times  {\bf det}^\prime \mathsf{A}^{\L}_n \sim  \L^2  + {\cal O}(\L^0),\quad\, \textit{The dashed red line cuts two arrows (singular cut).}\,\,~~~ \nonumber  
\end{align}
}
\vskip-0.1cm\noindent
Similarly, for an sGal-graph, we find that
{\small
	\begin{align}
		{\bf det}^\prime \mathsf{A}^\L_n \times {\bf det}^\prime \mathsf{A}^\L_n
		\sim \L^4 + {\cal O}(\L^2), \quad\, &\textit{The dashed red line cuts
		one or two arrows}\nonumber\\ &\textit{from each of the determinants.}\,\,~~~\nonumber \\
		{\bf det}^\prime \mathsf{A}^\L_n \times {\bf det}^\prime \mathsf{A}^\L_n
		\sim \L^2 + {\cal O}(\L^2), \quad\, &\textit{The dashed red line cuts
		one or two arrows}\nonumber\\ &\textit{from a single the determinant (singular cut).}\,\,~~~\nonumber \\
		{\bf det}^\prime \mathsf{A}^\L_n \times {\bf det}^\prime \mathsf{A}^\L_n
		\sim \L^0 + {\cal O}(\L^2), \quad\, &\textit{The dashed red line cuts
		no arrows (singular cut).}\,\,~~~\nonumber 
	\end{align}
}
We combine this with \cref{eq:fadeevLscaling,eq:measureLscaling}. For an NLSM-graph, there is no residue 
when more than four arrows are cut, and the configuration vanishes. When three or four arrows are cut,
the factor of $1/\L^4$ from the Faddeev-Popov determinants is canceled by the integrand, and we have a 
simple pole in $\L$. We can evaulate the contribution directly. However, when only two arrows are cut,
we do not have a simple pole, and we need to expand beyond leading order. We call this configuration 
a {\it singular cut}.
We summarize this in the second integration rule for an NLSM-graph;
\begin{itemize}
	\item {\bf Rule-II ({\it NLSM-graph}).} {\it If the dashed red line cuts fewer than three arrows over
		the NLSM-graph, the integrand must be expanded to next to leading order (singular cut).
		If the dashed red line cuts three or four arrows, the leading order expansion is sufficient.
	Otherwise, the cut is zero.}
\end{itemize}
We can perform a similar analysis for an sGal-graph. If one or two arrows from each of the 
determinants are cut, we have a simple pole and the contribution can be evaluated directly.
Otherwise, the cut is singular and we need to expand beyond leading order. This produces the 
second integraion rule for an sGal-graph;
\begin{itemize}
	\item {\bf Rule-II ({\it sGal-graph}).} {\it If the dashed red line cuts at least one arrow
			from each of the determinants, the leading order expansion is sufficient.
		Otherwise, the integrand must be expanded to next to leading order.}
\end{itemize}
In Ref.~\cite{Gomez:2016bmv}, this rule was called the $\L$-theorem. In general, we want to avoid singular
cuts. If the graph in question is regular (not singular), the following rule apply
\begin{itemize}
	\item {\bf Rule-IIIa ({\it NLSM- and sGal-graphs}).} {\it When the dashed red line cuts four arrows, the graph breaks into two smaller graphs (times a propagator). The off-shell puncture corresponds to a scalar particle. }
	\item {\bf Rule-IIIb ({\it NLSM- and sGal-graphs}).} {\it If the dashed red line cuts three arrows in a graph, there is an off-shell vector field (gluon) propagating among the two resulting graphs. The two resulting graphs must be glued by the identity, $\sum_{M}\epsilon^{M\, \mu} \, \epsilon^{M\,\nu}= \eta^{\mu \nu}$.} 
	\item {\bf Rule-IIIc} ({\it sGal-graph}). {\it If the dashed red line cuts two arrows, there is an off-shell spin-2 field (graviton) propagating between the two resulting smaller graphs. The two sub-graphs are glued together by the identity $\sum_M \, \epsilon^{M\, \mu\a}\epsilon^{M\, \nu\beta} = \eta^{\mu\nu}\eta^{\a\beta}$.}
\end{itemize}
When there are off-shell gluons or gravitons connecting the sub-graphs, we must replace the 
corresponding off-shell momentum by a polarization vector,
$P_i^\mu \rightarrow P_i^{M\,\mu}= \frac{1}{\sqrt{2}} \epsilon_i^{M\,\mu}$,
in the reduced determinants
\cite{Bjerrum-Bohr:2018jqe}.

Finally, we note that the integration rules are independent of the embedding,
\begin{itemize}
\item {\bf Rule-IV.} {\it The number of intersection points among the dashed red-line and the arrows is given mod 2.}
\end{itemize}
We can always find an embedding where the dashed red line cuts any arrow zero or one time.


\section{Three-Point Functions}\label{three-point}

Before we look at examples, it is useful to compute the three-point amplitudes that 
will work as building blocks for higher-point amplitudes.

We are using the objects defined in \cref{eq:NLSMnew,eq:Aij}.
For the non-linear sigma model, the fundamental three-point functions are given by the expressions 
\vspace{-0.3cm}
{\small
\begin{align}\label{3pt-Nor}
A^{\phi^3}(P_a,P_b,P_c)= \hspace{-0.65cm}	\parbox[c]{5.6em}{\includegraphics[scale=0.22]{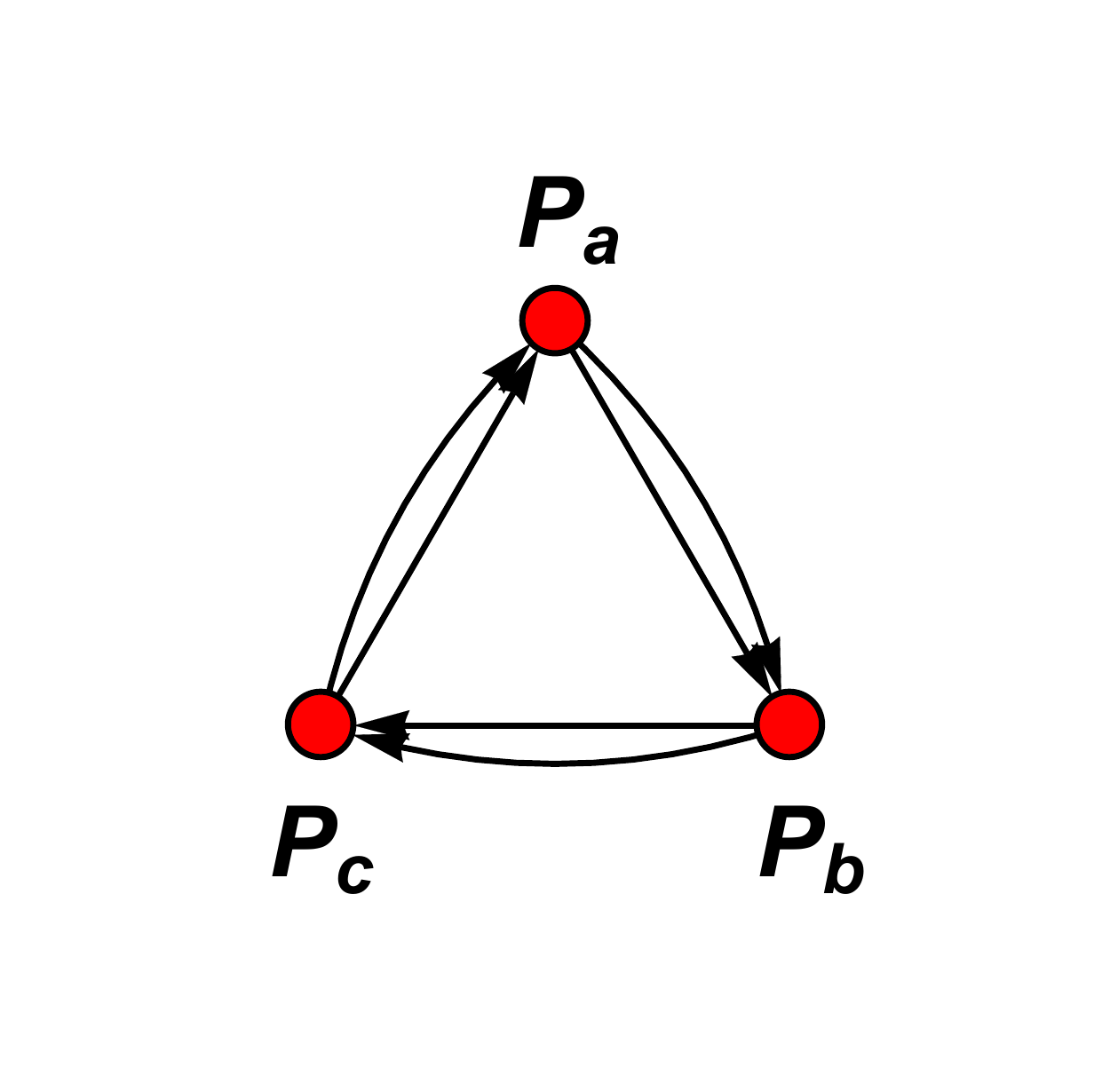}} 
&= \int d\mu_3^{\rm CHY} (\s_{P_a P_b} \s_{P_b P_c}\s_{P_cP_a})^2 \,\, {\rm PT}(P_a,P_b,P_c)^2
=
 \,1\, , 
\\
A^{\prime}_3(P_a ,P_b ,P_c )
=
\hspace{-0.65cm}
\parbox[c]{5.6em}{\includegraphics[scale=0.22]{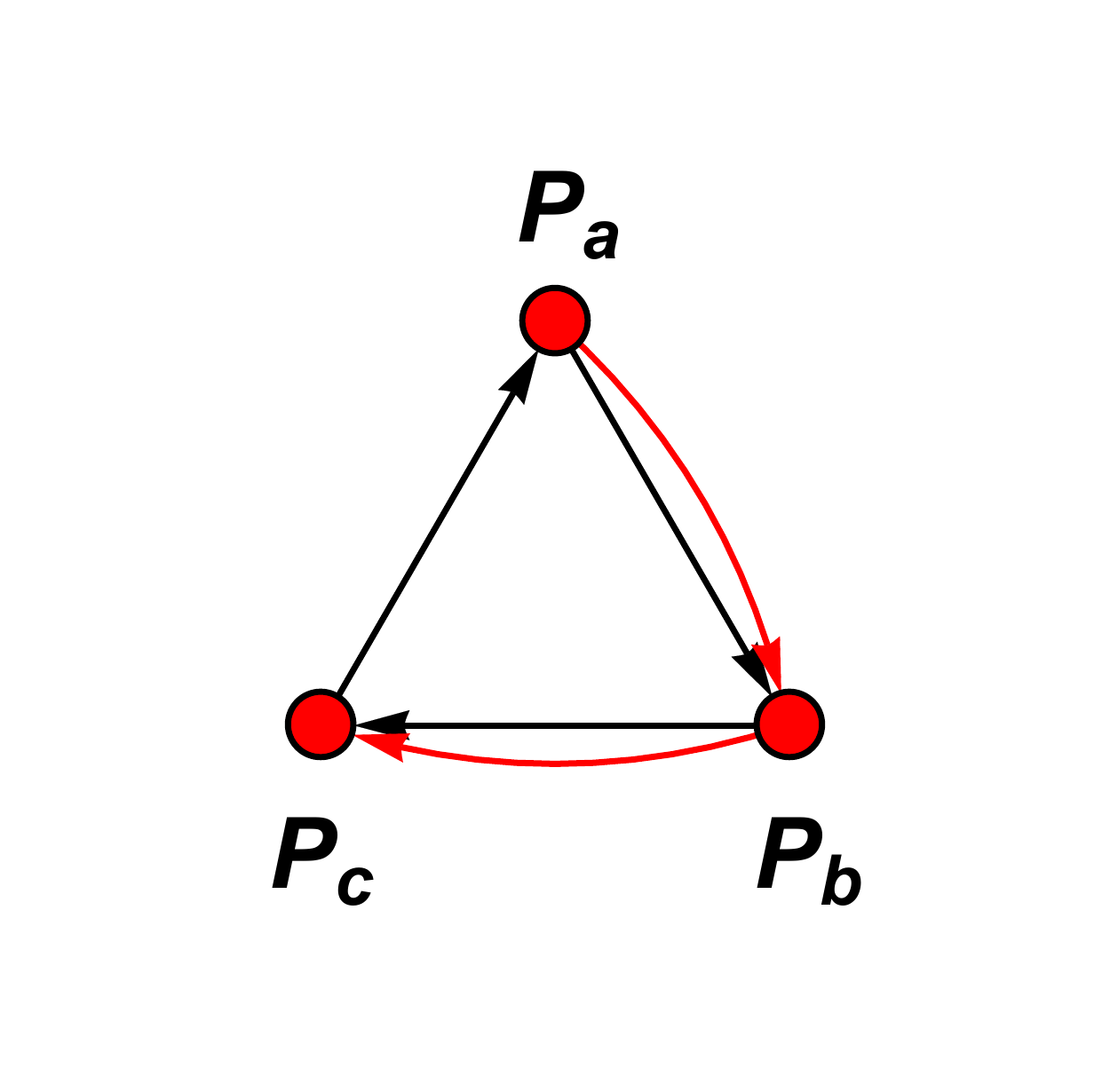}}
&=
\int d\mu_3^{\rm CHY} (\s_{P_a P_b} \s_{P_b P_c}\s_{P_cP_a})^2  \, {\rm PT}(P_a,P_b,P_c)\,
\frac{1}{\s_{P_aP_b}\, \s_{P_bP_c}}\, \frac{s_{P_cP_a}}{\s_{P_cP_a}}
\nonumber \\&= 
\,s_{P_cP_a}, 
\label{3pt-1}
\\
A^{(P_aP_b)}_3(P_a,P_b,P_c)
= 
\hspace{-0.65cm}
\parbox[c]{5.4em}{\includegraphics[scale=0.22]{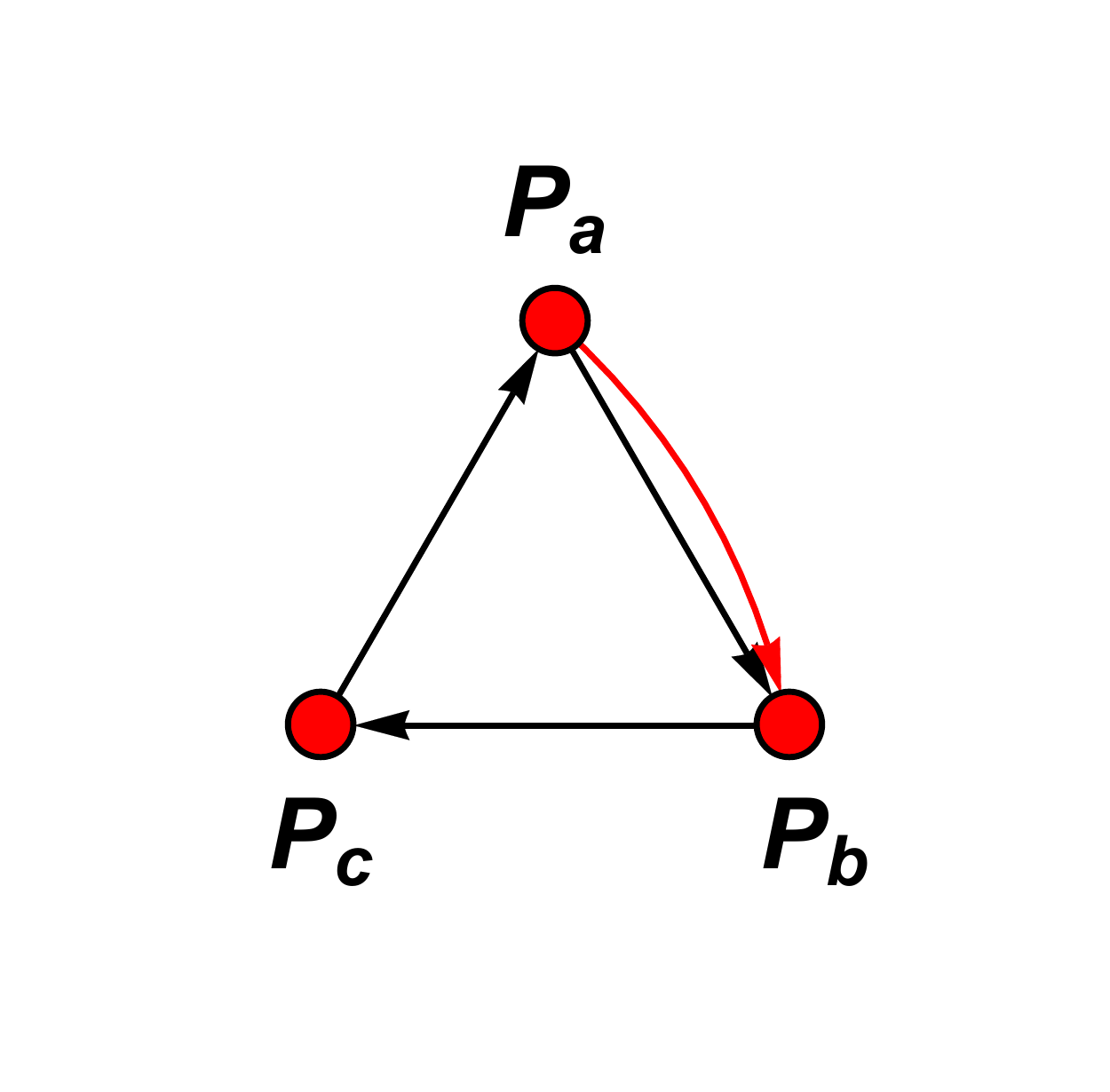}}
&=
\int d\mu_3^{\rm CHY} (\s_{P_a P_b} \s_{P_b P_c}\s_{P_cP_a})^2  \, {\rm PT}(P_a,P_b,P_c)\,
\nonumber \\
& \times \frac{(-1)}{\s_{P_aP_b}}\, {\rm det} 
\left[
\begin{matrix}
\frac{s_{P_bP_a}}{\s_{P_bP_a}} & \frac{s_{P_bP_c}}{\s_{P_bP_c}} \\
\frac{s_{P_cP_a}}{\s_{P_cP_a}} & 0\\ 
\end{matrix}
\right]  
= \,s_{P_bP_c} \,s_{P_cP_a}, \label{3point-BB} 
\end{align}
}
\vskip-0.5cm\noindent
where $P_{a}^\mu+P_{b}^\mu+P_{c}^\mu=0$ and all particles could be off-shell, {\it i.e.} $P^2_{i}\neq 0$.
Using momentum conservation, we reformulate the expressions as 
\begin{align}
	  A^{\prime}_3(P_a ,P_b ,P_c )&=s_{P_cP_a}=-(P_a^2-P_b^2+P_c^2) , \label{eq:3pointmass} \\
	 A^{(P_aP_b)}_3(P_a,P_b,P_c)&= \,s_{P_bP_c} \,s_{P_cP_a}\, 
	=(P_c^2-P_a^2+P_b^2)\times(P_a^2-P_b^2+P_c^2) \nonumber \\
	&= A_3^{\prime}(P_c,P_a,P_b)\times 
	 A^{\prime}_3(P_a ,P_b ,P_c ). \label{identity-3point}
\end{align}
We see that the three-point functions in \cref{eq:3pointmass,identity-3point} vanish when the particles are on-shell.

\section{Factorization Relations}\label{sec-Ex-NLSM}

We will presents three different prescriptions for computing NLSM amplitudes.
As we will see, they lead to three different factorization relations.

First, we start with the conventional NLSM prescription given in \cref{eq:NLSMold}  
(in the double-cover language). 
It is useful to remember that for an odd number of external particles,
the amplitude vanishes, 
\begin{equation}\label{Aodd}
	A_{2n+1} (1,\dots,\bm{P_i},\dots,\bm{P_j},\dots,n)=0.
\end{equation}
This relation holds even when the particles removed from the determinant 
by the choice $(i,j)$ are off-shell, {\it i.e.} when $P_i^2 \neq 0$ and/or $P_j^2 \neq 0$.

Secondly, we will use the alternative prescription given in \cref{eq:NLSMnew} with two different
gauge fixing choices,
resulting in two new factorization formulas.
Parts of the results were reported by us in Ref.~\cite{Bjerrum-Bohr:2018jqe}.

In general, we denote the sum of cyclically-consecutive external momenta (modulo the total number of 
particles) by $P_{i:j}\equiv k_i + \dots + k_j$. We also use the shorthand notation
$P_{i,j} \equiv k_i + k_j$ for two (not necessarily consecutive) momenta.
We also define the generalized Mandelstam variables 
$s_{i:i+j} \equiv s_{ii+1\dots i+j}$ and $s_{i:i+j,L} \equiv s_{ii+1\dots i+jL}$, with
$s_{i_1\dots i_p} \equiv \sum_{a\neq b,a,b=1}^p k_{i_a} \cdot k_{i_b}$.
\subsection{Four-Point}
\subsubsection{The Usual Integrand Prescription}\label{sec-NLSM-1}
 
Let us start by considering the four-point amplitude, $A_4(1,2,3,4)$. 
Without loss of generality, we choose the gauge fixing $(pqr|m)=(123 | 4)$. 
In order to avoid singular cuts (see \Cref{sec:integrationRules}), 
we remove the columns and rows $(i,j)=(1,3)$ for the determinant in \cref{eq:NLSMold}.
For notational simplicity, we define $\mathbb{I}_n=(1,\dots,n)$, 
$\mathbb{I}_n^{(ij)}=(1,\dots,{\bm i},\dots,\bm j,\dots,n)$, and
$\mathbb{I}_n^{(ijk)}=(1,\dots,{\bm i},\dots,{\bm j},\dots,{\bm k},\dots, n)$.
Graphically, the amplitude factorizes into
\vspace{-0.4cm}
{\small
\begin{eqnarray}\label{4pts-Gg-cuts}
	A_4(\mathbb{I}_4^{(13)} ) = 
\int d\mu_{4}^\L  
\hspace{-0.5cm}
\parbox[c]{5.2em}{\includegraphics[scale=0.21]{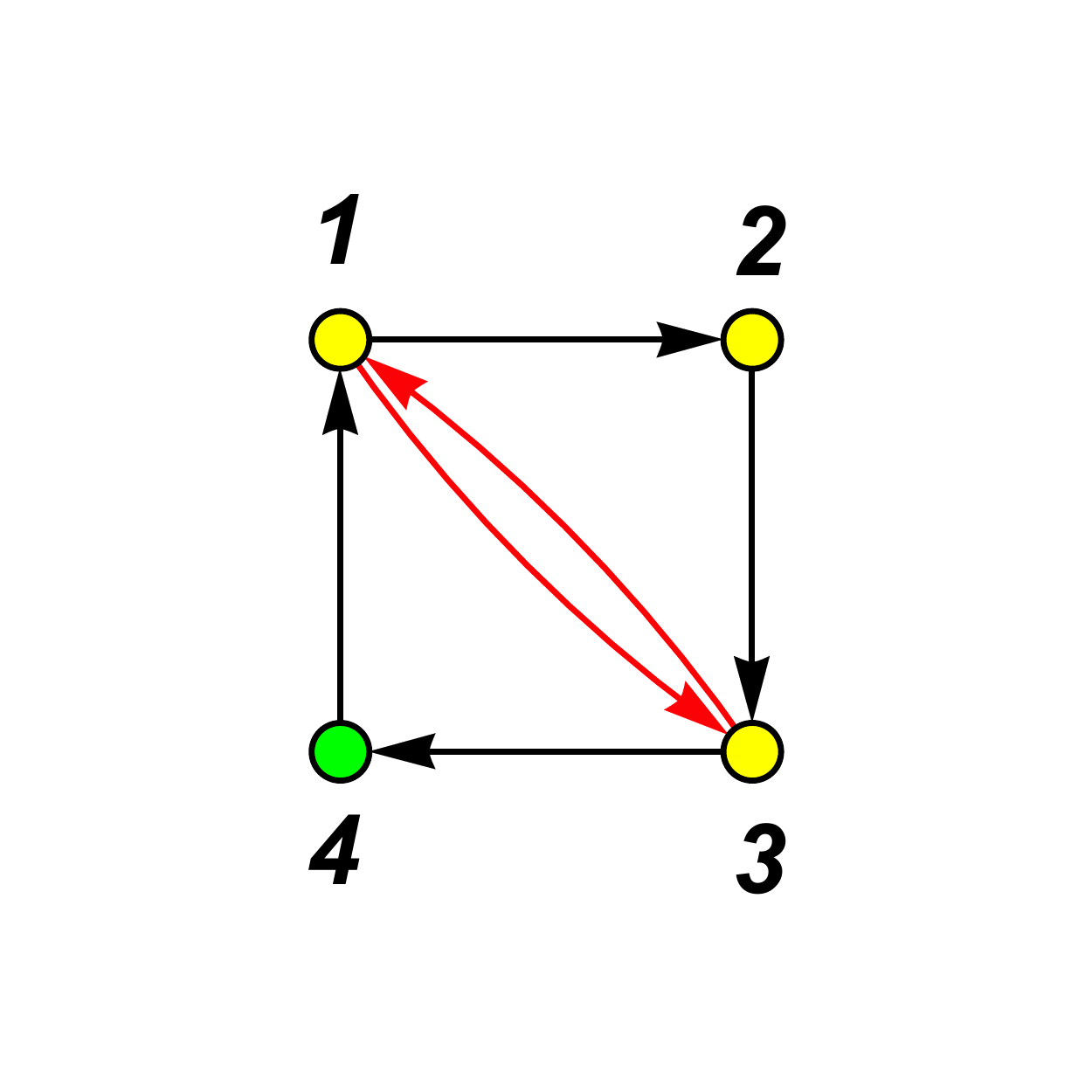}} =
\hspace{-0.5cm}
\parbox[c]{5.4em}{\includegraphics[scale=0.21]{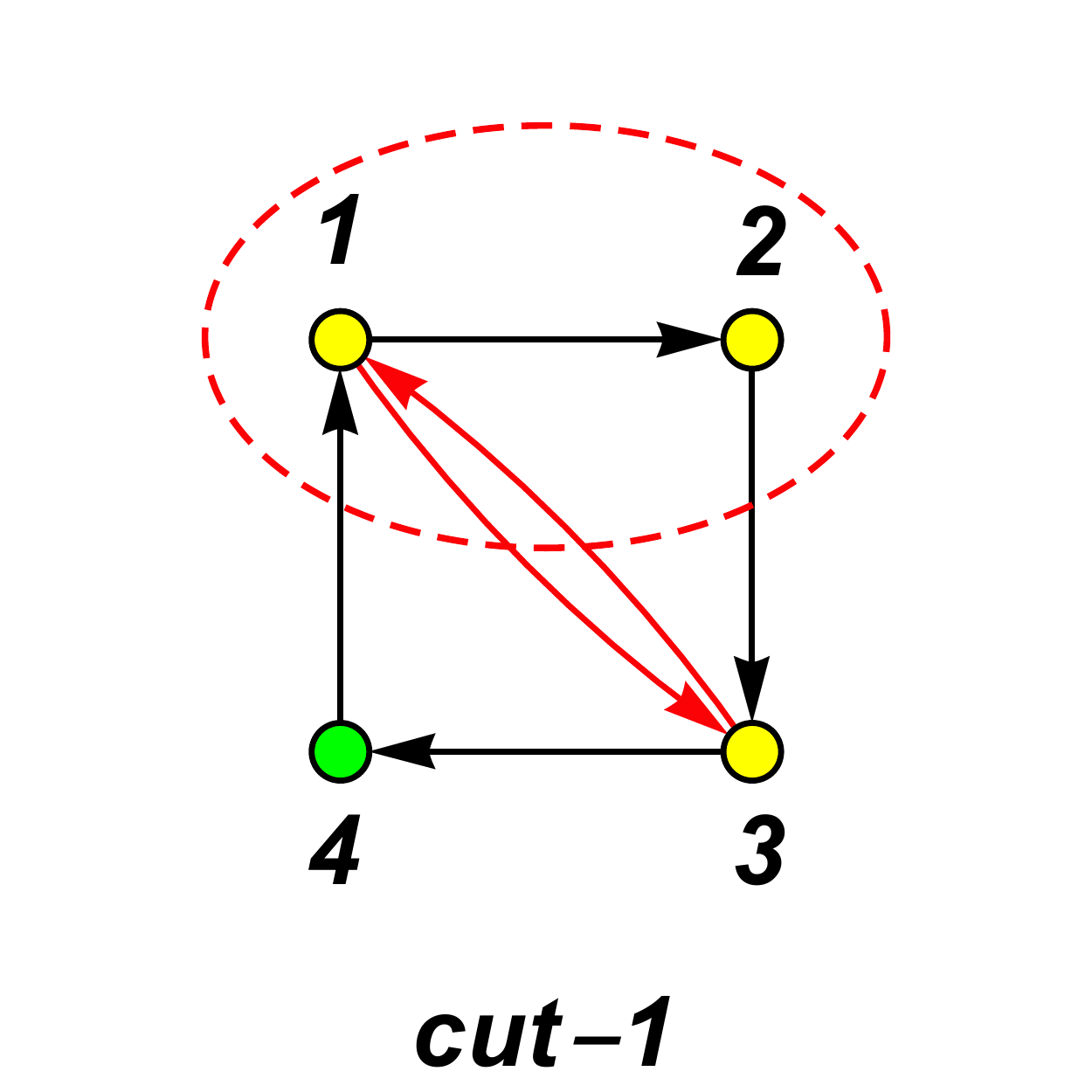}} +
\hspace{-0.5cm}
\parbox[c]{6.3em}{\includegraphics[scale=0.21]{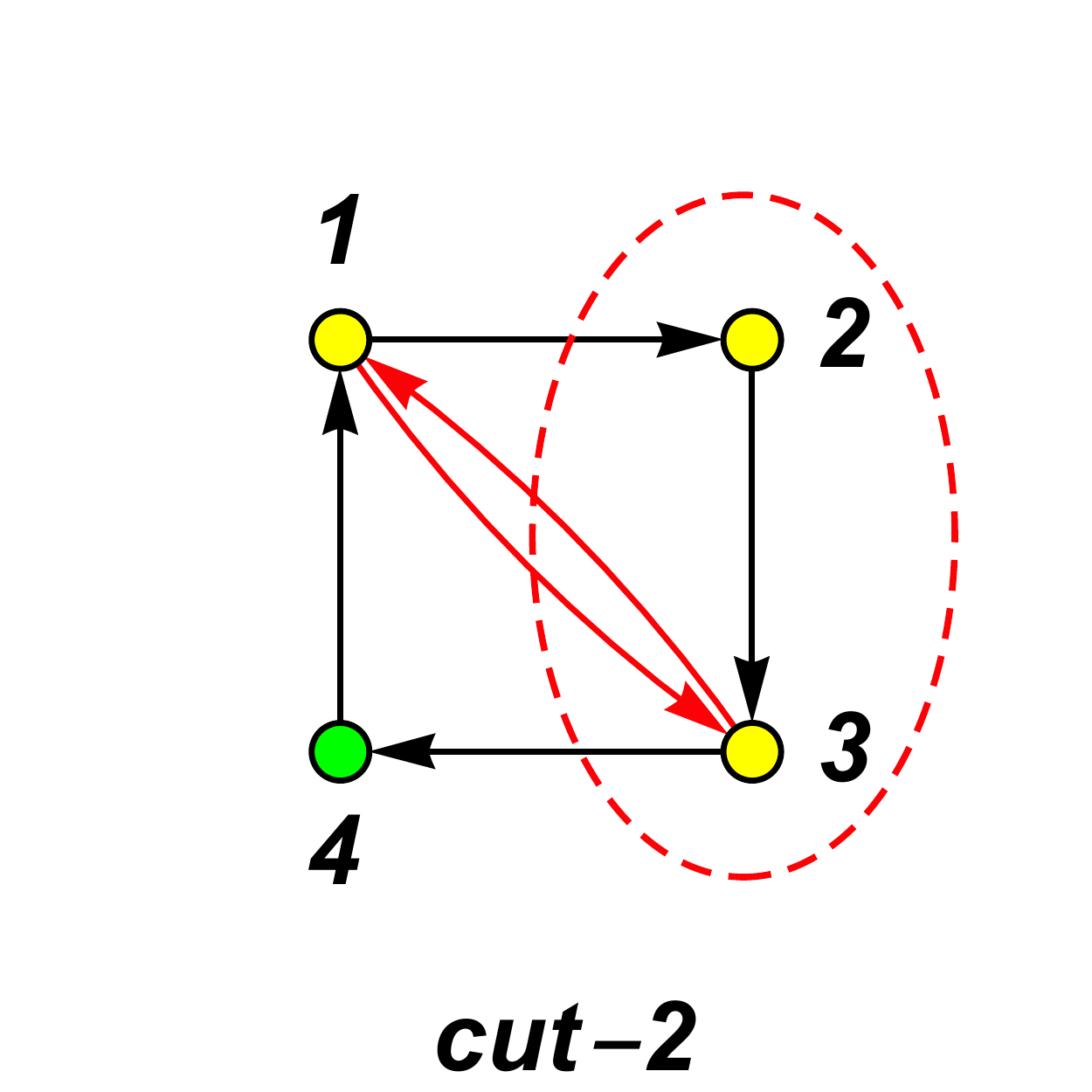}} +
\hspace{-0.44cm}
\parbox[c]{5.9em}{\includegraphics[scale=0.21]{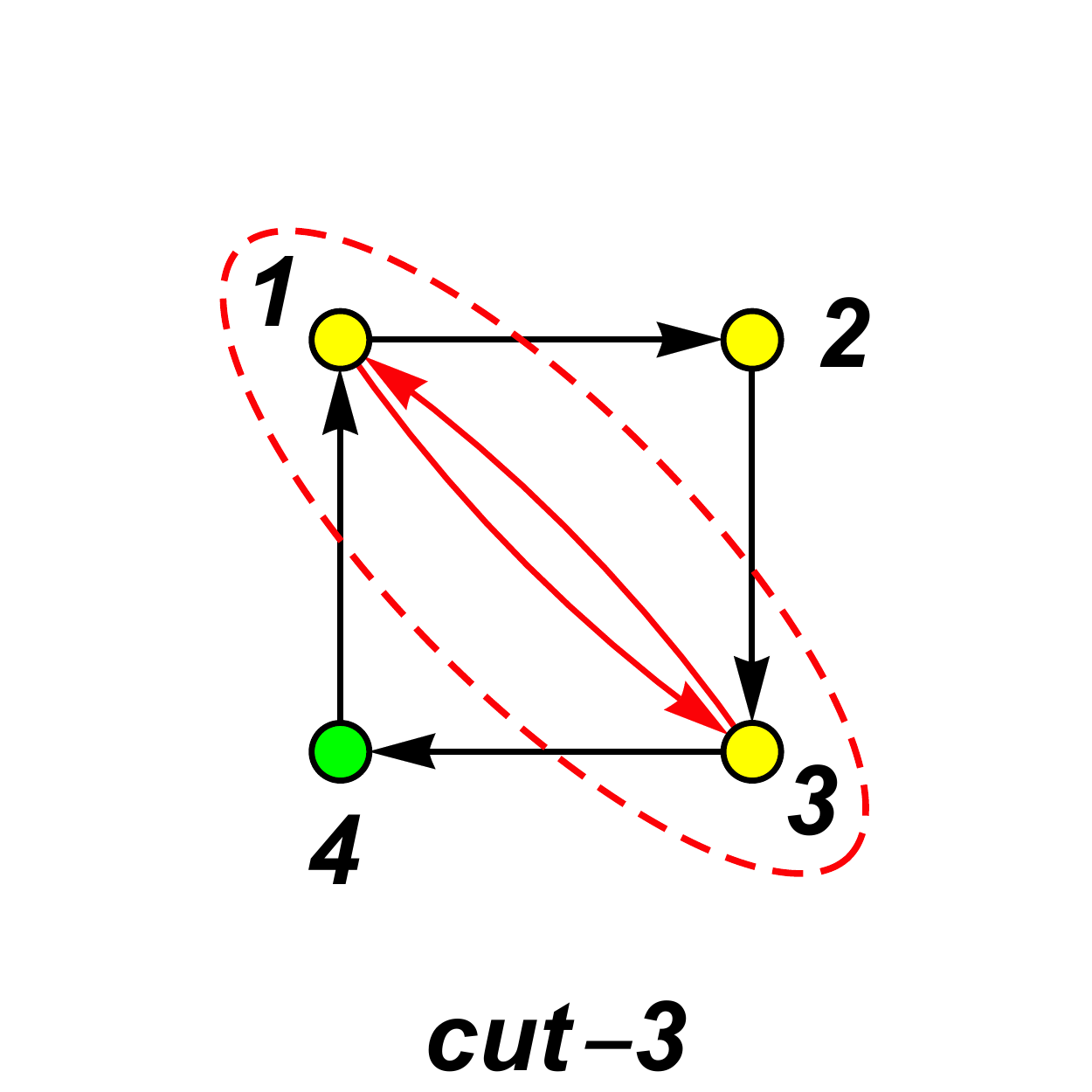}}  .~~
\end{eqnarray}
}
\vskip-0.3cm\noindent
By applying {\bf rule-III}, we can evaluate {\it cut-1}, finding 
\vspace{-0.3cm}
{\small
\begin{eqnarray}\label{4p-cut-1}
\parbox[c]{5.7em}{\includegraphics[scale=0.21]{4pts-Gg-cut1.pdf}} &=&
\hspace{-0.5cm}
\parbox[c]{5.8em}{\includegraphics[scale=0.23]{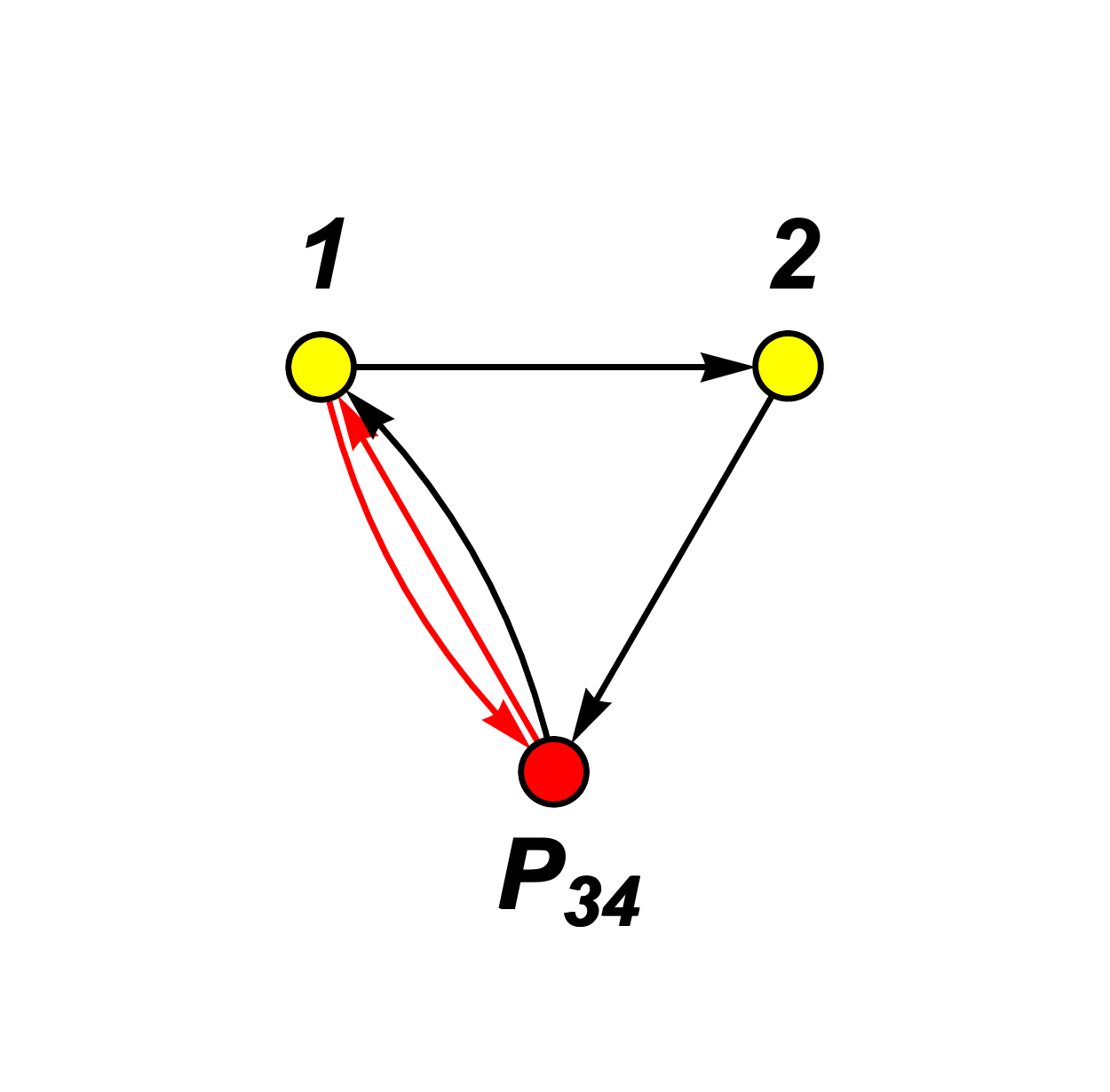}}
\times \left(\frac{1}{ s_{34}}\right)\times
\hspace{-0.7cm}
\parbox[c]{5.8em}{\includegraphics[scale=0.23]{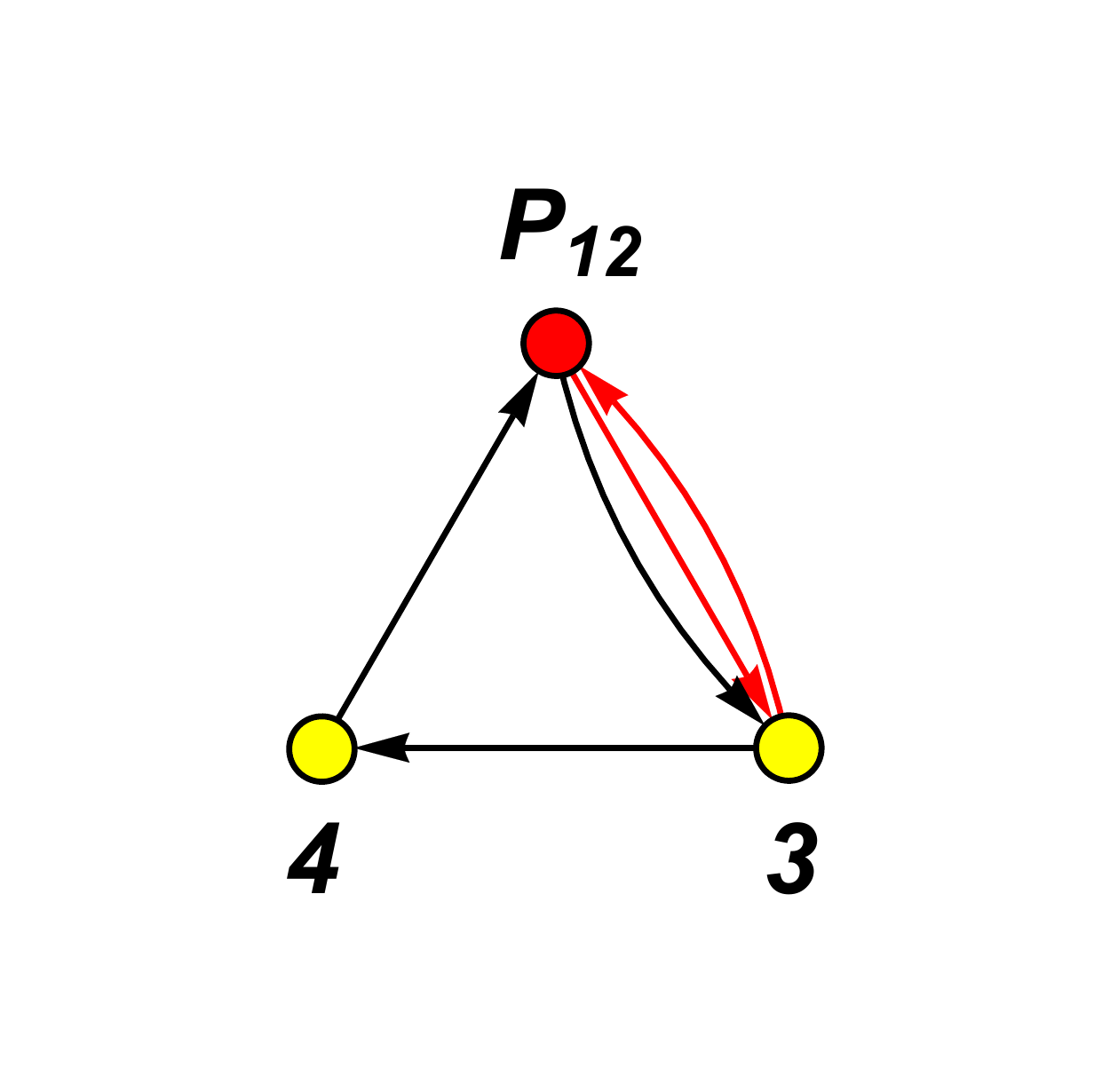}} 
 = \frac{A_3 (P_{34},1,2) \times A_3 (P_{12}, 3 ,4)} { s_{34}}~ =~0, \qquad~~~
\end{eqnarray}
}
\vskip-0.3cm\noindent
where we have used eq.~\eqref{Aodd}. {\it Cut-2} can be evaluated in a similar manner.
Finally, it is straightforward to see that the last cut ({\it cut-3}) is broken into
\vspace{-0.5cm}
{\small
\begin{eqnarray}\label{4p-cut-3}
\parbox[c]{5.8em}{\includegraphics[scale=0.21]{4pts-Gg-cut3.pdf}}=
\hspace{-0.55cm}
\parbox[c]{5.8em}{\includegraphics[scale=0.21]{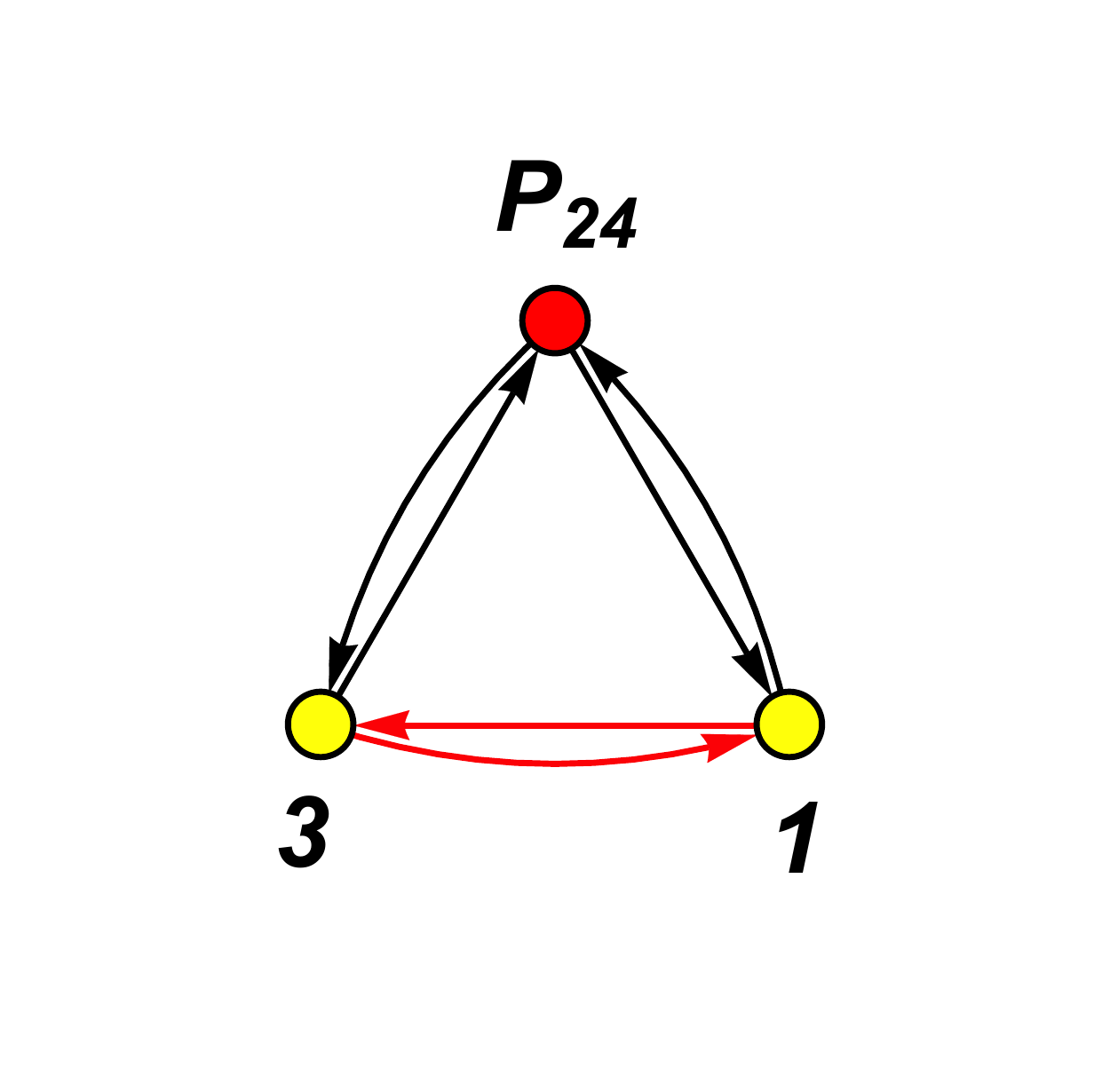}}
\times \left(\frac{1}{s_{24}}\right)\times
\hspace{-0.55cm}
\parbox[c]{6.5em}{\includegraphics[scale=0.21]{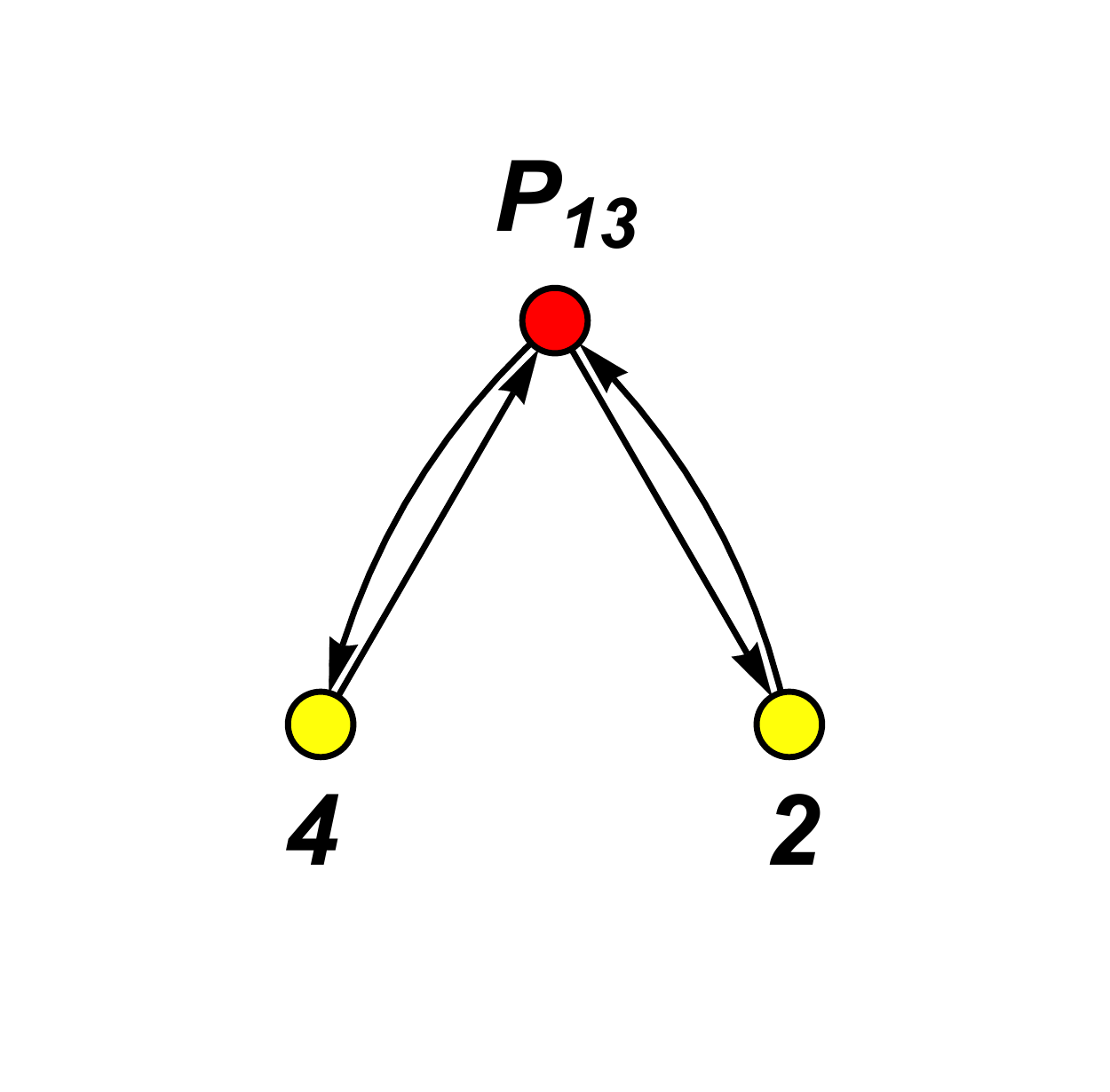}} .  
\end{eqnarray}
}
\vskip-0.2cm\noindent
From the normalization of the three-point function in \cref{3pt-Nor}, the first graph evaluates to $(-1)$, while the second is (using {\bf rule-III})
\vspace{-0.3cm}
{\small
\begin{eqnarray}\label{4p-cut-3-2}
 \hspace{-0.4cm}
   \parbox[c]{5.3em}{\includegraphics[scale=0.21]{4pts-cut2-2.pdf}} =\,\,\frac{(\s_{P_{13}2}\, \s_{24}\, \s_{4P_{13}} )^2}{(\s_{P_{13}2} \, \s_{2P_{13}}) \times (\s_{P_{13}4} \s_{4 P_{13}})}   \times {\rm det}
   \left[\displaystyle
   \begin{matrix}
    0 & \displaystyle\frac{ s_{24}}{\s_{24}}\\
   \displaystyle \frac{ s_{24}}{\s_{42}} & 0
   \end{matrix}
   \right]
 =   s_{24}^2\, . 
\end{eqnarray}
}
We can also rewrite the cut using matrix relations defined in \cref{Pf-properties},
\begin{align}
	\text{\it cut-3} = 
	-\frac{ A_3^{\prime}(P_{13}, 2, 4) A_3^{\prime}( 1, 3,P_{24})}{s_{24}}.
\end{align}
By evaluating the cuts, we have that
\begin{align}
	A_4(\mathbb{I}_4^{(13)}) =& 
	\frac{A_3(P_{34}, 1,2) A_3(P_{12}, 3,4)}{s_{34}} 
	+\frac{A_3(P_{23}, 1,4) A_3(3,P_{14}, 2)}{s_{23}} 
	\nonumber \\
	&-\frac{ A_3^{\prime}(P_{13}, 2, 4) A_3^{\prime}( 1, 3,P_{24})}{s_{24}}
	\nonumber \\
	=& -\frac{ A_3^{\prime}(P_{13},2, 4)  A_3^{\prime}( 1, 3,P_{24})}{s_{24}}
	= -\frac{\left(-s_{13}\right)\left(-s_{24}\right)}{s_{24}}=-s_{13}.
	\label{eq:fourpntold}
\end{align}
Here we have used \cref{Aodd,eq:3pointmass} when evaluating the amplitude.
Notice that the factorization channels with poles $s_{34}$ and $s_{23}$ vanish
because they factorize into an odd NLSM amplitude, see \cref{Aodd}.
The last contribution does not vanish, as it is not the usual NLSM prescription, but rather
an off-shell amplitude with the new prescription given in \cref{eq:NLSMnew}.
Of course, the subamplitudes would vanish if all particles, including intermediate particles, 
were on-shell. In particular if
$P_{24}$ was on-shell (collinear limit). We can see this reflected by the answer, which would vanish in that case.

\subsubsection{The New Integrand Prescription}

In the previous section, we expressed the factorized non-linear sigma model amplitude with the 
usual prescription in terms of lower-point amplitudes with the new prescription.
In this section we are going to do the calculations using the new prescription.

Let us consider the four-point amplitude, with gauge fixing $(pqr|m)=(123|4)$. 
In order to get a better understanding of the method, we are going to choose
two different reduced determinants, {\it i.e.} we consider removing columns and rows such that
$(ijk)=(123)$ in the first example, and $(ijk)=(134)$ in the second example.
In the first example, we have the graphical representation
\vspace{-0.5cm}
{\small
\begin{eqnarray}\label{new-exp1}
	A_4^{\prime}( \mathbb{I}_4 ) = 
\int d\mu_{4}^\L  
\hspace{-0.5cm}
\parbox[c]{5.2em}{\includegraphics[scale=0.21]{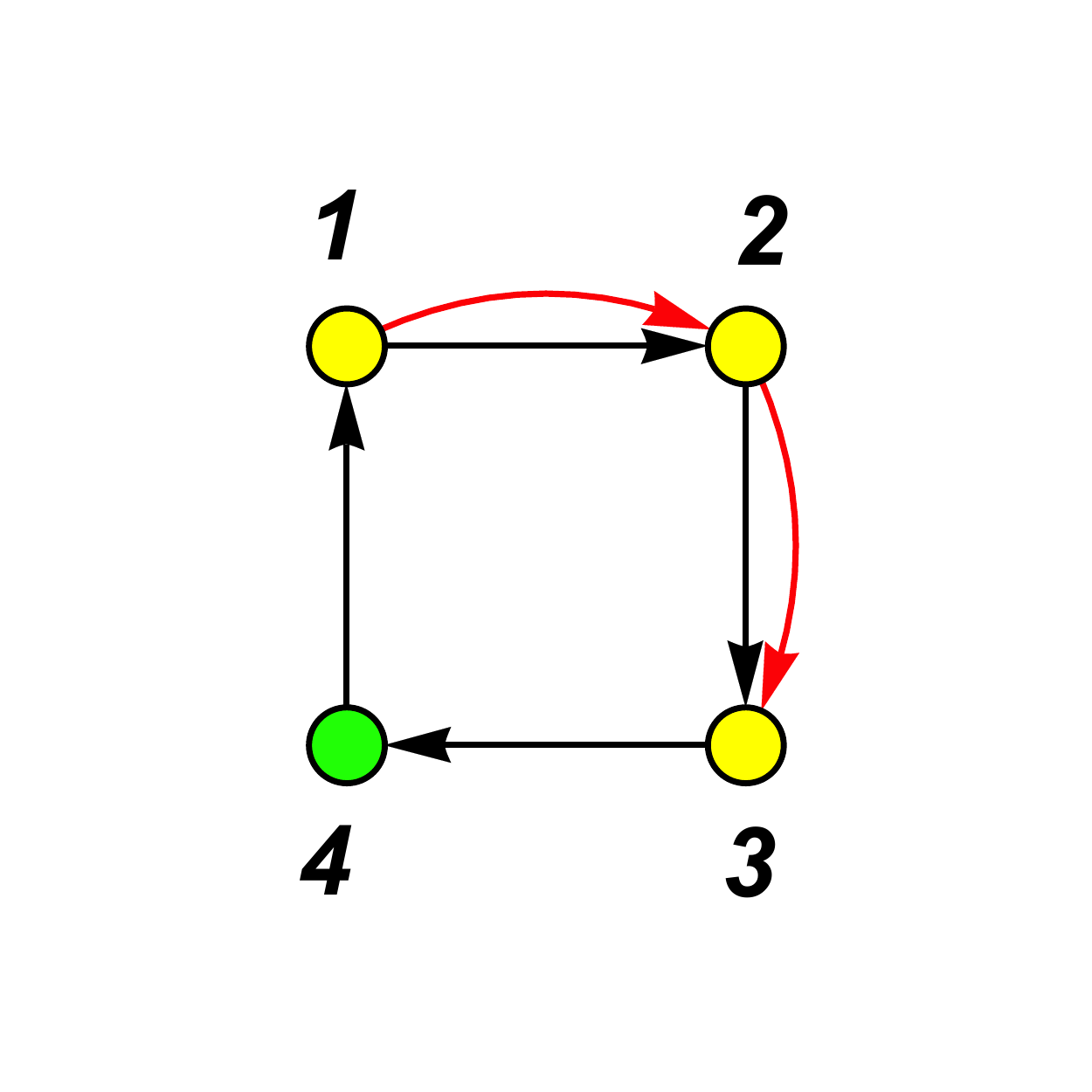}} =
\hspace{-0.5cm}
\parbox[c]{5.9em}{\includegraphics[scale=0.21]{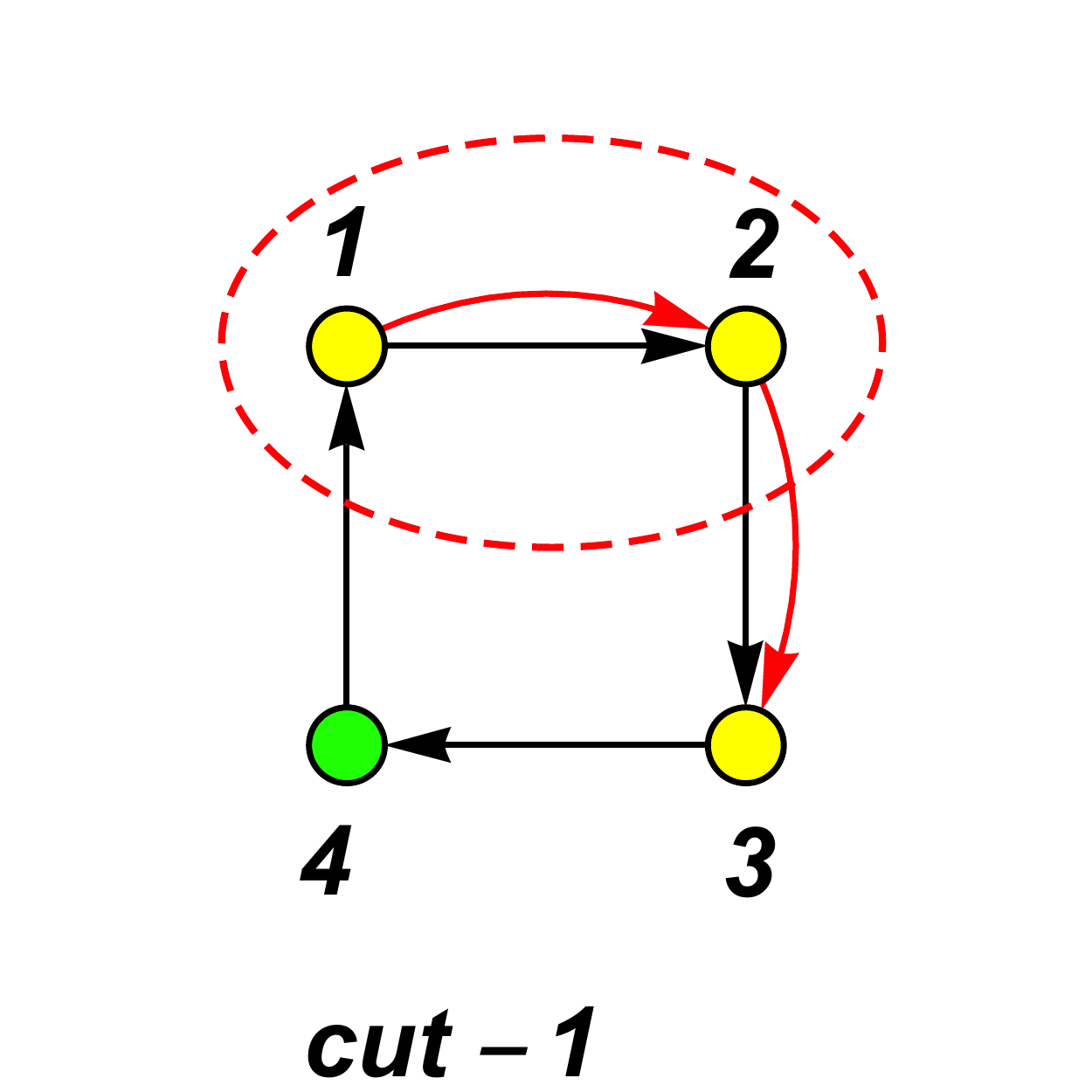}} +
\hspace{-0.5cm}
\parbox[c]{6.6em}{\includegraphics[scale=0.21]{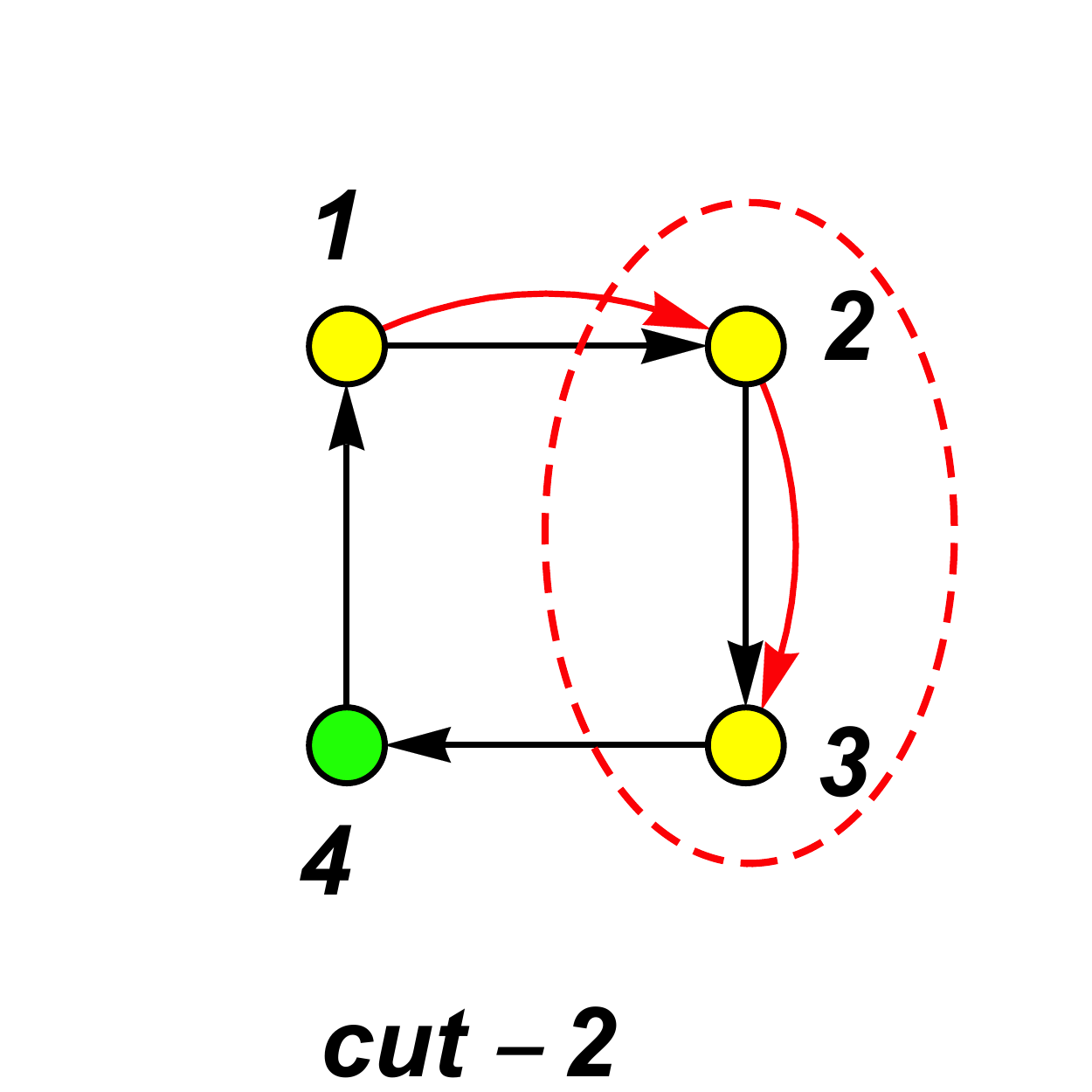}} .
\end{eqnarray}
}
\vskip-0.4cm\noindent 
The graphs can be evaluated as
\begin{align}
	\label{eq:fourpntnew1}
	 A_4^{\prime}\left(\mathbb{I}_4\right) = \sum_M \left[
		\frac{ A_3^{\prime}(  1, 2,P_{34}^M) A_3^{(P_{12}3)}(P_{12}^M,3,4)}{s_{34}}
	+ \frac{A_3^{(1P_{23})}(1,P_{23}^M,4)  A_3^{\prime}(P_{41}^M, 2, 3)}{s_{41}}\right].
\end{align}
We see that all factorization contributions are glued together by an off-shell vector field
(off-shell gluon). The notation $P_i^M$ means the replacement $P_i^\mu \rightarrow \frac{1}{\sqrt{2}}
\epsilon_i^{M\, \mu}$ in the $\mathsf{A}_n$ matrix. Also, the gluing relation is
\begin{align}
	\label{eq:epsM}
	\sum_M \epsilon_i^{M\, \mu} \epsilon_j^{M\, \nu} = \eta^{\mu\nu}.
\end{align}
Explicitly, the two factorization contributions become
\begin{align}
	\sum_M \frac{ A_3^{\prime}(  1, 2,P_{34}^M) A_3^{(P_{12}3)}(P_{12}^M,3,4)}{s_{34}}
	=\sum_M \frac{\left(\sqrt{2} \epsilon_{34}^M \cdot k_1\right) \times s_{34}\left(\sqrt{2}
	\epsilon_{12}^M \cdot k_4\right)}{s_{34}} = \frac{s_{14} s_{34}}{s_{34}} = s_{14},
\end{align}
and 
\begin{align}
	\sum_M \frac{ A_3^{(1P_{23})}( 1,P_{23}^M,4)  A_3^{\prime}(P_{41}^M, 2, 3)}{s_{23}}
	=\sum_M \frac{\left(\sqrt{2} \epsilon_{23}^M \cdot k_4\right) s_{41}\times \left(\sqrt{2}
	\epsilon_{41}^M \cdot k_3\right)}{s_{23}} = \frac{s_{14} s_{34}}{s_{23}} = s_{12}.
\end{align}
As a second example, consider
\vspace{-0.6cm}
{\small
\begin{eqnarray}\label{new-exp2}
	A_4^{\prime}( \mathbb{I}_4^{(134)} ) = 
\int d\mu_{4}^\L  
\hspace{-0.5cm}
\parbox[c]{5.2em}{\includegraphics[scale=0.21]{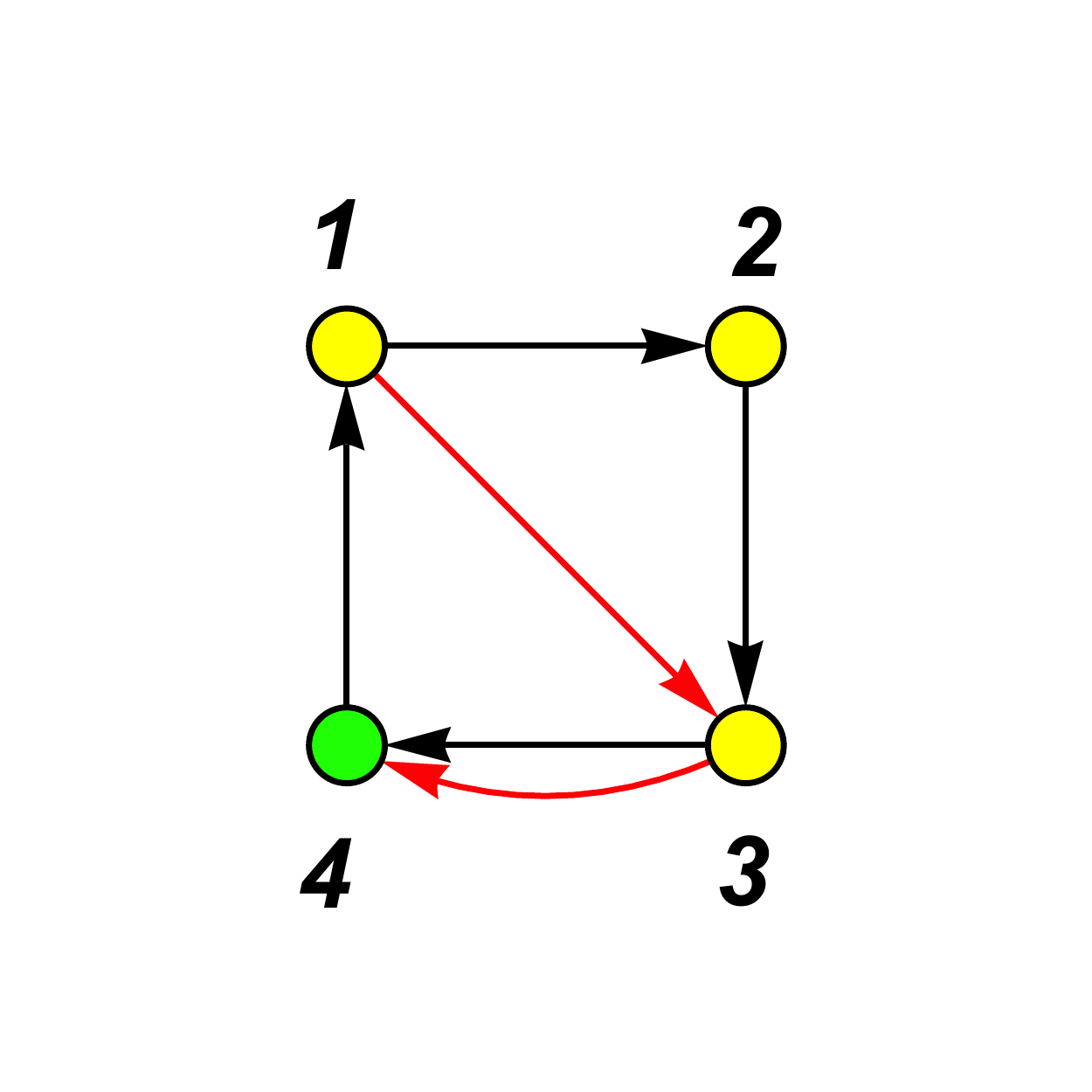}} =
\hspace{-0.5cm}
\parbox[c]{5.6em}{\includegraphics[scale=0.21]{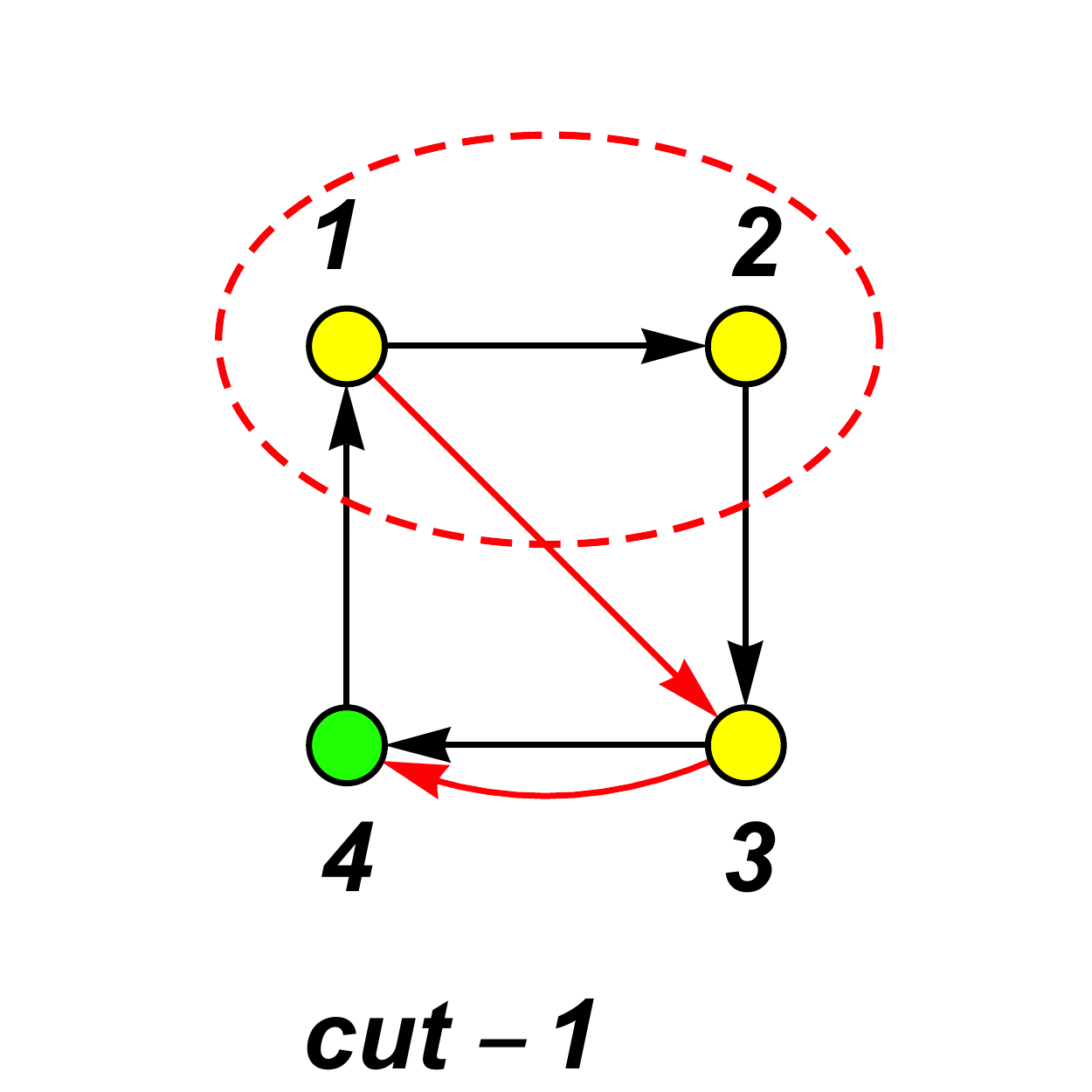}}  
+
\hspace{-0.5cm}
\parbox[c]{6.9em}{\includegraphics[scale=0.21]{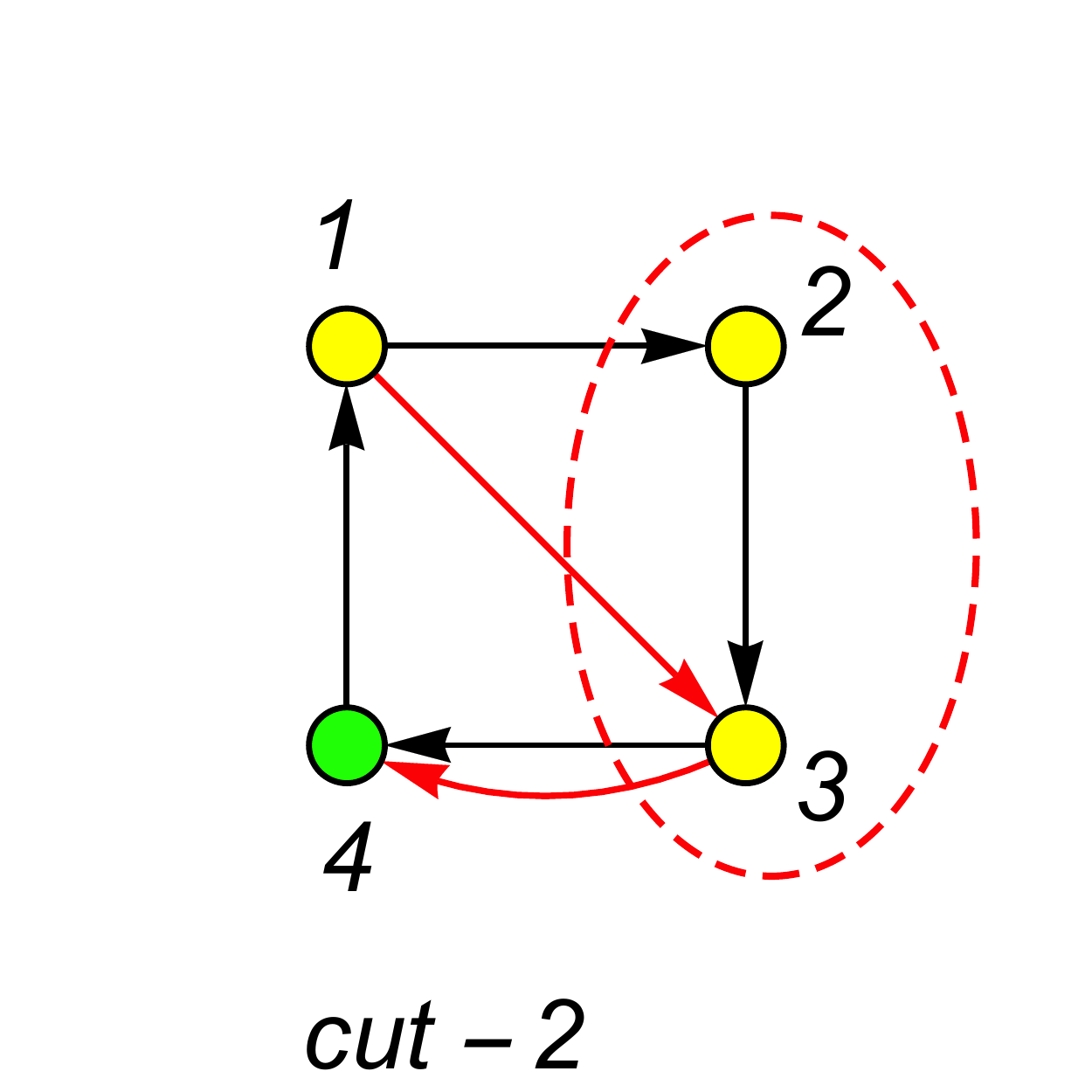}}  
.~~
\end{eqnarray}
}
\vskip-0.3cm\noindent
The graphs evaluate to
\begin{align}
	\label{eq:fourpntnew2}
	 A_4^{\prime}(\mathbb{I}_4^{(134)}) = 
	\sum_M \frac{ A_3^{(1P_{34})}(1,2,P_{34}^M)  A_3^{\prime}(P_{12}^M, 3, 4)}{s_{34}}
	+ \frac{ A_3^{\prime}( 1,P_{23}, 4) A_3( 3,P_{41},2)}
	{s_{23}}
\end{align}
Notice that only one of the factorization contributions ({\it cut-1}) is glued together by an off-shell gluon,
while the second contribution ({\it cut-2}) is a purely scalar contribution. Evaluating the contributions, we 
find that
\begin{align}
	\sum_M \frac{ A_3^{(1P_{34})}(1,2,P_{34}^M)  A_3^{\prime}(P_{12}^M, 3, 4)}{s_{34}}
	= \sum_M \frac{-\left(\sqrt{2} \epsilon_{34}^M \cdot k_2\right)s_{12} \times \left(\sqrt{2}
	\epsilon_{12}^M\cdot k_4\right)}{s_{34}} &= -\frac{s_{12}s_{24}}{s_{34}} = -s_{13},
\end{align}
and
\begin{align}
	\frac{ A_3^{\prime}( 1,P_{23}, 4) A_3( 3,P_{41},2)}
	{s_{23}} = \frac{P_{23}^2 \times 0}{s_{23}} = 0.
\end{align}
The scalar contribution vanishes, as an odd amplitude in the usual prescription vanishes, see 
\cref{Aodd}.

Summing the contributions, we obtain
\begin{align}
	 A_4^{\prime}(\mathbb{I}_4^{(123)}) = s_{14} + s_{12} = -s_{13}, \\
	 A_4^{\prime}(\mathbb{I}_4^{(134)}) = -s_{13} + 0 = -s_{13}.
\end{align}
This agrees with \cref{eq:fourpntold}.

\subsection{Six-Point}

Next, we compute the six-point amplitude using the double-cover formalism. We stick to the 
gauge fixing $(pqr|m)=(123|4)$, and to removing the columns and rows $(i,j)=(1,3)$.
Graphically, the amplitude factorizes into
\vspace{-0.3cm}
{\small
\begin{eqnarray}\label{cuts-6pts-1}
	A_6( \mathbb{I}_6^{(13)} ) =
\int d\mu_6^\L
\hspace{-0.25cm}
 \parbox[c]{6.1em}{\includegraphics[scale=0.21]{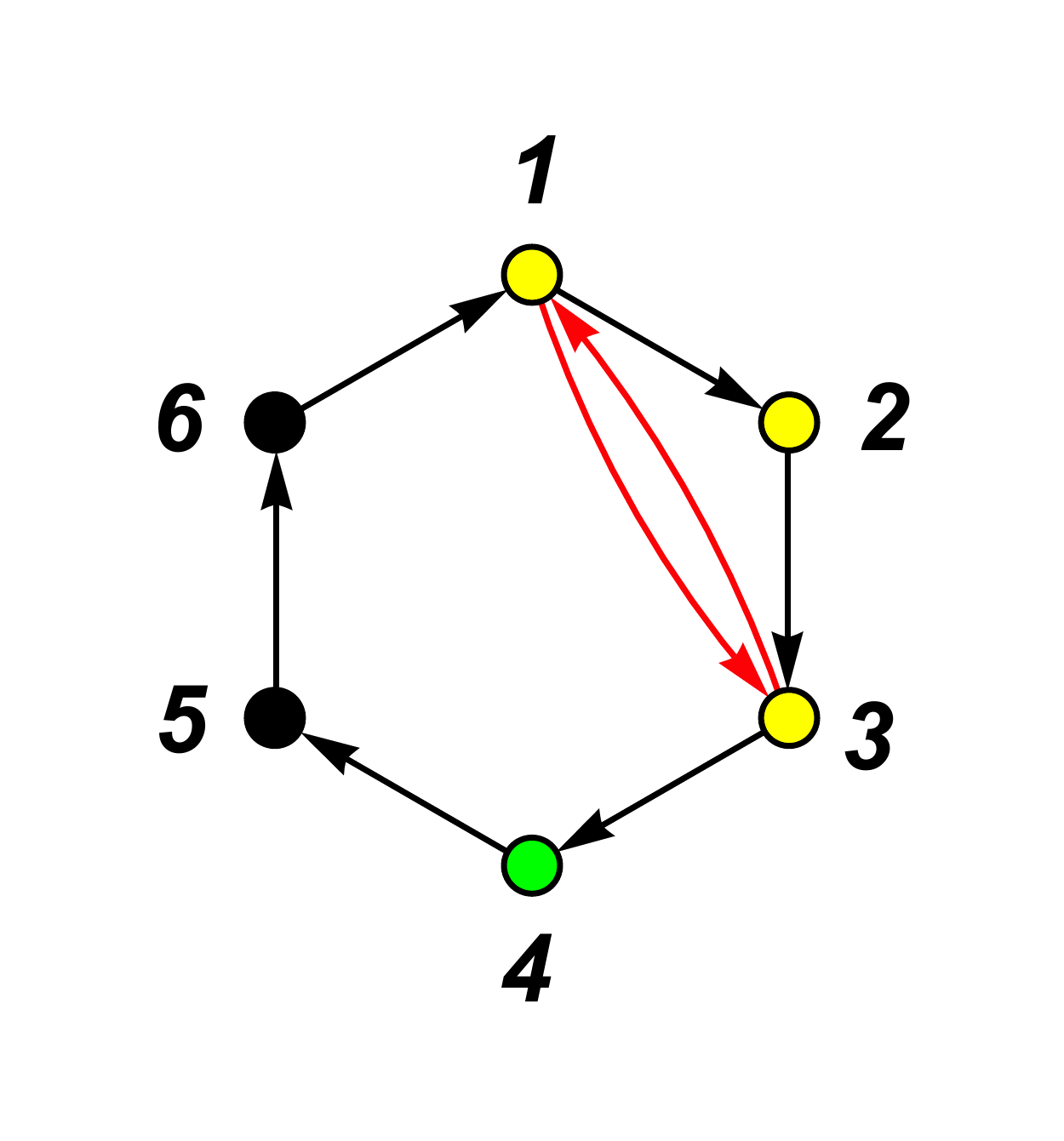}}
=
\hspace{-0.3cm}
 \parbox[c]{6.1em}{\includegraphics[scale=0.21]{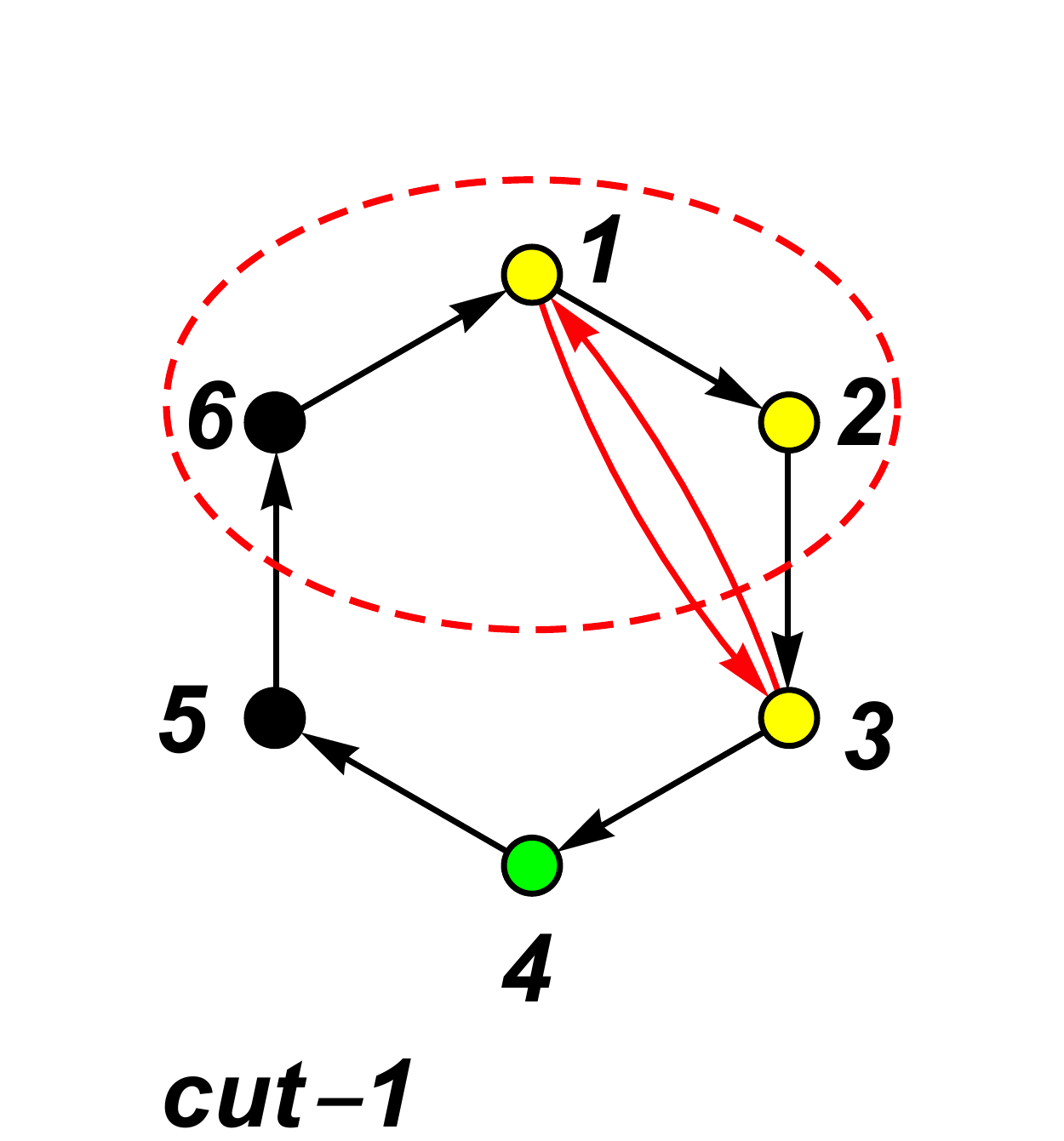}} +
 \hspace{-0.3cm}
  \parbox[c]{6.1em}{\includegraphics[scale=0.21]{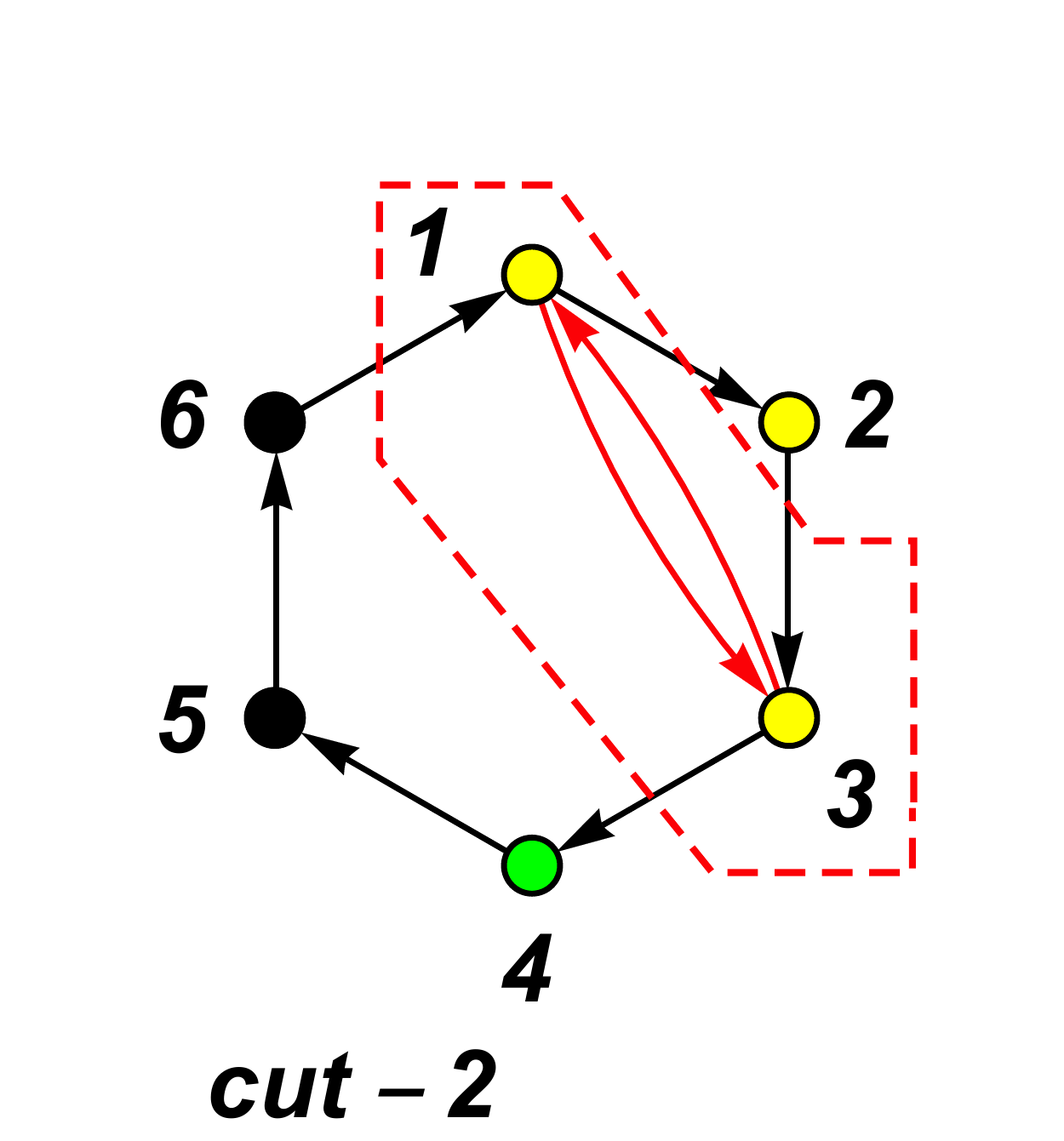}} +
  \hspace{-0.25cm}
   \parbox[c]{6.1em}{\includegraphics[scale=0.21]{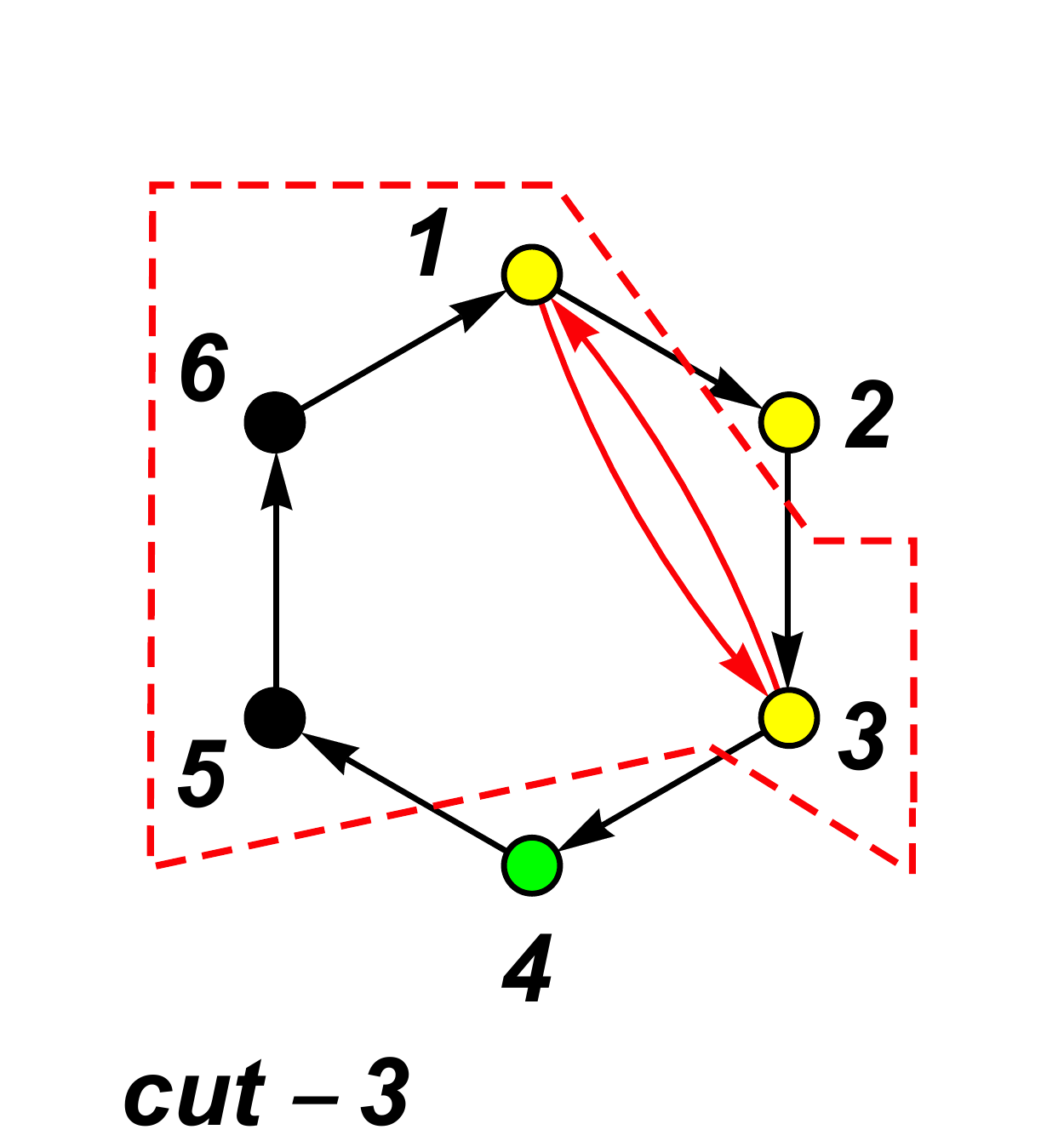}} ~ .
   \qquad
\end{eqnarray}
}
\vskip-0.15cm\noindent
We have omitted some factorizations, which evaluate to zero by analogy to the four-point case.
Note that, the  {\it cut-1} is straightforward to evaluate, as it factorizes into lower-point NLSM amplitudes.
However, {\it cut-2} and {\it cut-3} do not have straightforward interpretations
(which is why they sometimes are referred to as {\it strange-cuts}). Take {\it cut-2} as an 
example, it graphically takes the form
\vspace{-0.3cm}
{\small
\begin{eqnarray}
\parbox[c]{6.2em}{\includegraphics[scale=0.21]{gauge1-cut2.pdf}}=
\int d\mu_5^{\rm CHY}
 \hspace{-0.25cm}
  \parbox[c]{5.8em}{\includegraphics[scale=0.21]{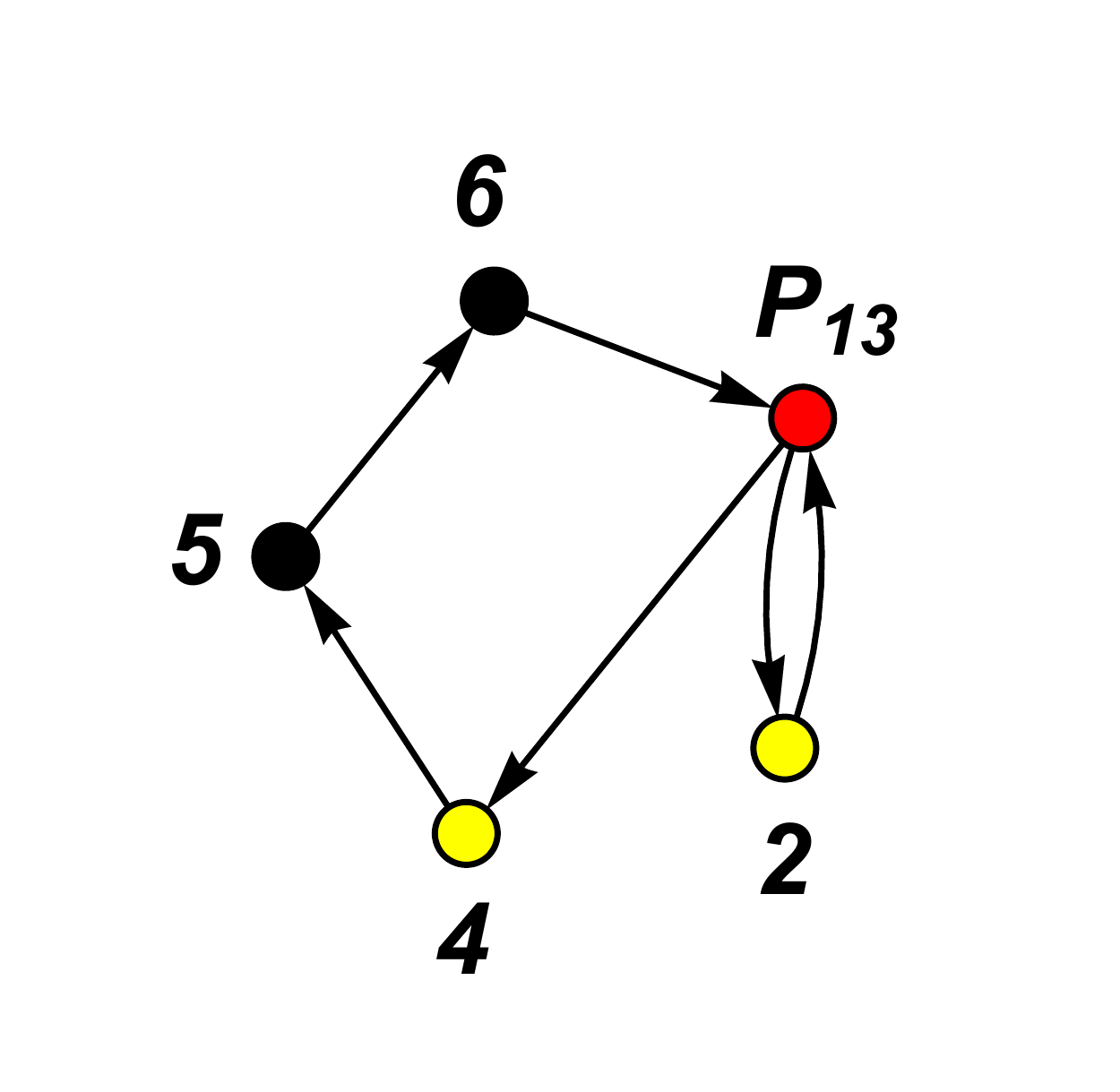}}
\times \left(  \frac{1}{s_{4:6,2}}  \right) \times 
\hspace{-0.65cm}
   \parbox[c]{6.0em}{\includegraphics[scale=0.21]{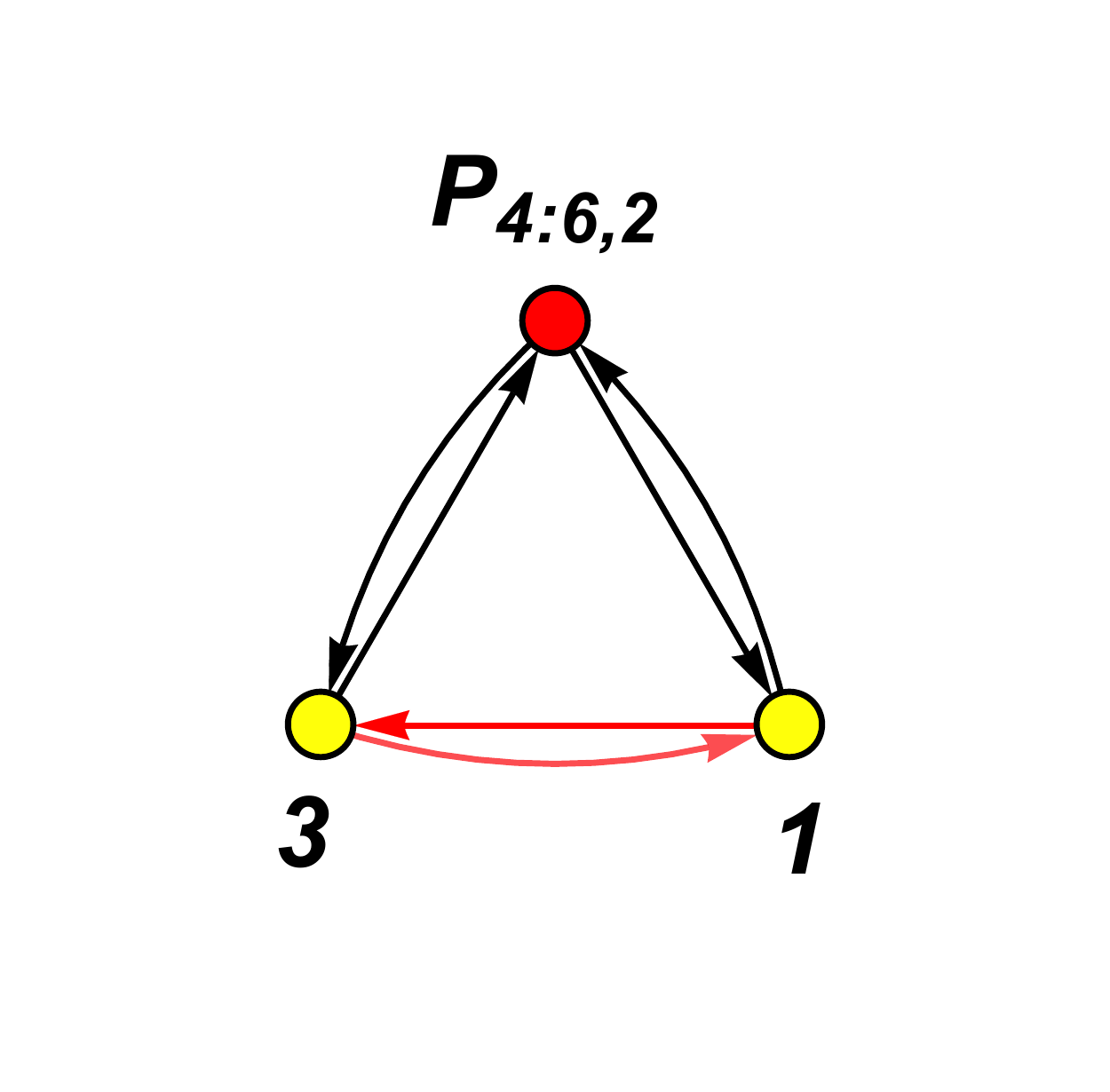}} .
\end{eqnarray}
}
\vskip-0.2cm\noindent
The first graph looks non-simple to be computed since there is no way to avoid the 
{\it singular cuts}. Nevertheless, such as in Yang-Mills theory, Ref.~\cite{Gomez:2018cqg}, 
this strange-cut can be rewritten in the following way  
\vspace{-0.3cm}
{\small
\begin{eqnarray}
\int d\mu_5^{\rm CHY}
 \hspace{-0.25cm} 
\parbox[c]{5.8em}{\includegraphics[scale=0.21]{G1-C2-RG1.pdf}}
\times 
\hspace{-0.65cm}
   \parbox[c]{5.8em}{\includegraphics[scale=0.21]{G1-C2-RG2.pdf}} 
& =& (-1)\,
A^{\prime}_5(P_{13},2,4,5,6) \times  A^{\prime}_3( 1,3,P_{4:6,2})  ,\,\, ~~     
\end{eqnarray}
}
\vskip-0.2cm\noindent
which comes from the matrix identities given in \cref{Pf-properties}.
We can do a similar rewriting for {\it cut-3}.  The full calculation is presented in \cref{sec:sixpointex3}.

Putting it all together, the six-point amplitude factorizes as
\begin{align}
	A_6(\mathbb{I}_6^{(13)}) =&
	\frac{A_4({\bm 1},2,\bm{P_{3:5}},6) A_4(P_{6:2}, 3,4,5)}
	{s_{3:5}}
	- \frac{ A_5^{\prime}(P_{13}, 2, 4,5,6) A_3^{\prime}( 1,
	 3,P_{4:6,2})}{s_{13}}
	\nonumber \\
	&- \frac{ A_3^{\prime}(P_{5:1,3}, 2, 4) A_5^{\prime}( 1,
	 3,P_{24},5,6)}{s_{24}}
	\label{eq:sixpntold}
	\nonumber \\
	=& \frac{s_{26}s_{35}}{s_{3:5}} + s_{13}\left[ \frac{s_{46}}{s_{4:6}} + \frac{s_{26}+s_{46}}{s_{56P_{13}}}\right] + s_{24}\left[\frac{s_{26}+s_{46}}{s_{56P_{24}}} + \frac{s_{26}
	+s_{36}+s_{46}}{s_{5:1}}\right].
\end{align}
By using momentum conservation, all unphysical poles cancel, and we match with the known result
\begin{align}
	A_6\left(\mathbb{I}_6\right) =& \frac{(s_{12}+s_{23})(s_{45}+s_{56})}{s_{123}} 
	+ \frac{(s_{23}+s_{34})(s_{56}+s_{61})}{s_{234}} 
	+ \frac{(s_{34}+s_{45})(s_{56}+s_{61})}{s_{345}} 
	\nonumber \\ 
	&- (s_{12} + s_{23} + s_{34} + s_{45} + s_{56} + s_{61}).
\end{align}

The six-point amplitude can also be computed using the new prescription.
The first example with the choice $(ijk)=(123)$ gives, graphically,
\vspace{-0.3cm}
{\small
\begin{eqnarray}\label{cuts-6pts}
	A_6^{\prime}(\mathbb{I}_6^{(123)} ) =
\hspace{-0.15cm}
 \parbox[c]{6.1em}{\includegraphics[scale=0.19]{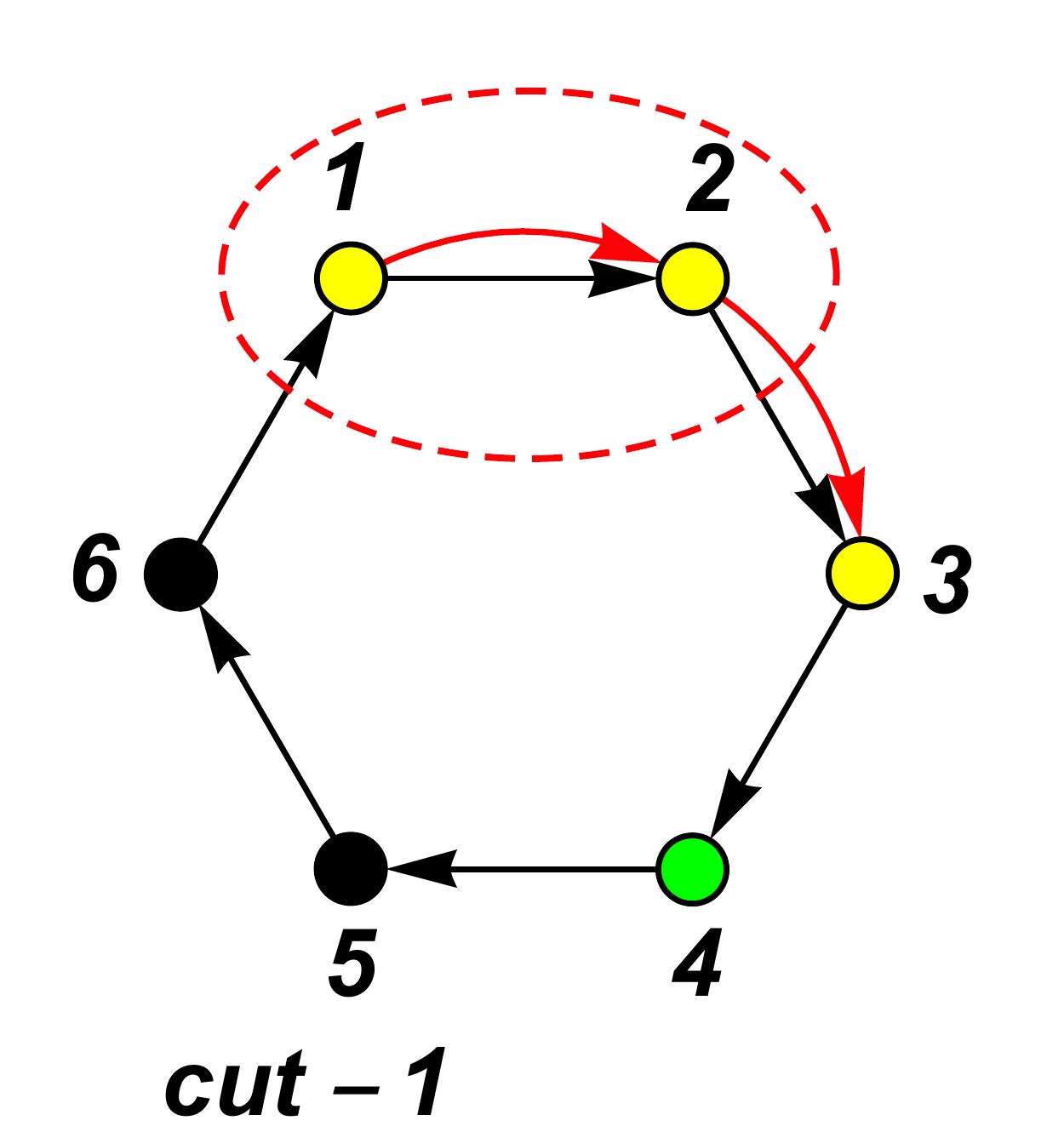}} +
 \hspace{-0.1cm}
  \parbox[c]{7.1em}{\includegraphics[scale=0.19]{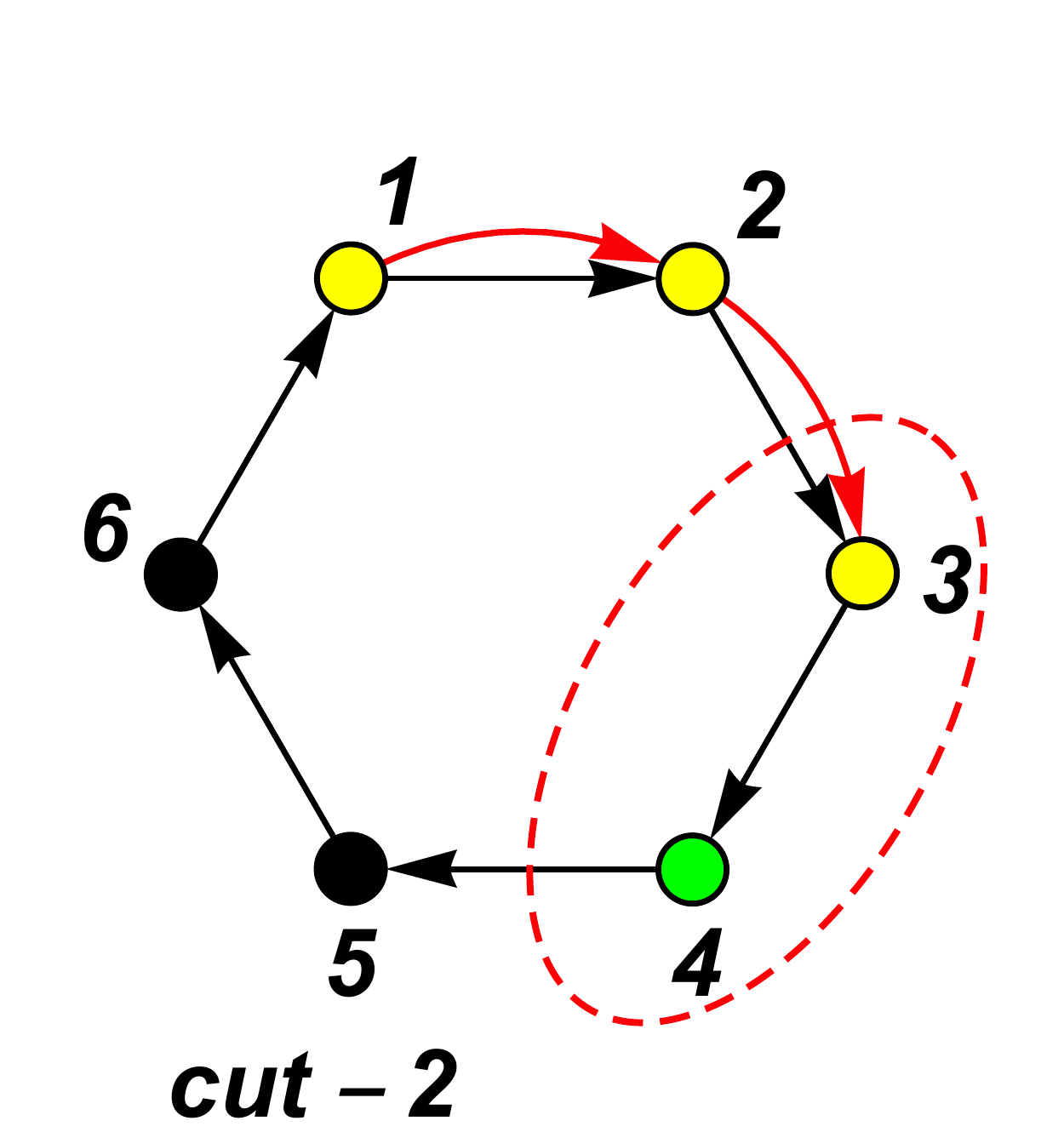}}
 \hspace{-0.36cm}  
  +
   \parbox[c]{6.1em}{\includegraphics[scale=0.19]{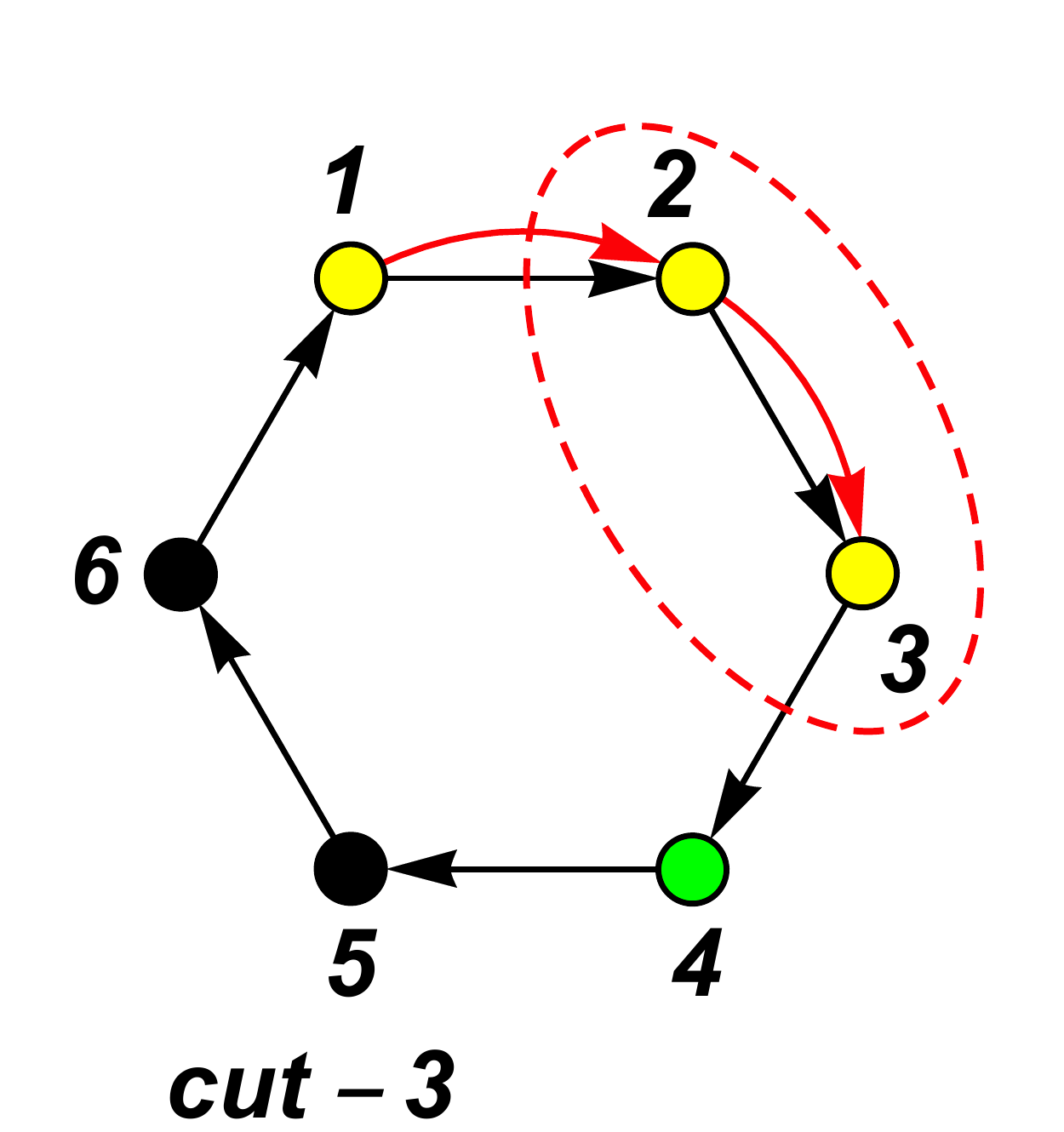}} 
  +
  \hspace{-0.05cm}
   \parbox[c]{6.1em}{\includegraphics[scale=0.19]{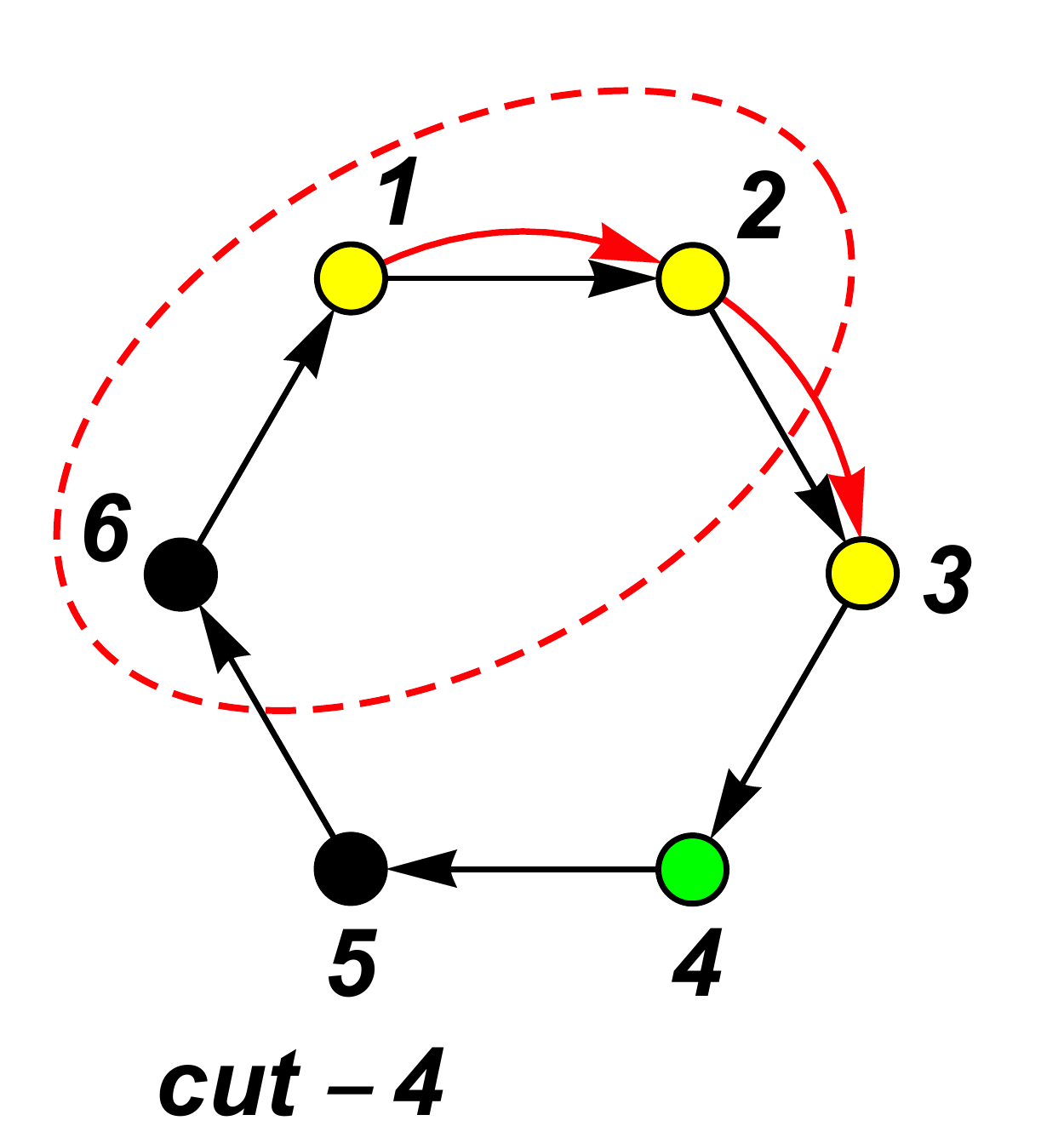}}    
   ~ .
   \qquad
\end{eqnarray}
}
\vskip-0.15cm\noindent
We have carried out the full calculation in \cref{sec:sixpointex1}.
The contributions unambiguously evaluate to
\begin{align}
	\label{eq:sixpntnew1}
	& A_6^{\prime}(\mathbb{I}_6^{(123)}) =\\& \sum_M \left[
		\frac{ A_3^{\prime}\left( 1, 2,P_{3:6}^M\right)
		A_5^{(P_{12}3)}(P_{12}^M,3,4,5,6)}{s_{3:6}} + 
		\frac{ A_5^{\prime}( 1, 2,P_{34}^M,5,6)
		A_3^{(P_{5:2}3)}(P_{5:2}^M,3,4)}{s_{34}} \right. \nonumber \\
		&+ \left. \frac{A_3^{\prime}\left(P_{4:1}^M, 2, 3\right)
		A_5^{(1P_{23})}(1,P_{23}^M,4,5,6)}{s_{4:1}} + 
		\frac{ A_4^{\prime}\left( 1, 2,P_{3:5}^M,6\right)
	A_4^{(P_{6:2}3)}\left(P_{6:2}^M,3,4,5\right)}{s_{3:5}}\right]. \nonumber
\end{align}
Graphically, the second example, with the choice $(ijk)=(134)$, is
\vspace{-0.3cm}
{\small
\begin{eqnarray}\label{cuts-6pts-2}
	A_6^{\prime}(\mathbb{I}_6^{(134)} ) =
\hspace{-0.15cm}
 \parbox[c]{6.1em}{\includegraphics[scale=0.19]{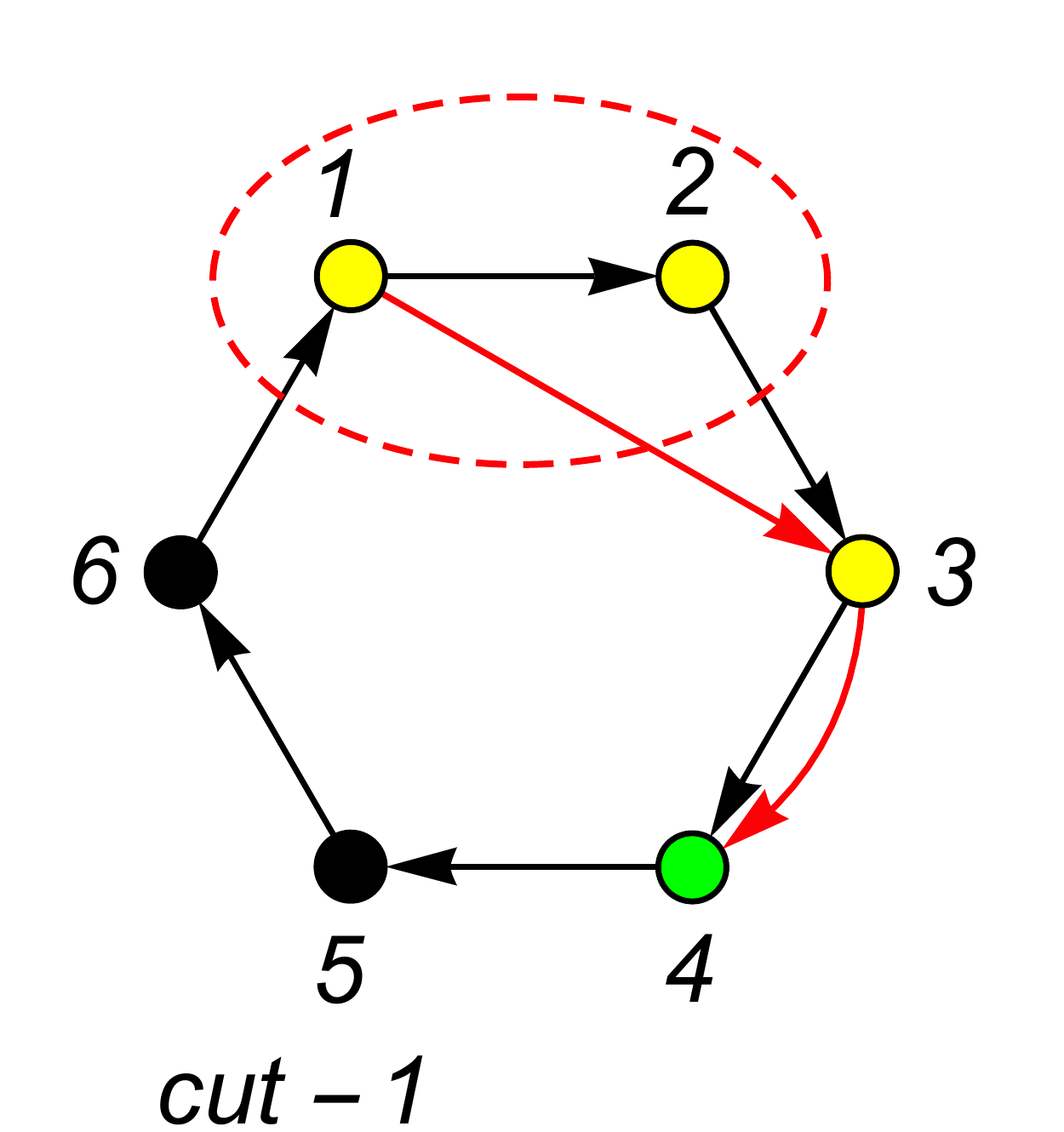}} +
 \hspace{-0.1cm}
  \parbox[c]{7.1em}{\includegraphics[scale=0.19]{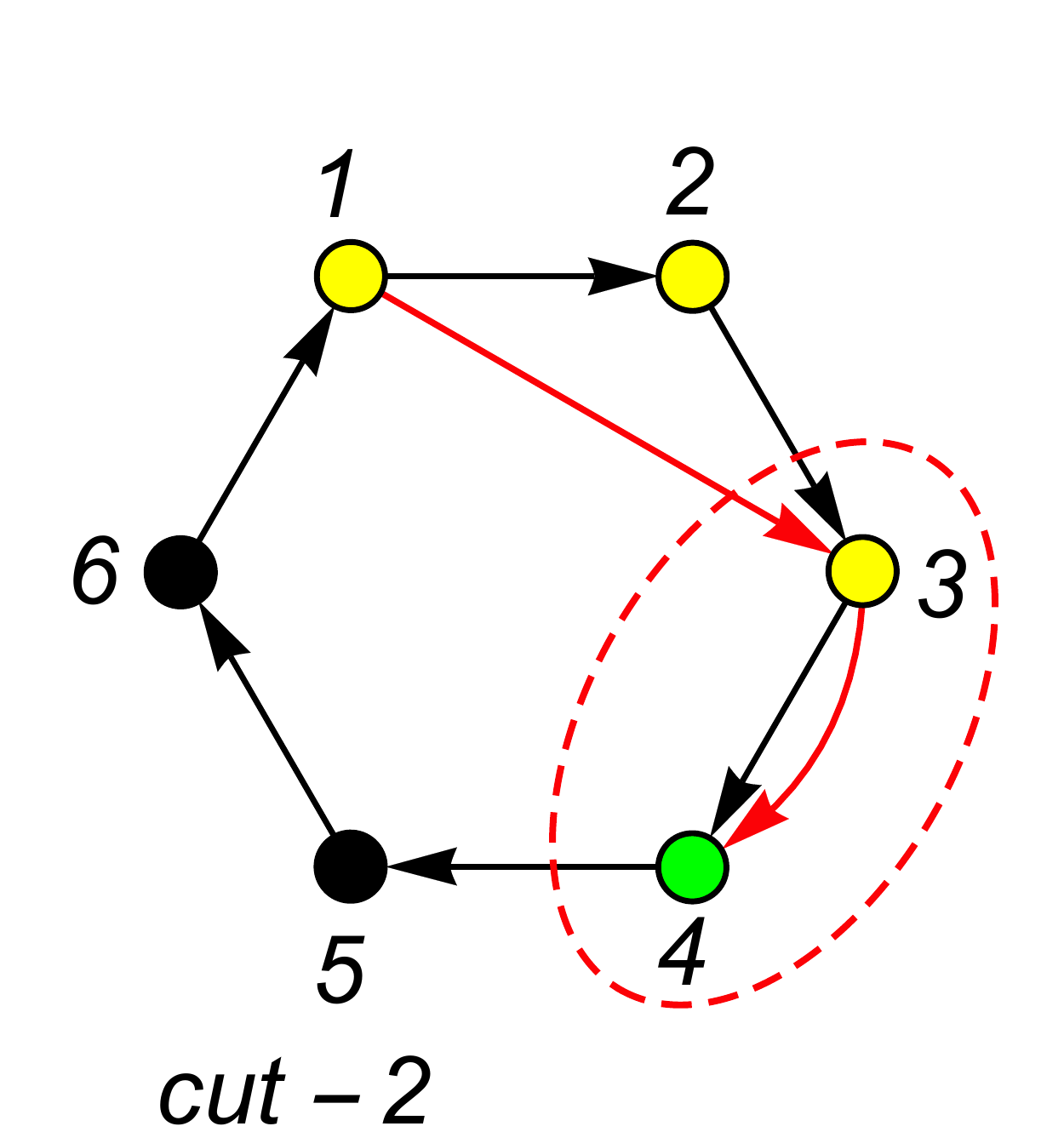}}
 \hspace{-0.36cm}  
  +
   \parbox[c]{6.1em}{\includegraphics[scale=0.19]{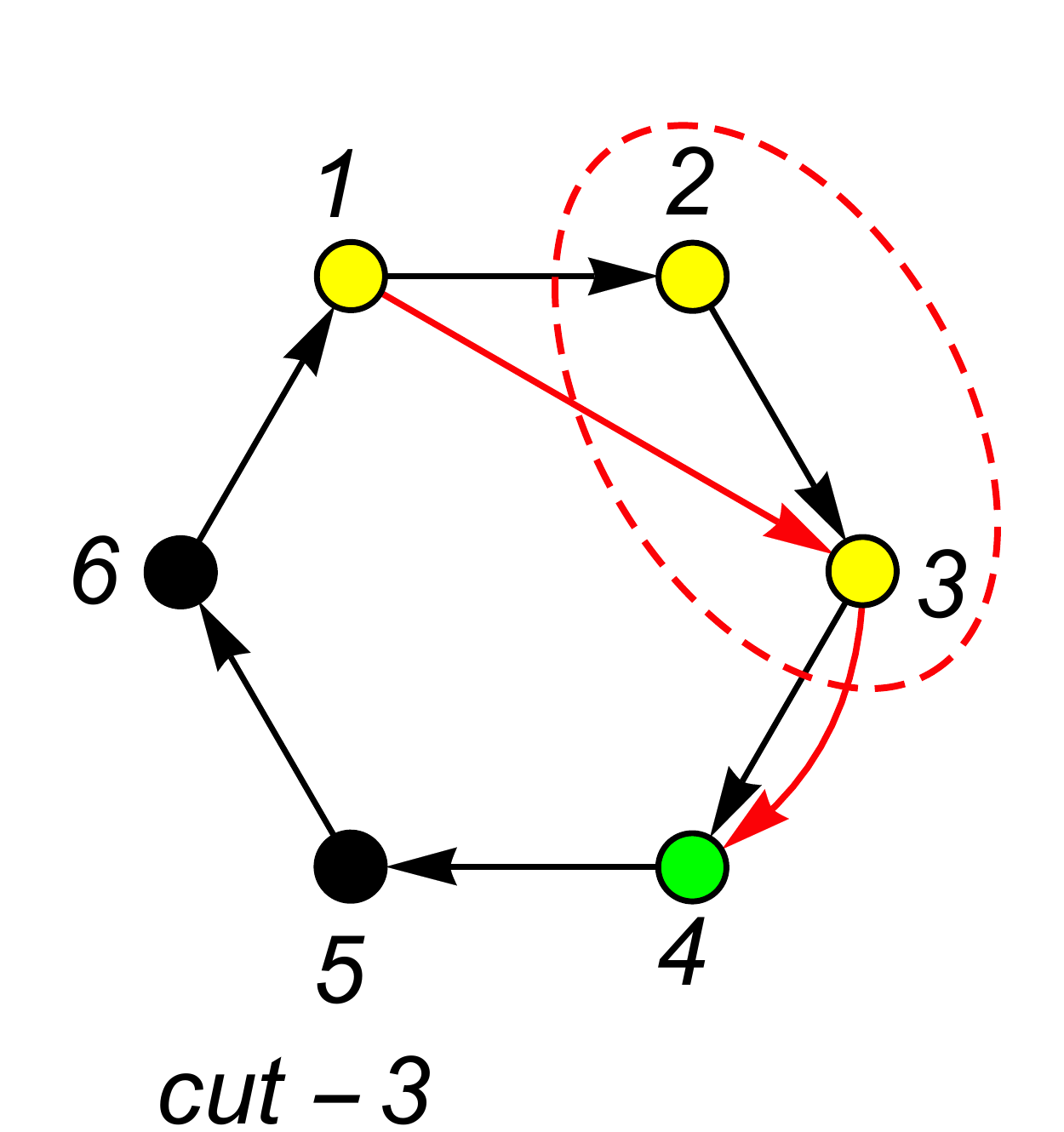}} 
  +
  \hspace{-0.05cm}
   \parbox[c]{6.1em}{\includegraphics[scale=0.19]{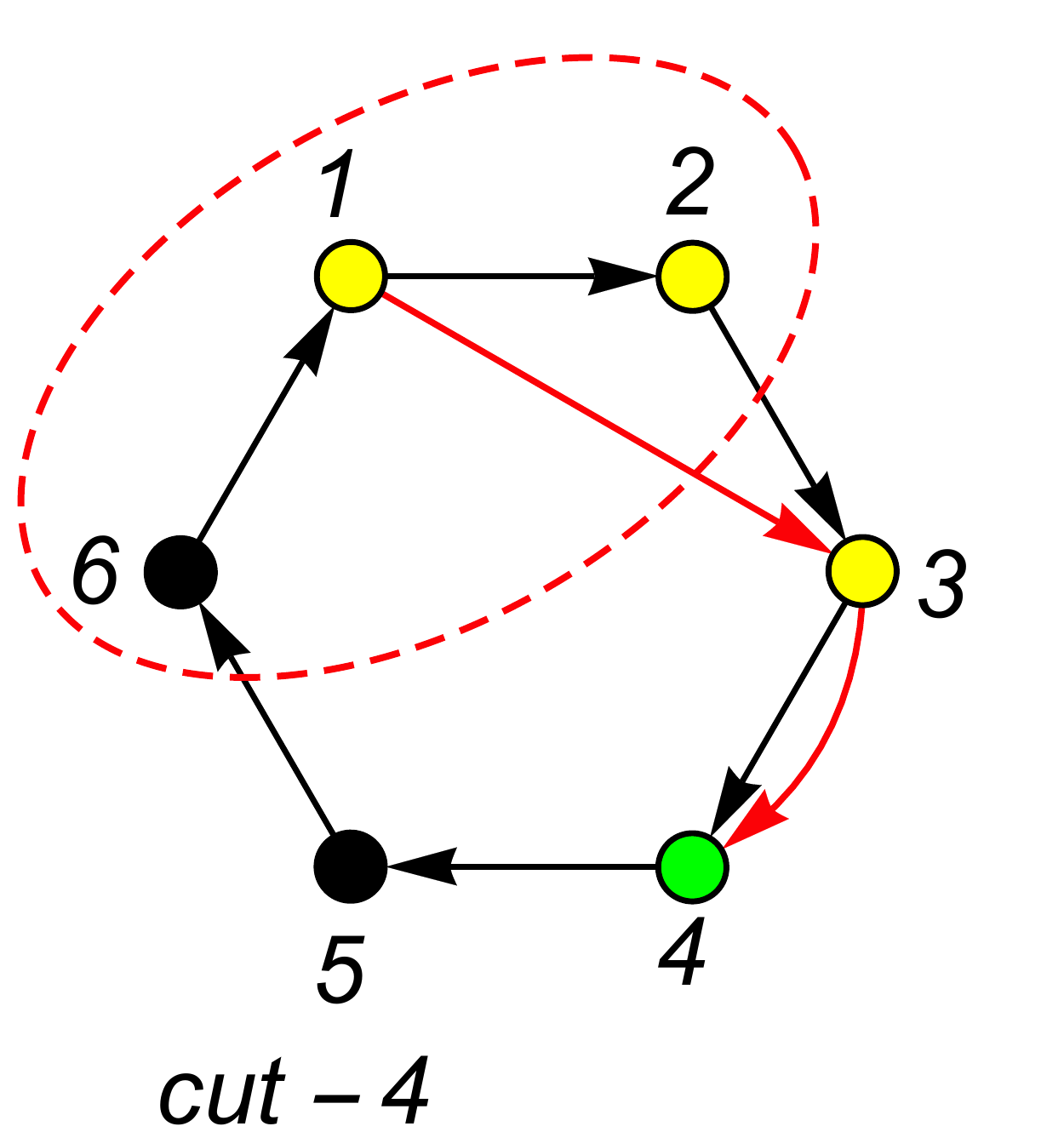}}    
   ~ ,
   \qquad
\end{eqnarray}
}
\vskip-0.15cm\noindent
which becomes (see \cref{sec:sixpointex2} to follow the full computation)
\begin{align}
	\label{eq:sixpntnew2}
	& A_6^{\prime}(\mathbb{I}_6^{(134)}) =\\& \sum_M \left[
		\frac{A_3^{(1P_{3:6})}\left(1,2, P_{3:6}^M\right)
		 A_5^{\prime}(P_{12}^M, 3, 4,5,6)}{s_{3:6}} + 
		\frac{A_5^{(1P_{34})}(1,2,P_{34}^M,5,6)
		 A_3^{\prime}(P_{5:2}^M, 3, 4)}{s_{34}} \right. \nonumber \\
		&+ \left. 
		\frac{ A_4^{(1P_{3:5})}\left( 1, 2,P_{3:5}^M,6\right)
	 A_4^{\prime}\left(P_{6:2}^M, 3, 4,5\right)}{s_{3:5}}\right]
		+\frac{ A_3\left( 3,P_{4:1},2\right)
		 A_5^{\prime}( 1,P_{23}, 4,5,6)}{s_{4:1}}.  \nonumber
\end{align}
Notice that the last contribution ({\it cut-3}) evaluates to zero.
We can check that both examples with the new integrand prescription reproduce the correct result. 
The full six-point calculation for both choices of gauge fixing is presented in \cref{6-pts-comp}.
Notice that in the first example,
all factorization contributions are glued together with off-shell gluons, while in the 
second example, three contributions involve off-shell gluons, and one contribution is purely
in terms of scalar particles.

So far we have seen three different kinds of factorization relations. The first kind, presented in 
\cref{eq:fourpntold,eq:sixpntold}, all particles were scalar. In the second case, given by \cref{eq:fourpntnew1,eq:sixpntnew1}, the 
intermediate particles were vector fields (off-shell gluons). Finally, in the last case,
\cref{eq:fourpntnew2,eq:sixpntnew2}, the factorization relation involved both intermediate scalar and vector fields.\footnote{Although in this case, the factorization contribution where the propagated particle is a scalar field vanishes, it is simple to find an example where this does not happen. For instance, let us choose the gauge, $(pqr|m)=(134|6)$, and the reduced $\mathsf{A}_n$ matrix with $(ijk)=(146)$. It is not hard to check that for this gauge fixing the amplitude, $ A_6^{\prime}( \mathbb{I}_6^{(146)} )$, has the two types of factorization contributions which are non-zero.}

\subsection{Longitudinal Contribution}\label{section-Lcontribution}

As the non-linear sigma model is a scalar theory, it is an interesting proposition to only
consider longitudinal contributions. An off-shell vector field can be decomposed in terms of 
transverse and longitudinal degrees of freedom. Let us consider only including the 
longitudinal degrees of freedom.

Practically, this means that instead of using the relation in \cref{eq:epsM}, we keep only
the longitudinal sector,
\begin{align}
	\label{eq:epsL}
	\sum_L \epsilon_i^{L\, \mu} \epsilon_j^{L\, \nu} = \frac{k_i^\mu k_j^\nu}{k_i \cdot k_j} = \overline{k}^\mu_{i} k_j^\nu, ~~{\rm with},~ 
	 k^\mu_i = -k_{j}^\mu, ~~ 
	 \overline{k}^\mu_{i}=-\left(\frac{ {k}^\mu_{i}}{k_i^2}\right).
\end{align}
Here we label the polarization vectors by a superscript $L$ instead of $M$ when keeping only longitudinal 
degrees of freedom.

In the four-point example, we have that 
\begin{align}
	& \sum_L \left[ \frac{ A_3^{\prime}( 1, 2, P_{34}^L)
		A_3^{(P_{12}3)}(P_{12}^L,3,4)}{s_{34}} + \frac{A_3^{(1P_{23})}(1,P_{23}^L,4)
	 A_3^{\prime}( P_{41}^L, 2, 3)}{s_{23}}\right] \nonumber \\
	&= -\frac{1}{2}\left[\frac{s_{12}^2}{s_{12}} + \frac{s_{14}^2}{s_{14}}\right] 
	= \frac{s_{13}}{2}
	= -\frac{1}{2} A_4(\mathbb{I}_4)
	\label{eq:long4pnt1}
\end{align}
and
\begin{align}
	& \sum_L  \frac{A_3^{(1P_{34})}(1,2,P_{34}^L)
		 A_3^{\prime}( P_{12}^L, 3, 4)}{s_{34}} 
		+ \frac{ A_3^{\prime}( 1,P_{23}, 4)
		A_3( 3, P_{41},2)}{s_{23}} \nonumber \\
	&= \frac{1}{2}\left[\frac{s_{12}^2}{s_{12}} + \frac{0}{s_{14}}\right] = \frac{s_{12}}{2}
	\neq  \rho A_4(\mathbb{I}_4)
	\label{eq:long4pnt2}
\end{align}
where is $\rho$ is a real constant. The sum of longitudinal contributions in \cref{eq:long4pnt1}
is proportional to the correct answer, while the sum of longitudinal contributions in 
\cref{eq:long4pnt2} is not.

Applying the same ideas to the six-point amplitude in \cref{eq:sixpntnew1}, we have that 
\begin{align}
	\label{eq:sixpntlong1}
	& \sum_L \left[
		\frac{ A_3^{\prime}\left( 1, 2, P_{3:6}^L\right)
		A_5^{(P_{12}3)}(P_{12}^L,3,4,5,6)}{s_{3:6}} + 
		\frac{ A_5^{\prime}( 1, 2,P_{34}^L,5,6)
		A_3^{(P_{5:2}3)}(P_{5:2}^L,3,4)}{s_{34}} \right. \nonumber \\
		&+ \left. \frac{ A_3^{\prime}\left( P_{4:1}^L, 2, 3\right)
		A_5^{(1P_{23})}(1,P_{23}^L,4,5,6)}{s_{4:1}} +
		(-1)\frac{ A_4^{\prime}\left( 1, 2, P_{3:5}^L,6\right)
	A_4^{(P_{6:2}3)}\left(P_{6:2}^L,3,4,5\right)}{s_{3:5}}\right] \nonumber \\
	&= -\frac{1}{2}A_6(\mathbb{I}_6).
\end{align}
Notice that the relative sign of the contribution from even subamplitudes (physical pole) was flipped
in order to reproduce the correct amplitude.\footnote{We have tested all possible sign
combinations, and this is the only one which is proportional to the correct amplitude.}
In the four-point example, all subamplitudes are odd, and no relative sign flip is needed.
All the longitudinal contributions are computed in \cref{appendix-Lcontribution}.

Now, let us focus on the factorization relation given in \cref{eq:sixpntnew2} and its longitudinal
contributions
\begin{align}
	\label{eq:sixpntlong2}
	& \sum_L \left[
		(-1)^{i_1}\frac{A_3^{(1P_{3:6})}\left(1,2, P_{3:6}^L\right)
		 A_5^{\prime}( P_{12}^L, 3, 4,5,6)}{s_{3:6}} + 
		(-1)^{i_2}\frac{A_5^{(1P_{34})}(1,2,P_{34}^L,5,6)
		A_3^{\prime}(P_{5:2}^L, 3, 4)}{s_{34}} \right. \nonumber \\
		&+ \left. 
		(-1)^{i_3}\frac{ A_4^{(1P_{3:5})}\left( 1, 2,P_{3:5}^L,6\right)
		A_4^{\prime}\left( P_{6:2}^L, 3, 4,5\right)}{s_{3:5}}\right]
		+\frac{ A_3\left( 3, P_{4:1},2\right)
		 A_5^{\prime}( 1, P_{23}, 4,5,6)}{s_{4:1}}  \nonumber\\
		& \neq \rho A_6(\mathbb{I}_6),
\end{align}
where the non-equality is preserved for all $2^3=8$ possible combinations of relative signs,
{\it i.e.} $(i_1,i_2,i_3) \in \{(0,0,0),(1,0,0),(0,1,0),(0,0,1),(1,1,0),(1,0,1),(0,1,1),(1,1,1)\}$.
Thus, like the four-point example, the amplitude with both off-shell gluons and scalars does 
not reproduce the full answer when only longitudinal contributions are kept.
Again, the longitudinal contributions are presented in \cref{appendix-Lcontribution}.

In summary, we have obtained examples of three different factorization relations, involving only 
intermediate scalars, off-shell gluons, or both scalars and off-shell gluons, respectively.
In the case where we have only off-shell gluons, we are able to reproduce the full answer by only
keeping the longitudinal degrees of freedom (with a relative sign flip between even and odd
factorization contributions).

\section{General Factorization Relations}
\label{sec:factorizationGeneral}

The factorization relations from the previous section can be generalized.
In this section, we present three different factorization formulas.
One formula is given in terms of exchange of off-shell vector fields,
while the other two formulas are given in terms of purely scalar fields.

First, let us consider the case, $A_{2n}(\mathbb{I}_{2n}^{(13)})$.
Thus, as in the section \ref{sec-NLSM-1}, we choose the gauge fixing $(pqr|m)=(123|4)$ and the reduced matrix with $(ij)=(13)$, 
namely $[\mathsf{A}_{2n}]^{13}_{31}$. 
Applying the integration rules, the amplitude becomes
\begin{align}\label{general-1}
	& A_{2n}(\mathbb{I}_{2n}^{(13)}) 
=
\sum_{i=3}^{n}
\frac{  A^{\prime}_{2(n-i+2)}\left(1, 2 , P_{\!3:2i-1} ,2i,..., 2n\right) 
\! \times \!  
 A^{\prime}_{2(i-1)}\left( P_{2i:2 }  ,3,4,..., 2i-1 \right) }{s_{3:2i-1}}  +  \nonumber \\
&(-1)
\sum_{i=3}^{n+1}
         \frac{  A^{\prime}_{2(n-i+2)+1}\left(  1,   3, P_{\!4:2i-2,2}  ,2i-1,..., 2n\right) 
\!\times \!   A^{\prime}_{2(i-1)-1}\left(   P_{2i-1:1,3},  2,  4, ..., 2i-2 \right) }{ s_{4:2i-2,2} } \, .
\end{align}
This formula has been check up to ten points.
In order to obtain the 
this relation, we used the matrix identities formulated in \cref{Pf-properties}.
In the first line, we used that
\begin{align}
&A_{2i} \left(...,\bm{P_p},... , \bm{P_q}, ...,   P_r,...\right)=
A_{2i} \left(...,\bm{P_p},... ,  P_q, ..., \bm{  P_r},...\right)
 \nonumber\\
&=
A_{2i} \left(...,P_p,... , \bm{  P_q}, ..., \bm{ P_r},...\right)
=  A^{\prime}_{2i} \left(...,\bm{P_p},... , \bm{P_q}, ..., \bm{P_r},...\right).
\end{align}
For the second line, we used properties {\bf I} and {\bf III} in \cref{Pf-properties}.

Thus, as the formula obtained in eq.~\eqref{general-1}, our second factorization relation, that was already presented in Ref.~\cite{Bjerrum-Bohr:2018jqe}, is supported on the double-cover formalism. In order to generalize the \cref{eq:fourpntnew1,eq:sixpntnew1},  we choose the same gauge fixing, $(pqr|m)=(123|4)$, and  the reduced matrix with, $(ijk)=(123)$,  ($\ie$ $[\mathsf{A}_{2n}]^{12}_{23}$). 
By the integration rules formulated in \cref{sec:integrationRules}, it is straightforward to see the 
amplitude turns into
\begin{align}\label{general-2}
	A_{2n}^{\prime}\left(\mathbb{I}_{2n}\right) 
&=
\sum_M \left[
\sum_{i=3}^{n}
\frac{ A^{\prime}_{2(n-i+2)}\left(1, 2 ,  P^{M}_{\!3:2i-1} ,2i,..., 2n\right) 
\! \times \!  A^{(P_{2i:2 } 3)}_{2(i-1)}\left(  P^{M}_{2i:2 } , 3, 4,..., 2i-1 \right) }{s_{3:2i-1}} \right.   \nonumber \\
&+
\sum_{i=3}^{n+1}
         \frac{ A^{\prime}_{2(n-i+2)+1}\left(  1,   2, P^{M}_{\!3:2i-2} ,2i-1,..., 2n\right) 
\!\times \!  A^{ (P_{2i-1:2} 3  )  }_{2(i-1)-1}\left(   P^{M}_{2i-1:2},  3,  4, ..., 2i-2 \right) }{ s_{3:2i-2} } \nonumber\\
&+
 \left. \frac{  A_{3}^{\prime} \left( {P}^{M}_{4:1},  2,  3\right) \!\times \!
 A^{(1P_{23})}_{2n-1}\left(1,{P}^{M}_{23},4,\ldots, 2n\right) }{ s_{4:1} }\right]  ,
\end{align}
where we use \cref{eq:epsM}. This second general formula has been verified  up to ten points.

On the other hand, from the results obtained in the \cref{eq:long4pnt1,eq:sixpntlong1} for four and six points, respectively,  we can generalize the idea  presented in section \ref{section-Lcontribution} to higher number of points.  Therefore, 
by considering just the longitudinal degrees of freedom in eq.~\eqref{general-2},  we conjecture the following factorization formula \cite{Bjerrum-Bohr:2018jqe}, 
\begin{align}\label{general-3}
 A_{2n}^{\prime}(\mathbb{I}_{2n}) 
&= 2\sum_L\left[
\sum_{i=3}^{n}
\frac{ A^{\prime}_{2(n-i+2)}\left(1, 2 ,   P^{L}_{\!3:2i-1} ,2i,..., 2n\right) \! \times \!  
A^{(P_{2i:2 } 3)}_{2(i-1)}\left(  P^{L}_{2i:2 } , 3, 4,..., 2i-1 \right) }{s_{3:2i-1}} \right.    \nonumber \\
&+(-1)
\sum_{i=3}^{n+1}
         \frac{  A^{\prime}_{2(n-i+2)+1}\left(  1,   2, P^{L}_{\!3:2i-2} ,2i-1,..., 2n\right) 
\!\times \!  A^{ (P_{2i-1:2} 3  )  }_{2(i-1)-1}\left(   P^{L}_{2i-1:2},  3,  4, ..., 2i-2 \right) }{ s_{3:2i-2} } \nonumber\\
&+(-1) \left.
 \frac{  A_{3}^{\prime} \left( {P}^{L}_{4:1} ,  2 ,  3\right) \!\times \!
 A^{(1P_{23})}_{2n-1}\left(1,{{P}^{L}_{23}},4,\ldots, 2n\right) }{ s_{4:1} } \right] ,
\end{align}
where we use \cref{eq:epsL}.
Finally, by applying the identities
{\small
\begin{eqnarray}
A^{(P_pP_q)}_{2i} \big(...,{\color{black}P_p},... , {\color{black}  P_q}, ..., {\color{black}  P_r},...\big )&=&
A^{(P_qP_r)}_{2i} \big(...,{\color{black}P_p},... , {\color{black}  P_q}, ..., {\color{black}  P_r},...\big ) \nonumber \qquad \quad \\
&=&
-(P^2_p+P^2_q+P^2_r)\!\times\! A^{\prime}_{2i} \big(...,{\bm P_p},... , {\bm  P_q}, ..., {\bm  P_r},...\big ),
\qquad \quad \nonumber
 \\
A^{(P_pP_q)}_{2i+1 } \big(...,{\color{black}P_p},... , {\color{black}  P_q}, ..., {\color{black}  P_r},...\big )&=&
A^{(P_qP_r)}_{2i+1} \big(...,{\color{black}P_p},... , {\color{black}  P_q}, ..., {\color{black}  P_r},...\big ) \nonumber \qquad \quad \\
&=&
(P^2_p-P^2_q-P^2_r)\!\times\!  A^{\prime}_{2i+1} \big(...,{\bm P_p},... , {\bm  P_q}, ..., {\bm P_r},...\big ),
\qquad \quad \label{even-odd-id}
\end{eqnarray}
}
\vskip-0.3cm\noindent
which are a consequence from the properties in appendix \ref{Pf-properties},
it is straightforward to see the eq.~\eqref{general-3} becomes
{\small
\begin{eqnarray}\label{general-3-2}
\hspace{-0.2cm}
 A_{2n}^\prime(\mathbb{I}_{2n}) 
&=&
\sum_{i=3 }^{n}
\frac{  A^{\prime}_{2(n-i+2)}\big(1,  2 ,   P_{3:2i-1} ,2i,..., 2n\big ) \! \times \!   A^{\prime}_{2(i-1)}\big( P_{2i:2 }  , 3, 4,..., 2i-1 \big) }{s_{3:2i-1}}    \nonumber \\
&+&
\sum_{i=3 }^{n+1}
         \frac{  A^{\prime}_{2(n-i+2)+1}\big(  1,  2, P_{3:2i-2} ,2i-1,..., 2n\big) 
\!\times \!   A^{\prime}_{2(i-1)-1}\big(   P_{2i-1:2},  3,  4, ..., 2i-2 \big) }{ s_{3:2i-2} } \nonumber\\
&+& (-1)\,
\frac{   A_{3}^{\prime} \big( P_{4:1} ,  2 , 3\big) \!\times \!  A^{\prime}_{2n-1}\big(1 , P_{23},4,\ldots, 2n\big) }{ s_{4:1} }  .
\end{eqnarray}
}
\vskip-0.3cm\noindent
This is our third general factorization formula.

\subsection{A New Relationship for the Boundary Terms}
\label{sec:boundaryterms}

As we argued in Ref.~\cite{Bjerrum-Bohr:2018jqe}, the amplitudes with an odd number of particles, 
$\ie$ amplitudes of the form $  A_{2m+1}^{\prime} (..., {\bm P_a},...)$ (odd amplitude), 
are proportional to $P_a^2$ since that them must vanish when all particles are on-shell. 
Thus, the poles  given by the odd contributions, namely expressions of the form  
$\frac{   A_{2m+1}^{\prime} (..., {\bm P_a},...)\times 
A_{2k+1}^{\prime} (..., {\bm P_b},...)  }{2 \,P_a\cdot P_b}$,  
are spurious and, therefore, those terms are on the boundary of any usual 
BCFW deformation \cite{Britto:2005fq}. In particular, under the BCFW deformation, 
\begin{equation}
k_2^\mu(z) = k_2^\mu +z\,q^\mu\, , \qquad k_3^\mu(z) = k_3^\mu-z\,q^\mu \, ,~~~ \text{with} ~~ q^2=0,
\end{equation}
all even contributions (physical poles), which are given by the sum
{\small
\begin{eqnarray}
\sum_{i=3 }^{n}
\frac{ A^{\prime}_{2(n-i+2)}\big(1,  2 ,  P_{3:2i-1},2i,..., 2n\big ) \! \times \!   
A^{\prime}_{2(i-1)}\big(  P_{2i:2 } , 3, 4,..., 2i-1 \big) }{P^2_{3:2i-1}(z) }    
\end{eqnarray}
}
\vskip-0.1cm\noindent
in eqs.~\eqref{general-1} and \eqref{general-3-2},  are localized over the $z$-plane at, $P^2_{3:2i-1}(z)=0$.  Thus,  by the above discussion,  all odd contributions  in eqs.~\eqref{general-1}  and \eqref{general-3-2} are localized at the point $z=\infty$ on the $z$-plane and, hence, we call those odd amplitudes the boundary terms.

Now, clearly, by comparing the factorization relations obtained in eqs.~\eqref{general-1}  and \eqref{general-3-2}, this is straightforward to see that one arrives to the identity 
{\small
\begin{eqnarray}
\hspace{-0.2cm}
&&
\sum_{i=3 }^{n+1}
         \frac{ A^{\prime}_{2(n-i+2)+1}\big(  1,   2, P_{3:2i-2} ,2i-1,..., 2n\big) 
\!\times \!  A^{\prime}_{2(i-1)-1}\big(   P_{2i-1:2},  3,  4, ..., 2i-2 \big) }{ s_{3:2i-2} }
+(2\, \leftrightarrow  \, 3)  \qquad \quad  \nonumber\\
&&
= 
\frac{   A_{3}^{\prime} \big( {P}_{4:1} ,  2 ,  3\big) \!\times \! 
 A^{\prime}_{2n-1}\big(1 , P_{23},4,\ldots, 2n\big) }{ s_{4:1} }  \, , \qquad \quad
\end{eqnarray}
}
\vskip-0.3cm\noindent
which lies on the boundary of any usual  BCFW deformation.  We have checked this identity up to $n=10$.

\section{A Novel Recursion Relation}
\label{sec:recursion}

In this section, we  are going to present a new recursion relationship, which can be used to write down  any 
NLSM amplitude in terms of the three-point building-block, $A_3^\prime(P_a,P_b,P_c)=-(P_a^2-P_b^2+P_c^3)$, given in \cref{3pt-1}.

Previously, in \cref{general-3}, we arrived at an unexpected factorization expansion, which, although it emerged accidentally from the integration rules, a formal proof is yet unknown.\footnote{It is important to remind ourselves  that the longitudinal contributions 
give the right answer only when, after applying the integration rules, all factorization channels are mediated by an off-shell vector field. This was exemplified in section \ref{section-Lcontribution}.} Thus, since applying the integration rules is an iterative process, we would like to know if the relationship in \cref{general-3} could be extended to off-shell amplitudes (both for an even and odd number of particles). Here, we are going to show how to do that. 

First, consider the four-point computation, $A^\prime_4(P_1,P_2,P_3,4)$, where the particles, $\{P_1,P_2,P_3\}$,  can be off-shell. By the integration rules, we obtain the same decomposition as in
\cref{eq:fourpntnew1},
\begin{align}
	\label{off-4pts}
	 & A_4^{\prime}\left( P_1,P_2,P_3,4 \right) = \\
	 &\sum_M \left[
		\frac{ A_3^{\prime}(  P_1, P_2,P_{34}^M) A_3^{(P_{12}P_3)}(P_{12}^M,P_3,4)}{s_{P_3P_4}}
	+ \frac{A_3^{(P_1P_{23})}(P_1,P_{23}^M,4)  A_3^{\prime}(P_{41}^M, P_2, P_3)}{s_{P_4P_1}}\right]=-s_{4P_2}. \nonumber 
\end{align}
Now, by using the longitudinal gluing relation given in \cref{eq:epsL}, \ie\, $\sum_{L}\eps_{34}^{\mu\,L}\eps_{12}^{\nu\,L} = \overline{P}_{34}^\mu P_{12}^\nu$ and $\sum_{L}\eps_{23}^{\mu\,L}\eps_{41}^{\nu\,L} =  P_{23}^\mu \overline{P}_{41}^\nu$, over the above factorized amplitude, one arrives at 
\begin{align}
	\label{off-L1-4pts}
	 &
	 (-2)\,\sum_L \left[
		\frac{ A_3^{\prime}(  P_1, P_2,P_{34}^L) A_3^{(P_{12}P_3)}(P_{12}^L,P_3,4)}{s_{P_3P_4}}
	+ \frac{A_3^{(P_1P_{23})}(P_1,P_{23}^L,4)  A_3^{\prime}(P_{41}^L, P_2, P_3)}{s_{P_4P_1}}\right] \nonumber\\
	&=\frac{ -(P_1^2-P_2^2+P_{34}^2) \, s_{4 P_{12}}  }{P_{34}^2} + \frac{    -(P_{41}^2-P_2^2+P_{3}^2)  \, s_{4 P_{23}}  }{P_{41}^2}  .
\end{align}
Clearly, since $\{ P_1,P_2,P_3 \}$ are off-shell,  the results found in
\cref{off-4pts,off-L1-4pts} do not match. However, there is a simple way to make them coincide. 
Instead of using the usual longitudinal identity, we  employ a generalized version where $\overline{P}^{\mu}_a$ is redefined as
{\small
\begin{eqnarray}
\overline{P}^\mu_{34}=\!- \!\left( \frac{P_{34}^\mu  }{P_{34}^2}   \right)  \rightarrow 
\overline{P}^\mu_{34}=\!- \!\left( \frac{P_{34}^\mu  }{P_1^2-P_2^2+P_{34}^2}   \right)\!, ~
\overline{P}^\mu_{41}=\!- \!\left( \frac{P_{41}^\mu  }{P_{41}^2}   \right)  \rightarrow 
\overline{P}^\mu_{41}=\!- \!\left( \frac{P_{41}^\mu  }{P_{41}^2-P_2^2+P_3^2}   \right). 
\nonumber
\end{eqnarray}
}
\vskip-0.1cm\noindent
It is straightforward to check that under this redefinition, the factored expression in \cref{off-L1-4pts} reproduces the same  result as in \cref{off-4pts}.  The generalization to a higher number of points is straightforward, so, 
 when the particles $\{ P_1,P_2,P_3 \}$ are off-shell, the longitudinal gluing relations that must be used in \cref{general-3} are given by 
{\small
\begin{eqnarray}
&& \sum_{L}  A^\prime_{2m+1}(\bm{P^{L}_{r}}, \ldots, \bm{P_{2}} , \dots , \bm{P_3} ,\ldots  ) \times 
A^{( P_1P_k) }_{2q+1}({ P_{1}}, \ldots,  P^{L}_{k} , \dots , )  ~  \rightarrow ~    
\sum_{L}\eps_{r}^{\mu\,L}\eps_{k}^{\nu\,L} = \overline{P}_{r}^\mu P_{k}^\nu , \nonumber
\\
&&
\sum_{L}  
A^{( P_1P_k) }_{2j}({ P_{1}}, \ldots,  P^{L}_{k} , \dots , ) \times
A^\prime_{2i}(\bm{P^{L}_{r}}, \ldots, \bm{P_{2}} , \dots , \bm{P_3} ,\ldots  ) 
  ~  \rightarrow    ~
\sum_{L}\eps_{k}^{\mu\,L}\eps_{r}^{\nu\,L} = \overline{P}_{k}^\mu P_{r}^\nu ,
\nonumber
\end{eqnarray}
}
\vskip-0.1cm\noindent
where, $P^\mu_r=-P_k^\mu$, and 
\begin{equation}\label{barPrPk}
\overline{P}_r^\mu= -\left( \frac{P^\mu_r}{P_r^2 -P_2^2+P_3^2 }\right), \qquad
\overline{P}_k^\mu= -\left( \frac{P^\mu_k}{P_1^2 +P_k^2  }\right) .
\end{equation}
Thus, by applying the identities in
eq.~\eqref{even-odd-id}, we obtain the following simple and compact expression 
{\small
\begin{align}\label{general-off-even}
	 A_{2n}^{\prime}&( P_1, P_2, P_3,4,...,2n) 
= \nonumber\\
&\sum_{i=3 }^{n}
\frac{  A^{\prime}_{2(n-i+2)}\big( P_1,  P_2,  P_{3:2i-1},2i,..., 2n\big ) \! 
\times \!   A^{\prime}_{2(i-1)}\big(  P_{2i:2} , P_3, 4,..., 2i-1 \big) }
{s_{3:2i-1}}    \nonumber \\
& +
\sum_{i=3 }^{n+1}
\frac{  A^{\prime}_{2(n-i+2)+1}\big(  P_1,   P_2, P_{3:2i-2} ,2i-1,..., 2n\big) 
\!\times \!   A^{\prime}_{2(i-1)-1}\big(   P_{2i-1:2}, P_3,  4, ..., 2i-2 \big)}
{P_1^2-P_2^2+ P^2_{3:2i-2} } \nonumber\\
& + (-1)\,
\frac{   A_{3}^{\prime} \big( P_{4:1},  P_2 ,  P_3\big) \!\times \! 
A^{\prime}_{2n-1}\big( P_1, P_{23}, 4,\ldots, 2n\big) }{ P^2_{4:1}-P_2^2+P_3^2 }  .
\end{align}
}
\vskip-0.1cm\noindent
Obviously,  when $\{ P_1,P_2,P_3\}$ become on-shell, we rediscover eq.~\eqref{general-3-2}.

In order to achieve a completed recursion-relationship, it is needed to get  a closed formula for the odd amplitude, $ A_{2n+1}^{\prime} ( P_1, P_2, P_3,4,...,2n+1)$. Therefore,  applying the integration rules over this amplitude, one obtains the following two types of combinations   
{\small
\begin{eqnarray}
{\bf I.}~~
&& 
\sum_{M}  A^\prime_{2m+1}(\bm{P^{M}_{r}}, \ldots, \bm{P_{2}} , \dots , \bm{P_3} ,\ldots  ) \times 
A^{( P_1P_k) }_{2j}({ P_{1}}, \ldots,  P^{M}_{k} , \dots , )   , \nonumber
\\
{\bf II.}~~
&&
\sum_{M}  
A^{( P_1P_k) }_{2q+1}({ P_{1}}, \ldots,  P^{M}_{k} , \dots , ) \times
A^\prime_{2i}(\bm{P^{M}_{r}}, \ldots, \bm{P_{2}} , \dots , \bm{P_3} ,\ldots  ).  \nonumber
\end{eqnarray}
}
\vskip-0.1cm\noindent
We found that, to land on the right result by using just longitudinal degrees of freedom, the combination ${\bf I}$ must be glued by the relation
{\small
\begin{eqnarray}
{\bf I.}~~
&& 
\sum_{L}\eps_{r}^{\mu\,L}\eps_{k}^{\nu\,L} =  (-1)(P_1^2-P_2^2+P_3^3) \times \overline{P}_{r}^\mu {\overline P}_{k}^\nu  , 
\end{eqnarray}
}
\vskip-0.2cm\noindent
where $\overline{P}_{r}^\mu$ and $ {\overline P}_{k}^\nu $ are defined in \cref{barPrPk},  while the combination ${\bf II}$ has to be discarded. Note that the overall factor, $(P_1^2-P_2^2+P_3^3)$, implies that when the off-shell external particles become on-shell,  the amplitude  $ A_{2n+1}^{\prime} $ vanishes trivially,   such as it is required.

To summarize, after applying the integration rules over an even or odd amplitude, such that the factorized subamplitudes are glued only by virtual vector fields, then, we can compute this process just by considering the longitudinal degrees of freedom and the rules given in the following box
\vspace{-0.1cm}
{\small
\begin{eqnarray}
\hspace{-0.2cm}
\begin{blockarray}{ccc}
  &    &    \\
\begin{block}{ccc}
A^\prime_{2m+1}(\bm{P^{\eps}_{r}}, \ldots, \bm{P_{2}} , \dots , \bm{P_3} ,\ldots  )\Big|_{\eps_r^{\mu} \rightarrow \overline{P}_r^\mu } &  \xLeftrightarrow[\qquad]{~\text{Product Allowed}~} & A^{( P_1P_k) }_{2q+1}({ P_{1}}, \ldots,  P^{\eps}_{k} , \dots ,  ) \Big|_{\eps_k^{\mu} \rightarrow {P}_k^\mu }  \\
\quad &      & \quad \\
    \,^{\rm Product}_{\rm Allowed}     \Bigg\Updownarrow  \times (-1)\, (P_1^2 - P_2^2 + P_3^2 )      &       &    \Bigg \Updownarrow  {\color{red} \,^{\rm Product}_{\rm Forbidden}   }   \\
\quad  &   & \quad \\
 A^{( P_1P_k) }_{2j}({ P_{1}}, \ldots,  P^{\eps}_{k} , \dots ,  ) \Big|_{\eps_k^{\mu} \rightarrow {\overline{P}}_k^\mu }    &  \xLeftrightarrow[\qquad]{~\text{Product Allowed}~} & A^\prime_{2i}(\bm{P^{\eps}_{r}}, \ldots, \bm{P_{2}} , \dots , \bm{P_3} ,\ldots  )\Big|_{\eps_r^{\mu} \rightarrow {P}_r^\mu }     \\
\end{block}
\end{blockarray}  
\nonumber
\end{eqnarray}
}
\vskip-0.1cm\noindent
where $\overline{P}_{r}^\mu$ and $ {\overline P}_{k}^\nu $ are given in \cref{barPrPk}. Notice that the horizontal rules on the box work over the  even amplitudes, \ie\,  $ A_{2n}^{\prime} ( P_1, P_2, P_3,4,...,2n)$, while the vertical rules  work over the odd ones, $ A_{2n+1}^{\prime} ( P_1, P_2, P_3,4,...,2n+1)$. 

Finally, by employing the identities in eq.~\eqref{even-odd-id} and the above box, we are able to write down a compact formula for $ A_{2n+1}^{\prime} ( P_1, P_2, P_3,4,...,2n+1)$,
{\small
\begin{align}\label{general-off-odd}
	& A_{2n+1}^{\prime}( P_1,P_2, P_3,4,...,2n+1) 
=
\left(  P_1^2 -P_2^2 + P_3^2  \right) \times \left[\, 
\sum_{i=3 }^{n+1}   \left(  \frac{1}{P_1^2-P_2^2+ P^2_{3:2i-1}} \right)   \right.
 \nonumber \\
&
\hspace{1.7cm}
\times
\frac{  A^{\prime}_{2(n-i+2)+1}\big(  P_1,  P_2, P_{3:2i-1} ,2i,..., 2n+1\big) 
\!\times \!   A^{\prime}_{2(i-1)}\big(  P_{2i:2},  P_3, 4, ..., 2i-1 \big) }
{ s_{3:2i-1} } 
 \nonumber\\
& 
\hspace{1.7cm}
+ \left.
  \left(  \frac{1}{P^2_{4:1}-P_2^2+P_3^2} \right) \times  
  \frac{   A_{3}^{\prime} \big( P_{4:1},  P_2 ,  P_3\big) \!\times \! 
  A^{\prime}_{2n}\big( P_1 , P_{23}, 4,\ldots, 2n+1 \big) }{ s_{4:1} } 
\right] .
\end{align}
}
\vskip-0.1cm\noindent
Evidently, the formulas, \cref{general-off-even,general-off-odd}, give us a novel recursion relation,
which we have checked against known results for up to ten points.
The big advantage with this relation is that it is purely algebraic, as any non-linear sigma model
amplitude can be decomposed to off-shell three-point amplitudes (without solving any scattering equations).

\section{The Soft Limit  and a New Relation  for $A_n^{\rm NLSM\oplus \phi^3}$}
\label{sec:softLimit}

The soft limit for the ${\rm U}(N)$ non-linear sigma model in its CHY representation  was already studied by Cachazo, Cha and Mizera  (CCM) in Ref.~\cite{Cachazo:2016njl}. One of the main results  is given by the expression (at leading order)
\begin{align}\label{CCM}
A_n (1,\ldots , n) = \epsilon \sum_{a=2}^{n-2} 2\, \tilde k_{n}\cdot k_a\,  A_{n-1}^{\rm NLSM \oplus \phi^3} (1,\ldots, n-1||n-1,a,1) + {\cal O}(\eps^2), 
\end{align}
where $k^\mu_n=\epsilon \, \tilde k_n^\mu$ and $\eps\rightarrow 0$. 

In this section  we  carry out, in detail, the soft limit behaviour at six-point, but using the new recursion relation proposed in section \ref{sec:recursion}. Although 
the generalization to a higher number of points is not straightforward,  it is not complicated. We will not take into account the general case in this work.

Let us consider the amplitude, $A_6 (1,2,3,4,5,6)={A}_6^{\prime} ( {5},{6},{1},2,3,4) $, where the soft particle  is,
$k_6^\mu = \epsilon \, \tilde k_6^\mu$, with $\eps\rightarrow 0$.  From \cref{general-off-even}, we have
{\small
\begin{eqnarray}
{A}_6^{\prime} ( {5},{6},{1},2,3,4) &=&
\frac{  {A}_3^{ \prime} ( { 5},  { 6} , { P_{1:4}} ) \times { A}_5^{\prime} ( { P_{56} },  { 1} , { 2 },3,4 )   }{  P_{56}^2 } - 
\frac{  { A}_3^{\prime} ( { P_{2:5}},  { 6} , { 1 } ) \times { A}_5^{\prime} ( { 5 },  { P_{61} } , {2 },3,4 )   }{  P_{61}^2 }  \nonumber \\
&&
+
\frac{  {A}_3^{\prime} ( { P_{3:6}},  { 1} , { 2 } ) \times { A}_5^{\prime} ( { 5 },  { 6 } , { P_{12} },3,4 )   }{  P_{12}^2 }  
+
\frac{  { A}_4^{\prime} ( { 5},  { 6} , { P_{1:3} },4 ) \times { A}_4^{\prime} ( { P_{4:6} },  { 1 } , {2 },3)   }{  P_{1:3}^2 }  ~~ \nonumber \\
&&
\hspace{-3.0cm}
=
- { A}_5^{\prime} ( { P_{56} },  { 1} , { 2 },3,4 )  + 
 { A}_5^{\prime} ( { 5 },  { P_{61} } , { 2 },3,4 )   
- { A}_5^{\prime} ( { 5 },  { 6 } , { P_{12} },3,4 ) 
-
\frac{   2\,\eps \, \tilde k_6\cdot k_4\,   \times { A}_4^{\prime} ( { P_{456} },  {1 } , { 2 },3)   }{  s_{45} + 2\eps \, \tilde k_6\cdot P_{45} },  \nonumber \\ 
\end{eqnarray}
}
\vskip-1.0cm\noindent
where the three-point building-blocks in \cref{3point-BB} have been used. Applying the off-shell formula proposed in \cref{general-off-odd}, it is not hard to check that, at leading order, the above five-point amplitudes become
{\small
\begin{eqnarray}
&& -{ A}_5^{\prime} ( { P_{56} },  { 1} , { 2 },3,4 )  =    (2\,\eps\, \tilde k_6\cdot k_5)\left[
\frac{  { A}_4^{\prime} ( { P_{51} },  { 2} , { 3 },4 )   }{ s_{51}  } +
\frac{  { A}_4^{\prime} ( { 5 },  { P_{12}} , { 3 },4 )   }{ s_{12}  }
\right],
 \\
 && { A}_5^{\prime} ( { 5 },  { P_{61} } , { 2 },3,4 )   =   (2\,\eps\, \tilde k_6\cdot k_1)\left[
\frac{  {A}_4^{\prime} ( { P_{51} },   2 , { 3 },4 )   }{ s_{51}  } +
\frac{{ A}_4^{\prime} ( { 5 },  {P_{12} } , { 3 },4 )   }{ s_{12}  } 
\right],
 \\
&&- { A}_5^{\prime} ( { 5 },  { 6 } , { P_{12} },3,4 ) =
-(2\,\eps\, \tilde k_6\cdot P_{125} )\times
\frac{  { A}_4^{\prime} ( { 5 },  { P_{12}} , { 3 },4 )   }{ s_{12}  } -
2\,\eps\, \tilde k_6\cdot k_{4}
.
\end{eqnarray}
}
\vskip-0.2cm\noindent
Therefore,  the six-point amplitude at leading order in $\epsilon$ is given by 
{\small
\begin{eqnarray}\label{softDC}
 A_6 (1,2,3,4,5,6) &=&    (2\,\eps\, \tilde k_6\cdot k_2)\left[-
\frac{  { A}_4^{\prime} ( { P_{51} },  { 2} , { 3 },4 )   }{ s_{15}  } -
\frac{  { A}_4^{\prime} ( { 5 },  { P_{12}} , { 3 },4 )   }{ s_{12}  }
\right]\nonumber
 \\
 &&  +
    (2\,\eps\, \tilde k_6\cdot k_3)\left[-
\frac{  { A}_4^{\prime} ( { P_{51} },  { 2} , { 3 },4 )   }{ s_{15}  } 
\right]\nonumber
 \\
&&
+(2\,\eps\, \tilde k_6\cdot k_4 ) \left[
-
\frac{  { A}_4^{\prime} ( { P_{51} },  { 2} , { 3 },4 )   }{ s_{15}  }  -
\frac{  { A}_4^{\prime} ( { P_{45} },  { 1 } , { 2 },3)   }{  s_{45}  } -
1 \right]
.
\end{eqnarray}
}
\vskip-0.2cm\noindent
Now, from the CCM formula in \cref{CCM} one has 
{\small
\begin{eqnarray}\label{softSC}
 A_6 (1,2,3,4,5,6) &=&  (2\,\eps\, \tilde k_6\cdot k_2)\times  A_5^{\rm NLSM \oplus \phi^3} (1,2,3,4,5||5,2,1)  \nonumber
 \\
 &&  +
    (2\,\eps\, \tilde k_6\cdot k_3)  \times  A_5^{\rm NLSM\oplus \phi^3} (1,2,3,4,5||5,3,1)  \nonumber
 \\
&&
+(2\,\eps\, \tilde k_6\cdot k_4 ) \times  A_5^{\rm NLSM \oplus \phi^3} (1,2,3,4,5||5,4,1) 
.
\end{eqnarray}
}
\vskip-0.5cm\noindent
Although at first glance, the eqs.~\eqref{softDC} and \eqref{softSC}  do not seem to be the same, notice that by choosing the gauge,  $(pqr|m)=(512|3)$, the amplitude $A_5^{\rm NLSM \oplus \phi^3} (1,2,3,4,5||5,2,1)$ turns into
\vspace{-0.3cm}
{\small
\begin{eqnarray}\label{A5-NLSMp3-1}
 &&
A_5^{\rm NLSM \oplus \phi^3} (1,2,3,4,5||5,2,1) =
\int d\mu_5^{ \L}
 \hspace{-0.15cm} 
\parbox[c]{6.3em}{\includegraphics[scale=0.21]{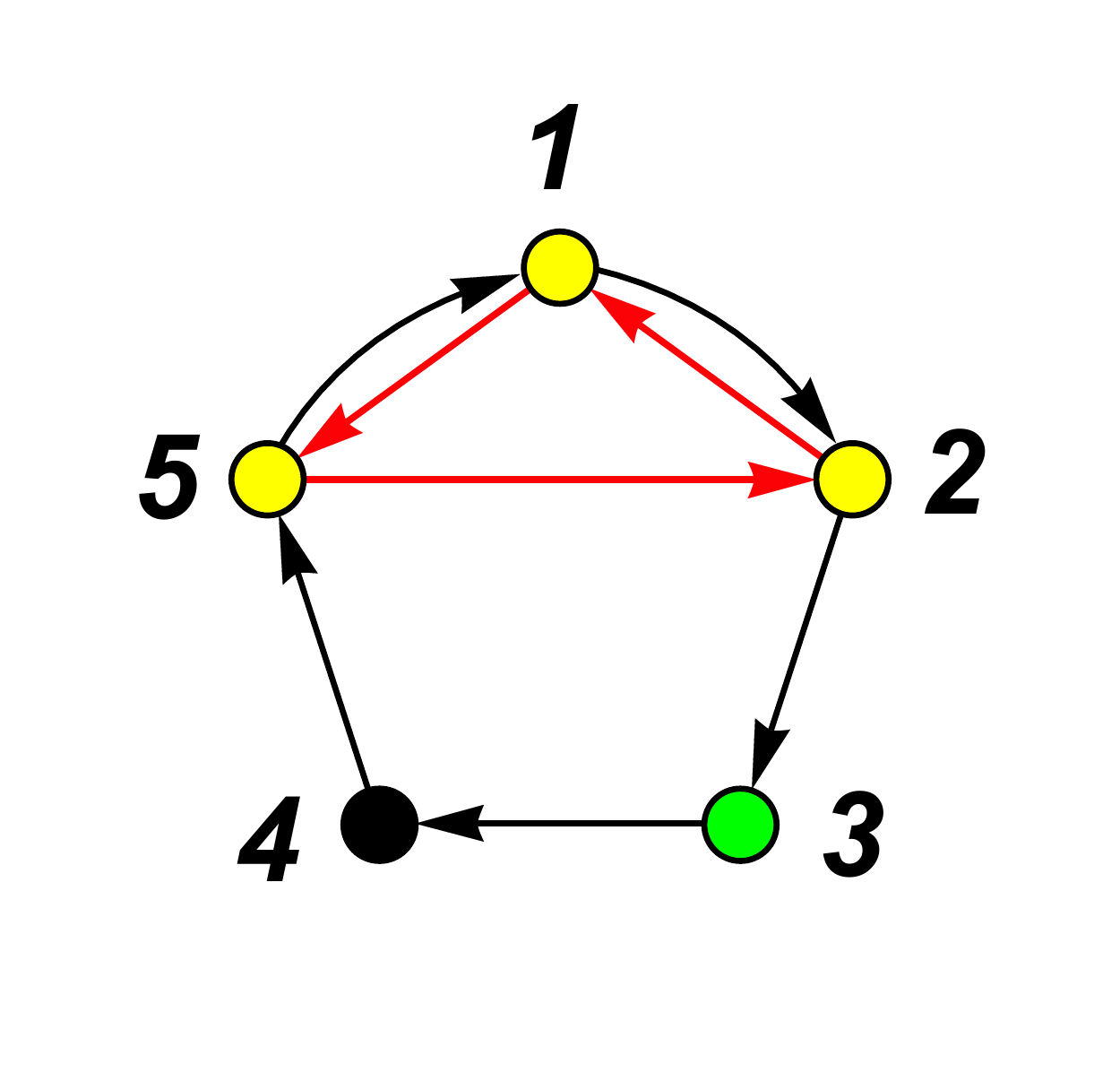}}
=
\hspace{-0.25cm}
   \parbox[c]{6.3em}{\includegraphics[scale=0.21]{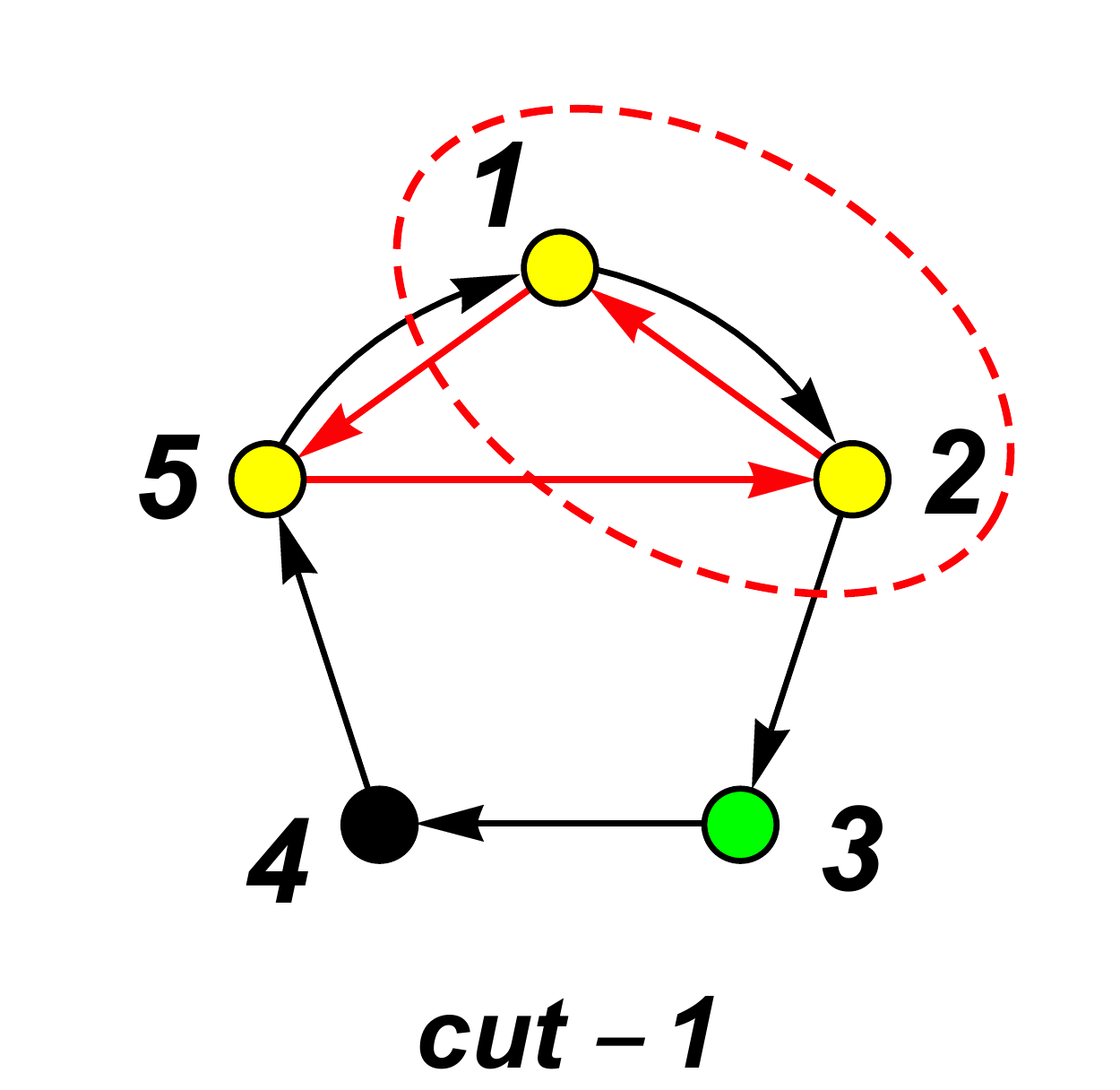}} +
 \hspace{-0.25cm}
  \parbox[c]{5.8em}{\includegraphics[scale=0.21]{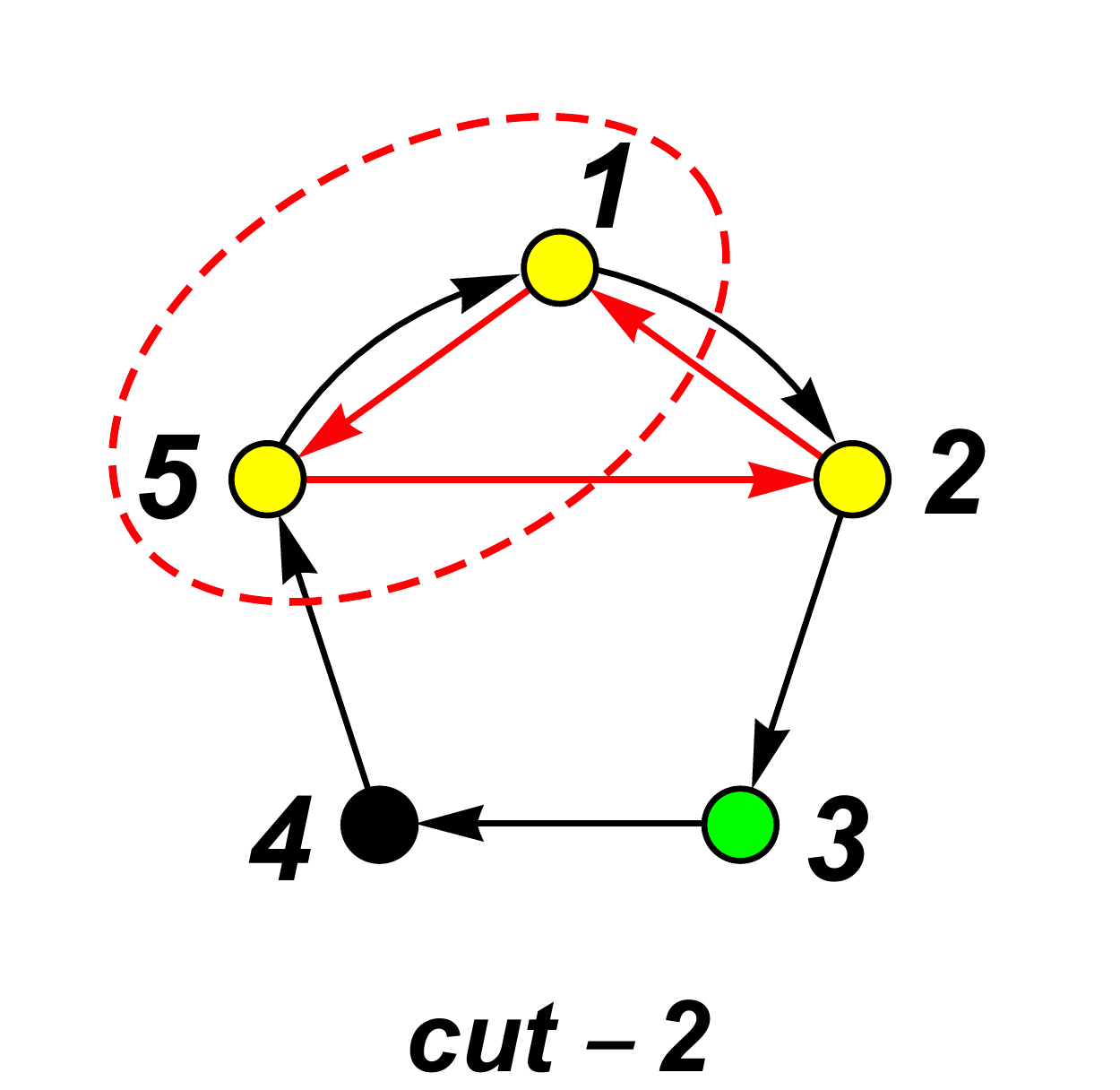}}   \nonumber
 \\
&&
=
-
\frac{   { A}_3^{\phi^3} ( 1,2,  { P_{3:5}} )   \times  { A}_4^{\prime} ( { 5 },  { P_{12}} , { 3 },4 )   }{ s_{12}  }   
-
\frac{  { A}_3^{\phi^3} ( 1,  { P_{2:4}} ,5 )   \times  { A}_4^{\prime} ( { P_{51} },  { 2} , { 3 },4 )   }{ s_{15}  }  \nonumber\\
&&
=
-
\frac{  { A}_4^{\prime} ( { 5 },  { P_{12}} , { 3 },4 )   }{ s_{12}  }   
-
\frac{  { A}_4^{\prime} ( { P_{51} },  { 2} , { 3 },4 )   }{ s_{15}  } ,
\end{eqnarray}
}
\vskip-0.5cm\noindent
where we employed the integration rules, the building-block, ${ A}_3^{\phi^3} ( P_1, P_2,  { P_{3}} )=1 $,
and the second property from the appendix \ref{Pf-properties}.
Following the same procedure, it is straightforward to see
{\small
\begin{eqnarray}\label{A5-NLSMp3-2}
A_5^{\rm NLSM \oplus \phi^3} (1,2,3,4,5||5,3,1) 
=
-
\frac{  { A}_4^{\prime} (  { P_{51}} ,2, { 3 },4 )   }{ s_{15}  }   .
\end{eqnarray}
}
\vskip-0.1cm\noindent
Clearly,
the first two lines in eqs.~\eqref{softDC} and \eqref{softSC}  match perfectly,  however, to compare the last lines we must take care. By direct computation, it is not hard to show that, in fact, the third lines in eqs.~\eqref{softDC} and \eqref{softSC} produce the same result, but, we can extract more information from them. For example, under the gauge fixing,  $(pqr|m)=(512|3)$, the amplitude $A_5^{\rm NLSM \oplus \phi^3} (1,2,3,4,5||5,4,1)$ is given by the cuts
\vspace{-0.3cm}
{\small
\begin{eqnarray}
A_5^{\rm NLSM \oplus \phi^3} (1,2,3,4,5||5,4,1) &=&
\int d\mu_5^{ \L}
 \hspace{-0.15cm} 
\parbox[c]{6.3em}{\includegraphics[scale=0.21]{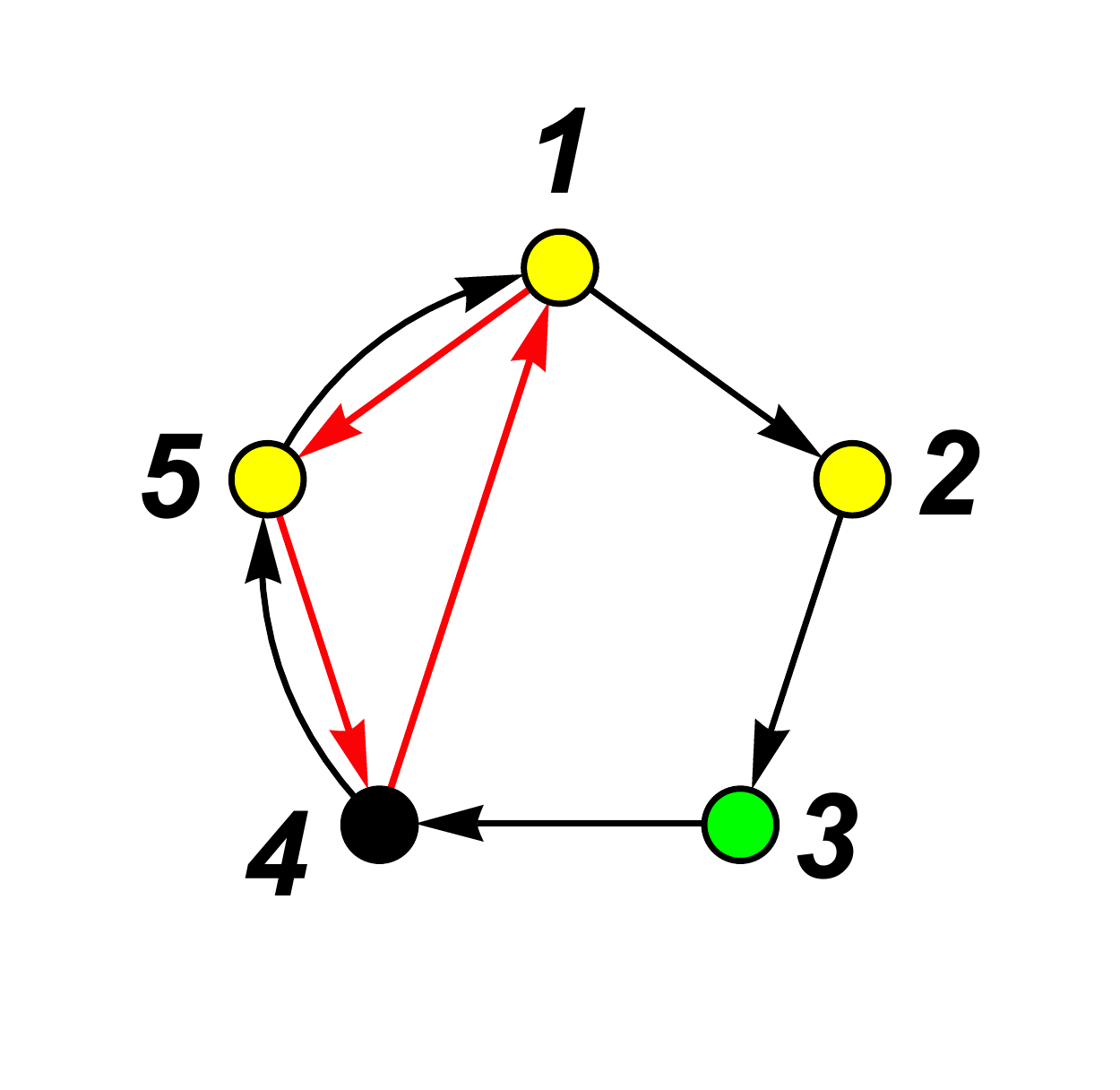}}
=
\hspace{-0.25cm}
   \parbox[c]{6.3em}{\includegraphics[scale=0.21]{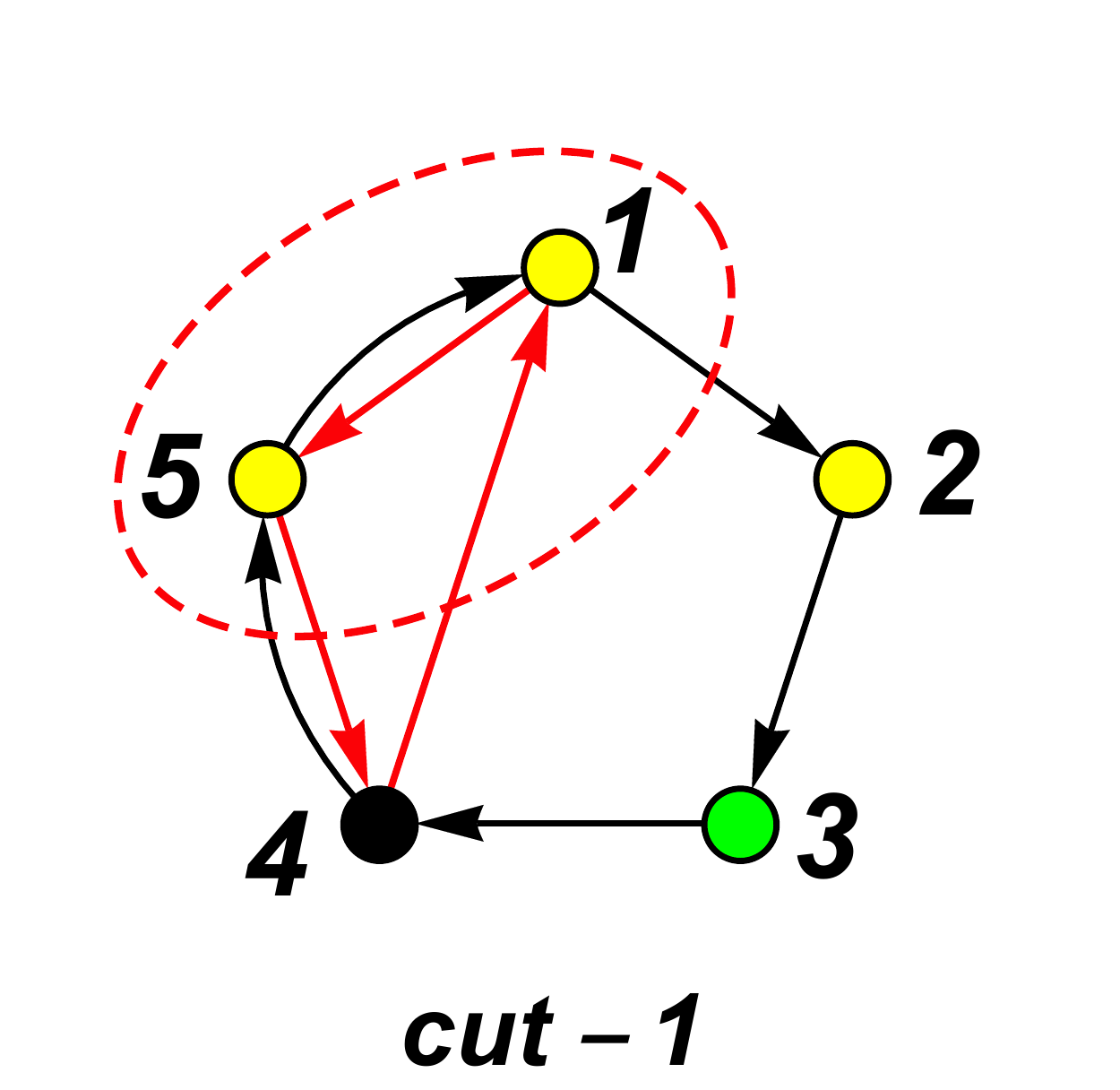}} +
 \hspace{-0.25cm}
  \parbox[c]{5.8em}{\includegraphics[scale=0.21]{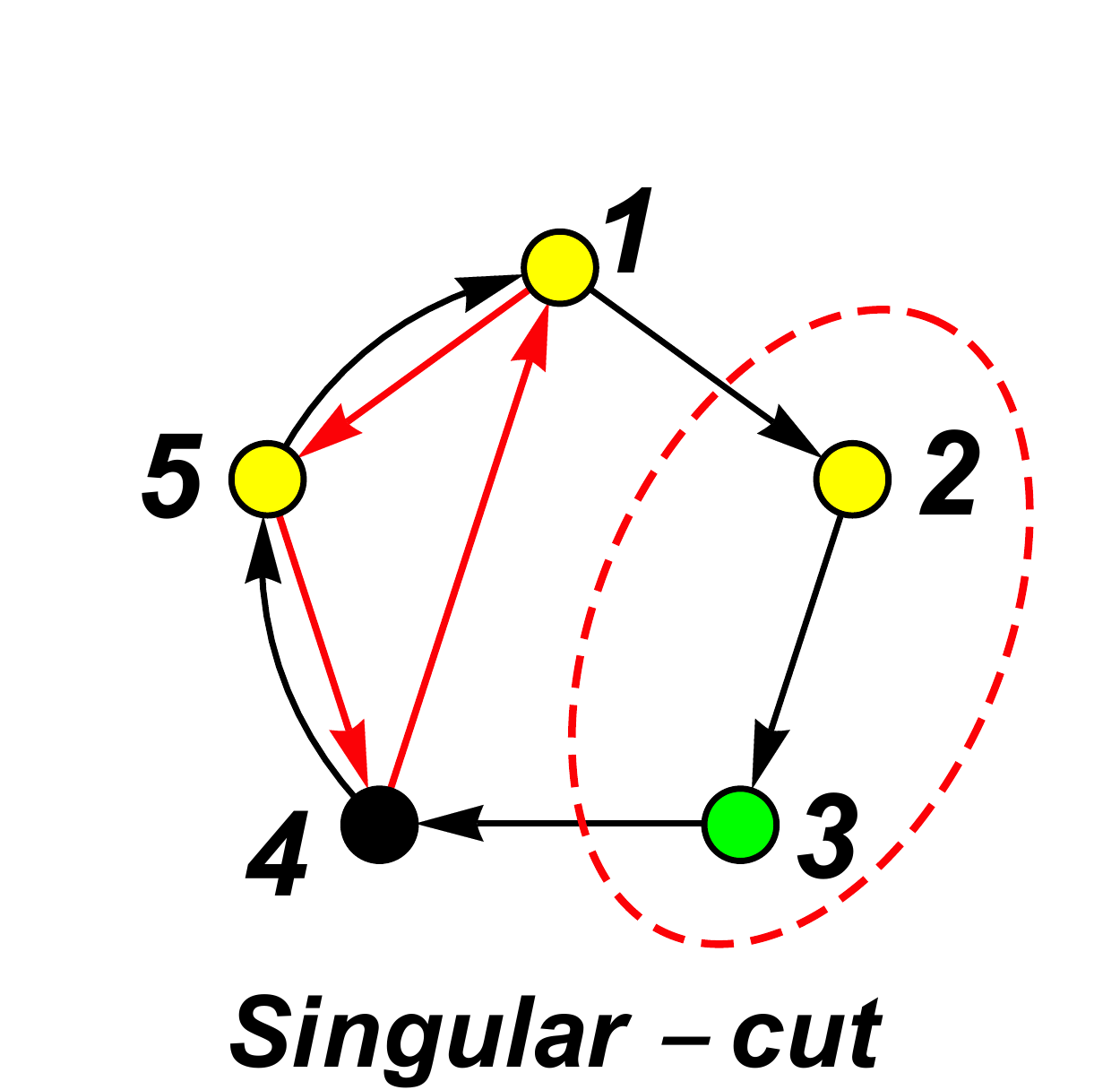}}   \nonumber
 \\
&
=&
-
\frac{   { A}_4^{\prime} ( { P_{51} },  { 2} , { 3 },4 )   }{ s_{15}  } + \, \text{Singular-cut}  .
\end{eqnarray}
}
\vskip-0.5cm\noindent
Clearly, by comparing the above expression with the last line in eq.~\eqref{softDC}, we arrive at
\begin{equation}\label{SC}
\text{Singular-cut}  = -
\frac{  { A}_4^{\prime} ( { P_{45} },  { 1 } , { 2 },3)   }{  s_{45}  } -
1,
\end{equation} 
which is a simple but strong result.  As it has been argued several times \cite{Gomez:2018cqg,Gomez:2016bmv} (see section \ref{sec:integrationRules}), the integration rules, 
which were obtained by expanding at leading order the $\L$ parameter of the double cover representation, can not be applied over singular cuts. In order to achieve an extension of these rules to singular cuts, one must expand beyond leading order  the $\L$  parameter and find a pattern, which is a highly non-trivial task.  Nevertheless, eq.~\eqref{SC}  tells us that the soft limit behaviour could help us to figure out this issue. This is an interesting subject to be studied in a future project.

 \subsection{A New Relation  for $A_n^{\rm NLSM\oplus \phi^3}$}

In the previous section, we  observe that, using the recursion relation proposed in  section  \ref{sec:recursion},   the soft limit  behaviour of the six-point amplitude, $ A_6 (1,2,3,4,5,6)$,  gives a factorized formula for  $ A_5^{\rm NLSM \oplus \phi^3} (1,2,3,4,5||5,a,1)$ in terms of off-shell NLSM amplitudes. In this section, we are going to show a new factorization formula for the general amplitude, $A_n^{\rm NLSM \oplus \phi^3} (1,\ldots ,n||n,a,1) $. 

First, let us consider the gauge fixing $(pqr|m)=(1an|2)$, so, we can suppose that the set of particles, $\{P_1,P_a,P_n\}$, are off-shell (here $a$ is a label between $2<a<n$).  Since the $A_n^{\rm NLSM \oplus \phi^3} (1,\ldots ,n||n,a,1) $  amplitude vanishes trivially when $n$ is even, then, it is enough to define, $n=2m+1$. Thus, applying the integration rules with the previous setup the amplitude is factorized into
\vspace{-0.1cm}
{\small
\begin{eqnarray}
&& A_n^{\rm NLSM \oplus \phi^3} (1,\ldots ,a-1,a,a+1,\ldots, n||n,a,1)=  \\
&&
\sum_{i=2}^{ \lfloor \frac{a}{2} \rfloor} \frac{A^\prime_{2i} ( P_{2i:n} ,  1,2,\ldots, 2i-1   ) \times A^{\rm NLSM \oplus \phi^3}_{2(m-i)+3} (  P_{1:2i-1}, 2i, \dots,  a, \ldots,  n||n,a,P_{1:2i-1}        ) }{s_{1:2i-1}} +  \nonumber \\
&&
\sum_{i={  \lceil \frac{a}{2}\rceil }}^{m} \frac{A^\prime_{2i} ( \bm{P_{2i+1:1}}  ,   \bm{2},.., \bm{a},...,   2i   ) \times A^{\rm NLSM \oplus \phi^3}_{2(m-i)+3} (1,  P_{2:2i}, 2i+1, \dots   ,  n||n,P_{2:2i},1        ) }{s_{2:2i}} , \nonumber
\end{eqnarray}
}
\vskip-0.2cm\noindent
where $ \lfloor x \rfloor$ and ${ \lceil x \rceil}$ are the Floor and  Ceiling  functions, respectively. 
Notice that when $a=3$, the first line doesn't contribute because of the properties of the Floor
function.

In the particular case when $a=2$,  we choose the gauge fixing $(pqr|m) = (12n|3)$, and the factorization relation becomes 
\vspace{-0.1cm}
{\small
\begin{eqnarray}
&& A_n^{\rm NLSM \oplus \phi^3} (1,2,\ldots, n||n,2,1)=  \\
&&
 \frac{A^\prime_{2m} ( n, P_{12},3,\ldots, n-1 ) \times A^{\rm NLSM \oplus \phi^3}_{3} (  P_{3:n},1,2||P_{3:n},2,1        ) }{s_{3:n}} +  \nonumber \\
&&
\sum_{i=2}^{m} \frac{A^\prime_{2i} ( { P_{2i+1:1}}  ,   { 2},3, \dots,   2i   ) \times A^{\rm NLSM \oplus \phi^3}_{2(m-i)+3} (1,  P_{2:2i}, 2i+1, \dots   ,  n||n,P_{2:2i},1        ) }{s_{2:2i}} . \nonumber
\end{eqnarray}
}
\vskip-0.2cm\noindent
Clearly, when $n=2m+1=5$, the relations obtained above are in agreement with the ones in 
eqs.~\eqref{A5-NLSMp3-1} and \eqref{A5-NLSMp3-2}.

\section{Special Galileon Theory}
\label{sec:sGalileon}

In Ref.~\cite{Cachazo:2014xea}, Cachazo, He and Yuan proposed  the CHY prescription to compute the $S$-Matrix of a special Galileon theory (sGal). The Galileon theories arise as effective field theories in the decoupling limit of massive
gravity \cite{Hinterbichler:2011tt,Kampf:2014rka,Dvali:2000hr}.  The special Galileon theory was discovered in Refs.~\cite{Cachazo:2014xea,Cheung:2014dqa} as a  special class of theory with soft limits that vanish particularly fast.

As discussed previously (for more details, see Ref.~\cite{Cachazo:2014xea}), the CHY prescription of the sGal is given by the integral
\begin{eqnarray}
	A_n^{\rm sGal}  =\int {\rm d}\mu_n^{\rm CHY} \, (z_{pq}z_{qr}z_{rp})^2 \times \left [{\rm det}^{\prime} \mathsf{A}_n \times
{\rm det}^{\prime} \mathsf{A}_n \right].
~~
\end{eqnarray}
From this expression, it is straightforward to  see the sGal is the square of the NLSM,  where the product is by means of the field theory Kawai-Lewellen-Tye  (KLT) kernel \cite{Kawai:1985xq}. Schematically, one has
\begin{eqnarray}\label{DCsGal}
A_n^{\rm sGal} = A_n \,
{\small
\begin{matrix}
\otimes\vspace{-0.8cm}\\
^{\small \rm KLT} 
\end{matrix}
}
 \, A_n   ,
\end{eqnarray}
where the KLT matrix, usually denoted as $S[\a|\b]$, is the inverse matrix of the double-color partial amplitude for the bi-adjoint  $\phi^3$ scalar theory \cite{Cachazo:2013gna,Cachazo:2013iea}.  Notice that, from this double copy formula, we can use the whole technology developed for NLSM and apply it in sGal. 
Nevertheless, since our main aim is to show how the integration rules work, we will not use eq.~\eqref{DCsGal}.

\subsection{A Simple Example}

In this section, we will show how the integration rules work in a theory without partial ordering. 
As a simple example, we will calculate the four-point amplitude for sGal.

From eq.~\eqref{sGal}, the sGal in the double cover representation is given by the integral  
{\small
\begin{eqnarray}
A_n^{\rm sGal}  =\!\! \int {\rm d}\mu_n^\L \frac{(-1) \Delta(pqr)\Delta(pqr|m)  }{ S_m^\tau} \times
\left[ {\bf det}^\prime \mathsf{A}^\L_n  
  \times 
{\bf det}^\prime \mathsf{A}^\L_n \right]   ,
~~~~
\end{eqnarray}
}
\vskip-0.1cm\noindent
where we have defined, ${\bf det}^\prime \mathsf{A}^\L_n =  \prod_{a=1}^n   \frac{(y\s)_a}{y_a}    \times  {\rm det}^{\prime} \mathsf{A}^\L_n$.
After choosing a gauge fixing, by the {\bf rule-I} in \cref{sec:integrationRules} we know that the Faddeev-Popov factor goes as, $\frac{(-1) \Delta(pqr)\Delta(pqr|m)  }{ S_m^\tau}  \sim \L^{-4} + {\cal O}(\L^{-2})$,
(\cref{eq:fadeevLscaling}). Thus,  in order to cancel this $\L^{-4} $ factor,  at leading order, a cut-contribution in the special  Galileon theory must cut at least one arrow of each reduced determinant, this fact comes from \cref{intruleTable}. This is summarized in {\bf Rule-II}. For example, for the four-point amplitude, $A_4^{\rm sGal} (1,2,3,4)$,  let us consider the following  four different setups
\vspace{-0.2cm}
{\small
\begin{eqnarray}\label{setups}
\hspace{-0.15cm}
 \parbox[c]{6.1em}{\includegraphics[scale=0.19]{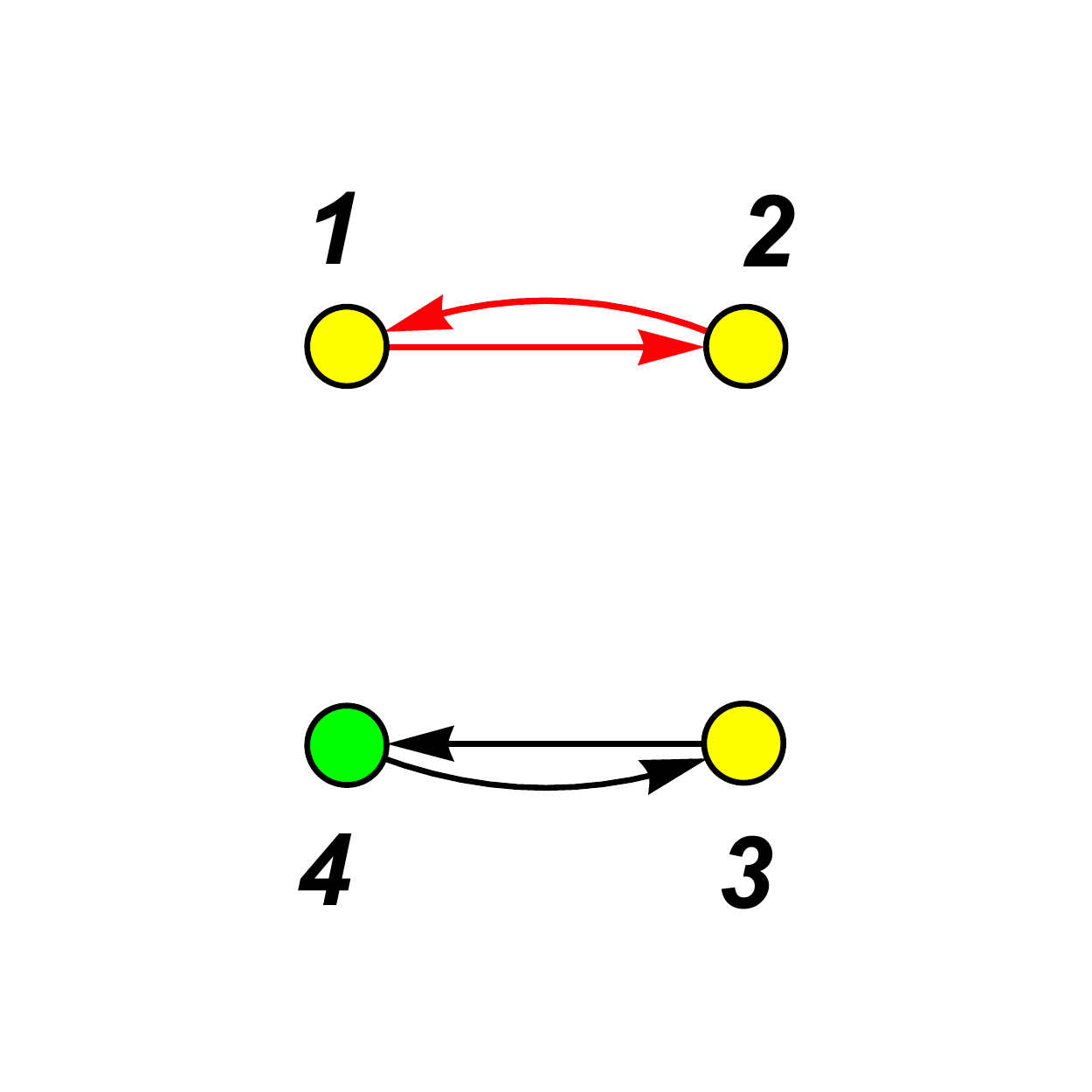}} ,
 \hspace{-0.1cm}
  \parbox[c]{7.1em}{\includegraphics[scale=0.19]{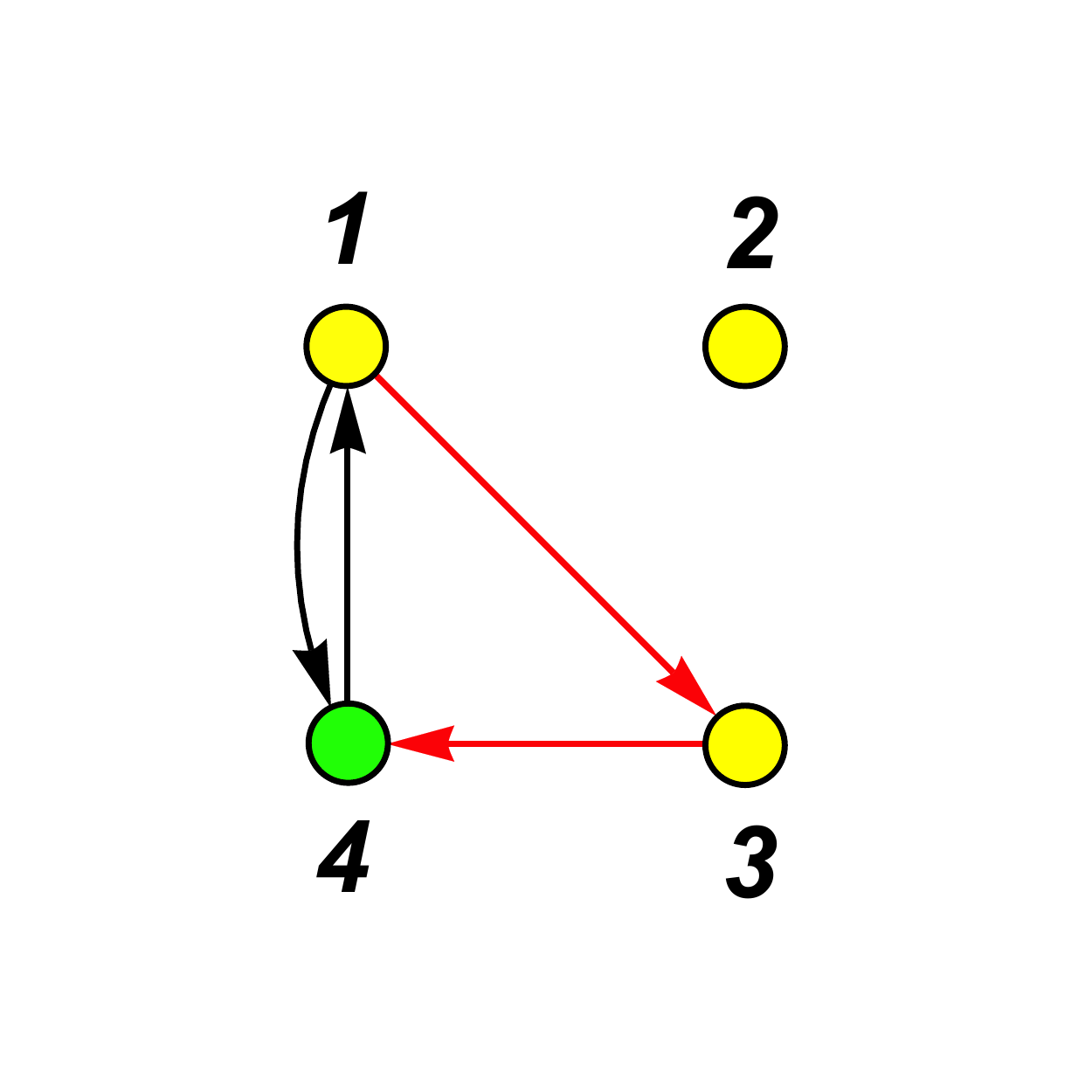}}
 \hspace{-0.36cm}  
  ,
   \parbox[c]{6.1em}{\includegraphics[scale=0.19]{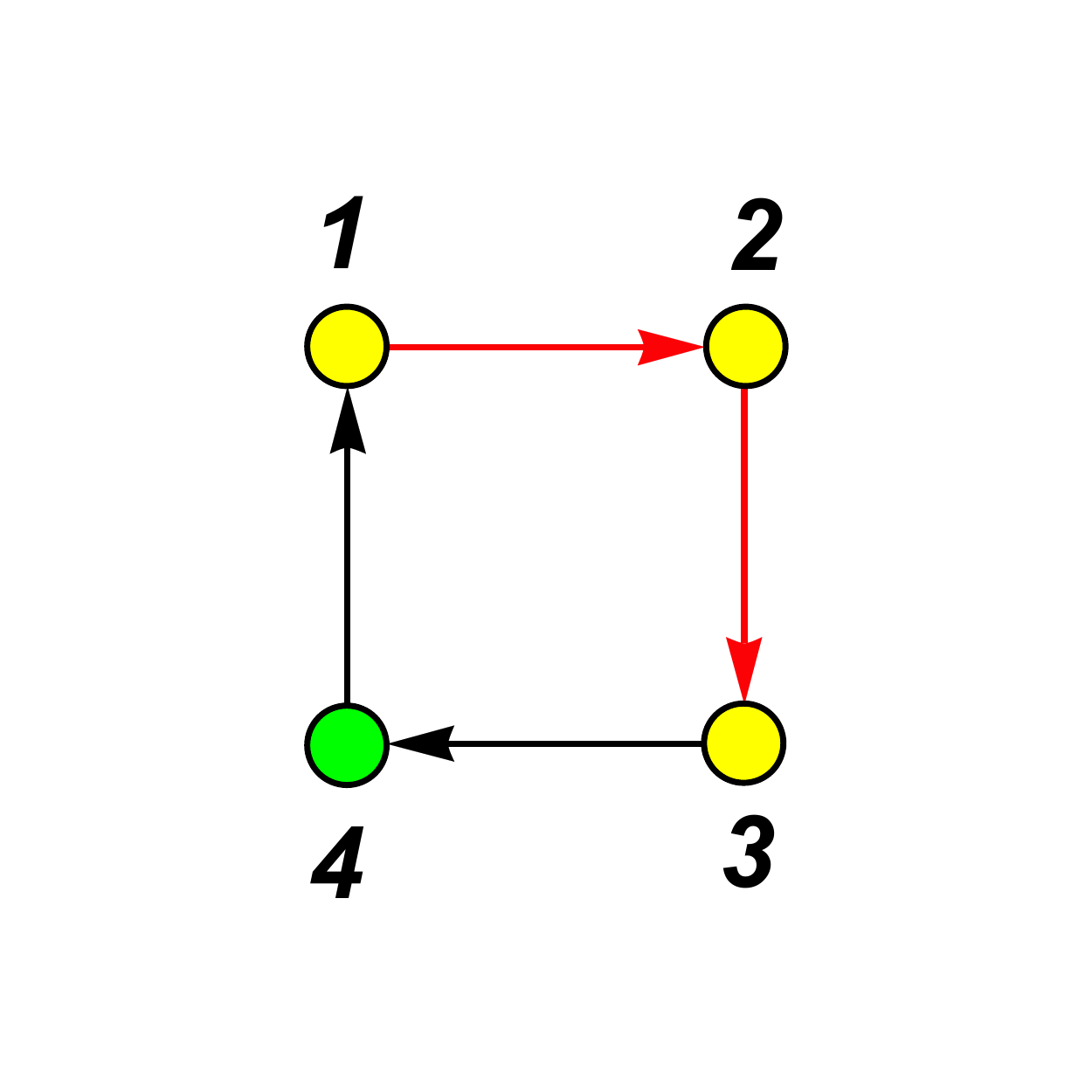}} 
  ,
  \hspace{-0.05cm}
   \parbox[c]{6.1em}{\includegraphics[scale=0.19]{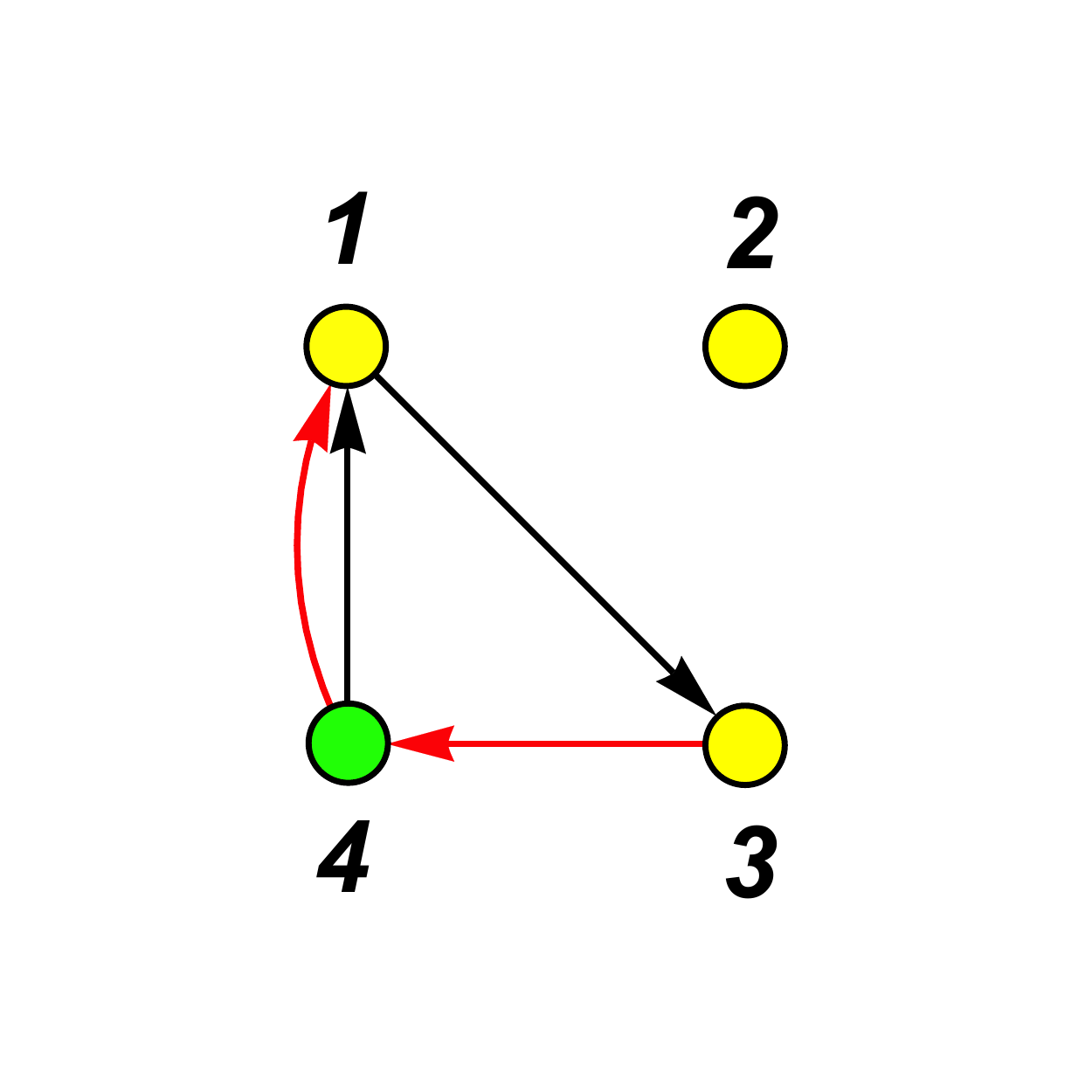}}    
   ~ ,
   \qquad
\end{eqnarray}
}
\vskip-0.3cm\noindent
where the red/black arrows denote a given reduced determinant. Clearly, the first two graphs with reduced matrices, 
$(\mathsf{A}^\L_4)^{12}_{12} \times (\mathsf{A}^\L_4)^{34}_{34}$ and $(\mathsf{A}^\L_4)^{13}_{34} \times (\mathsf{A}^\L_4)^{14}_{14}$, respectively,  have the following singular cuts 
\vspace{-0.2cm}
{\small
\begin{eqnarray}
\hspace{-0.25cm}
 \parbox[c]{4.8em}{\includegraphics[scale=0.19]{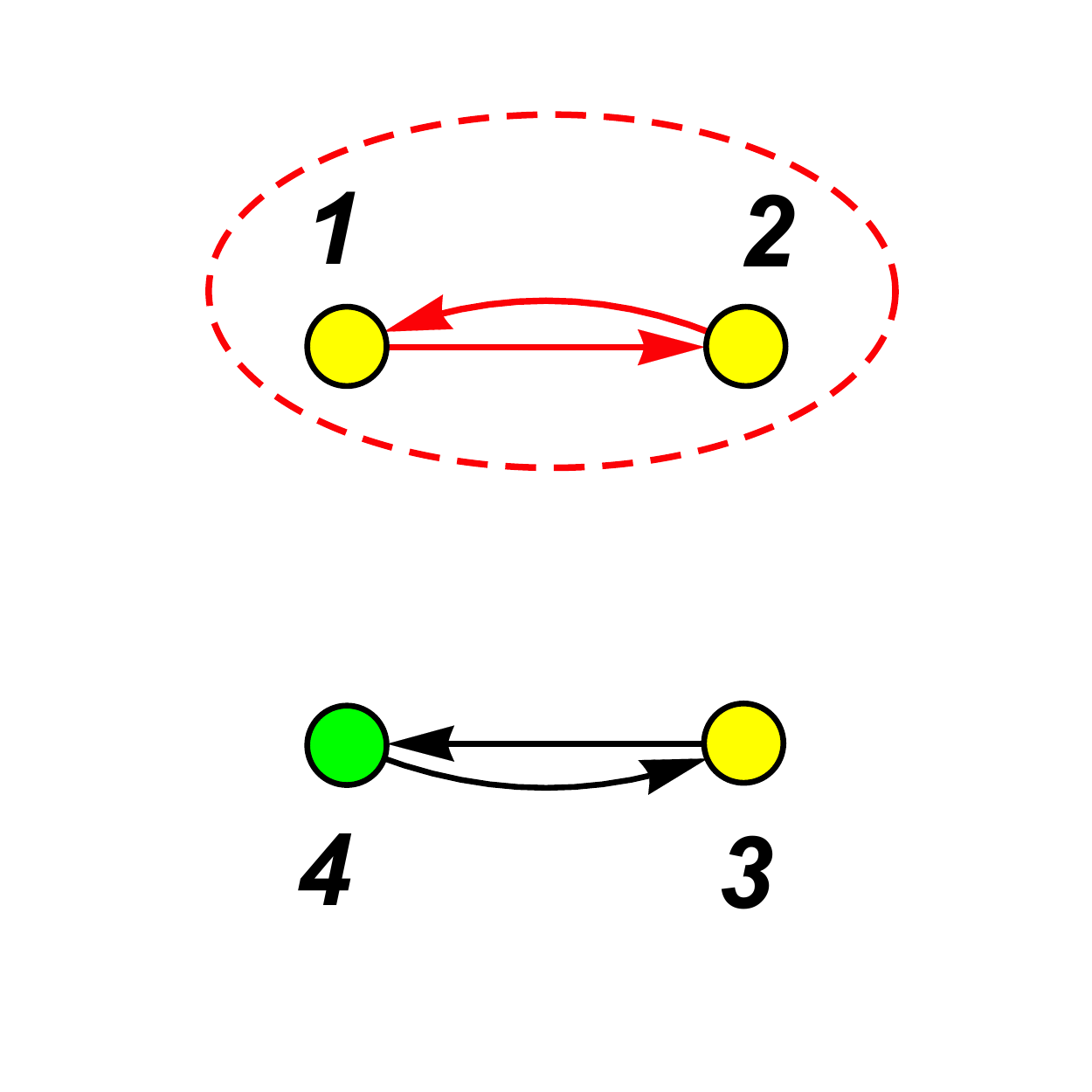}} \rightarrow \,
{\bf det}^\prime \mathsf{A}^\L_4 \times {\bf det}^\prime \mathsf{A}^\L_4 \Big|_{34}^{12} \sim \L^0\, , \quad 
 \parbox[c]{4.6em}{\includegraphics[scale=0.19]{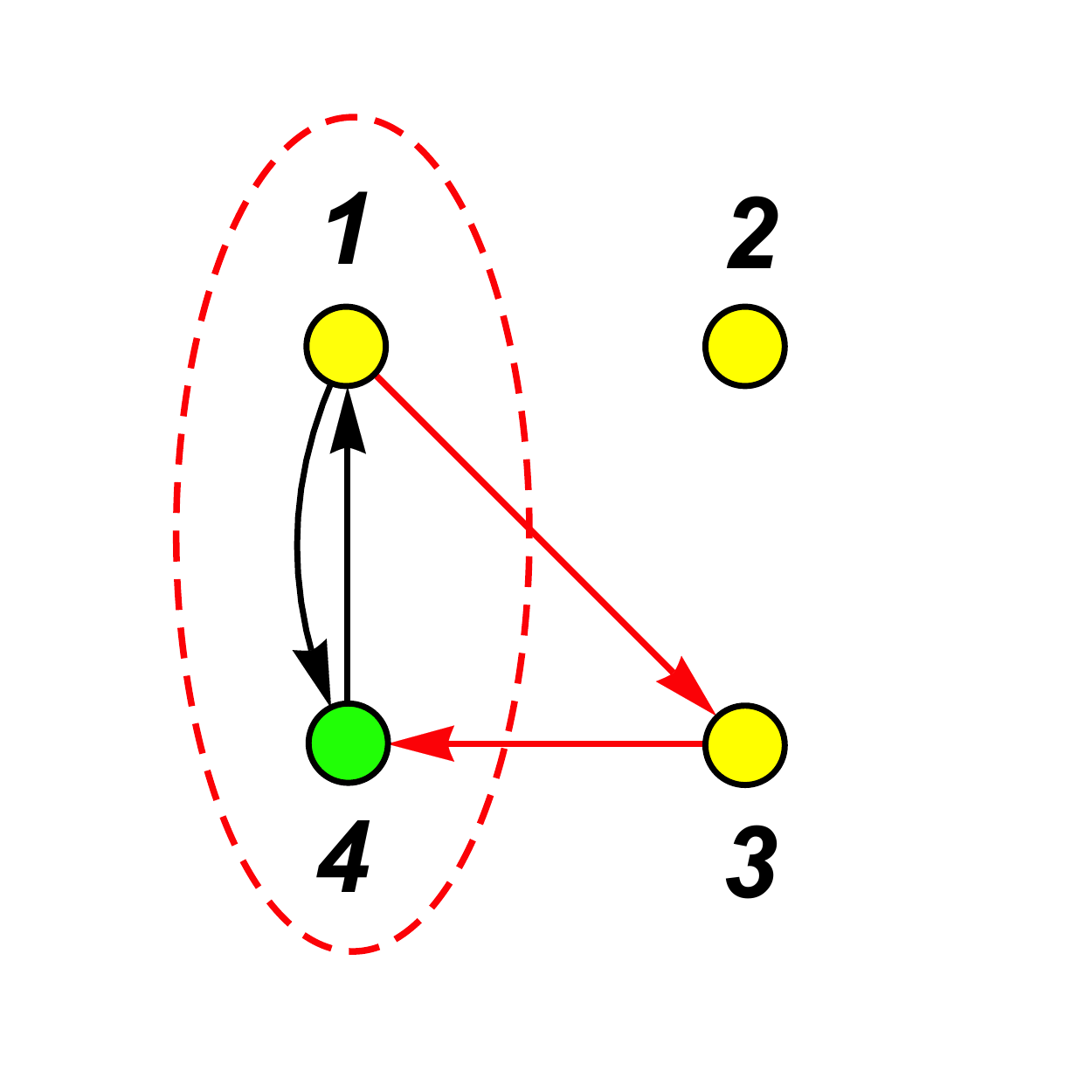}} \rightarrow \,
{\bf det}^\prime \mathsf{A}^\L_4 \times {\bf det}^\prime \mathsf{A}^\L_4 \Big|^{41}_{23} \sim \L^2\, . \nonumber
\end{eqnarray}
}
\vskip-0.3cm\noindent
On the other hand, the third and fourth graphs do not have any singular cuts, therefore, we can apply the integration rules over them.

\subsubsection{The Four-Point Computation}

To carry out the four-point sGal amplitude, we choose the fourth setup in eq.~\eqref{setups}. Thus, from the integration rules, we have three cut contributions given by
\vspace{-0.2cm}
{\small
\begin{eqnarray}
A_4^{\rm sGal}(1,2,3,4)=
\int d\mu_4^\L
\hspace{-0.45cm}
 \parbox[c]{5.1em}{\includegraphics[scale=0.19]{sGalG4.pdf}} =
 \hspace{-0.2cm}
  \parbox[c]{5.2em}{\includegraphics[scale=0.19]{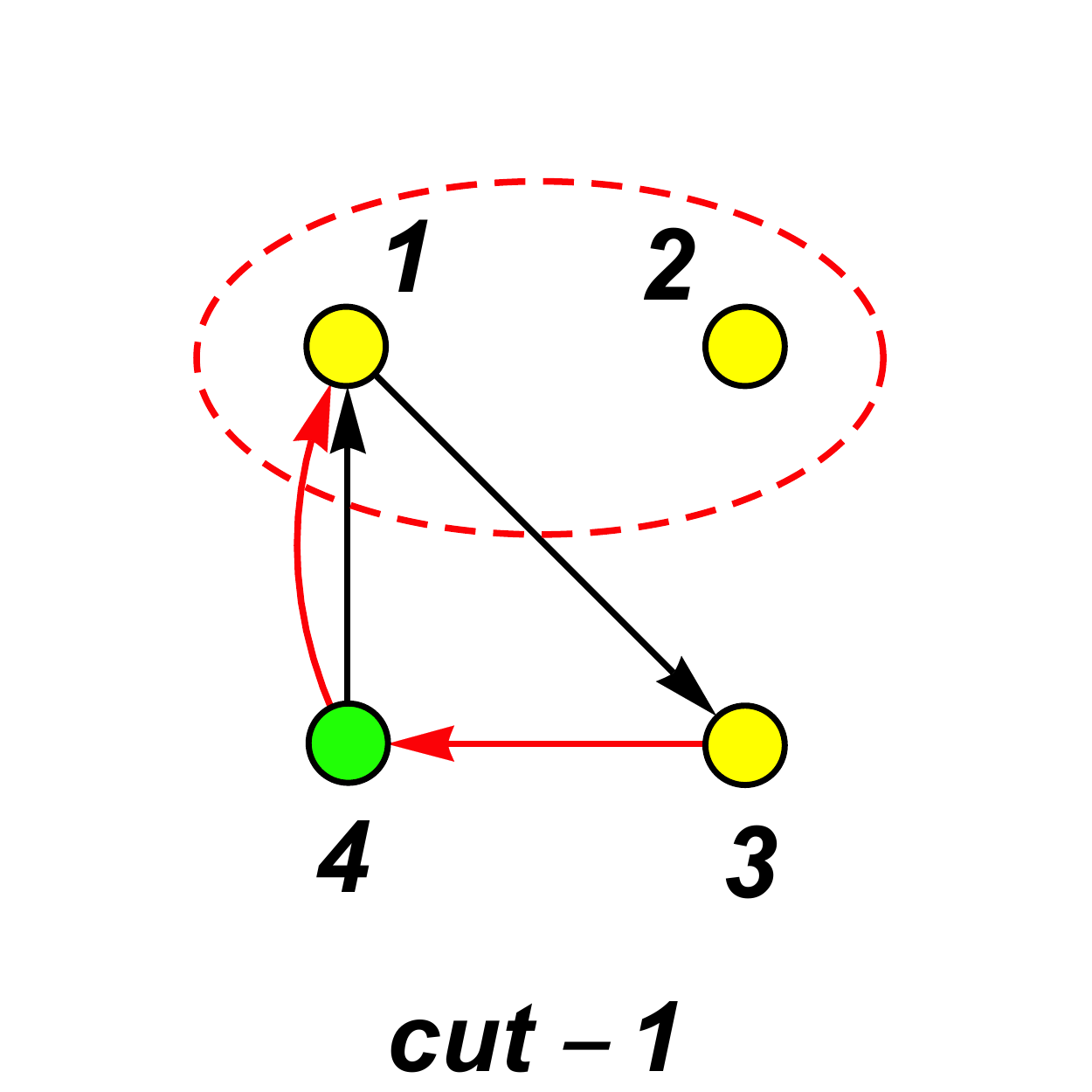}} +
 \hspace{-0.21cm}  
   \parbox[c]{5.2em}{\includegraphics[scale=0.19]{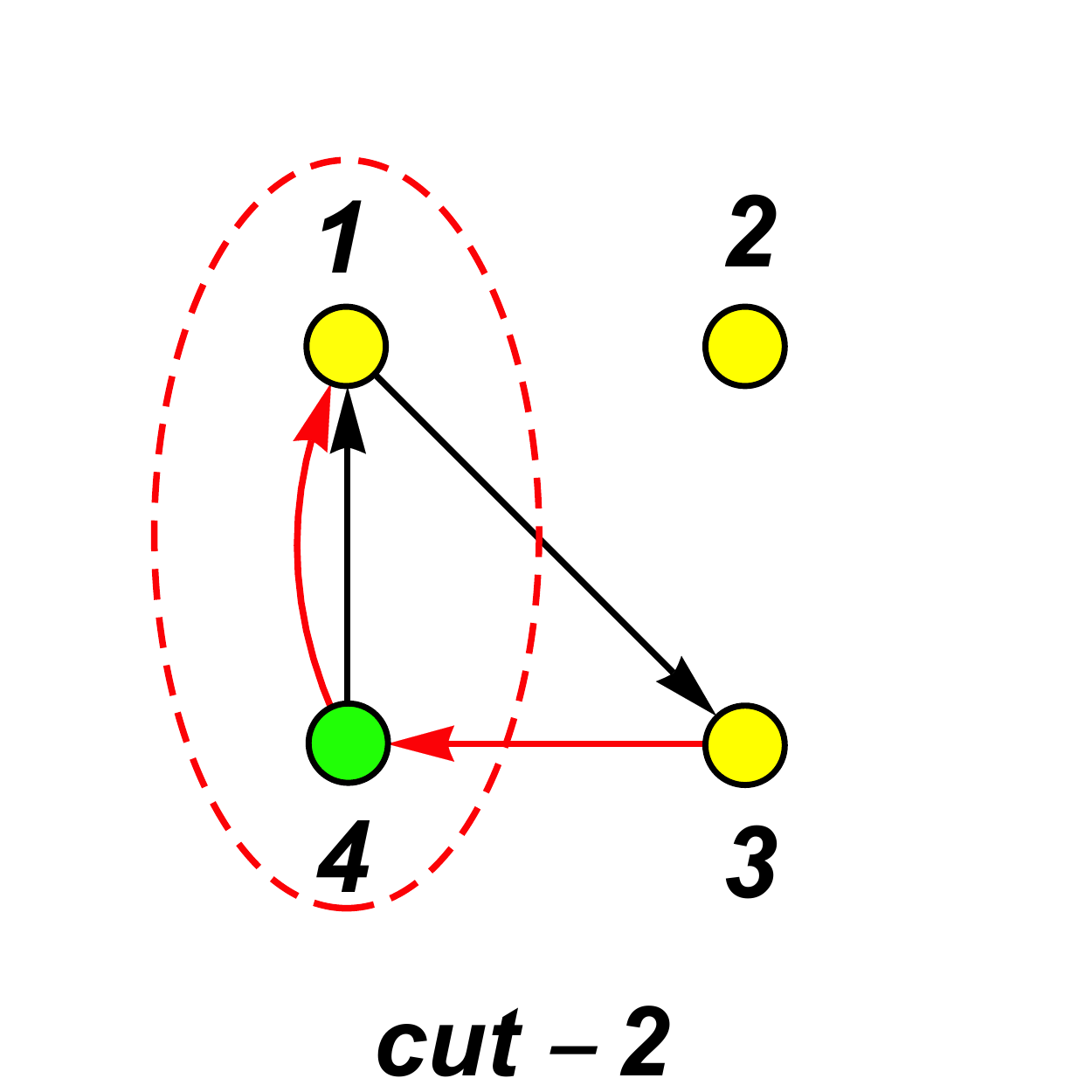}} 
  +
  \hspace{-0.21cm}
   \parbox[c]{6.1em}{\includegraphics[scale=0.19]{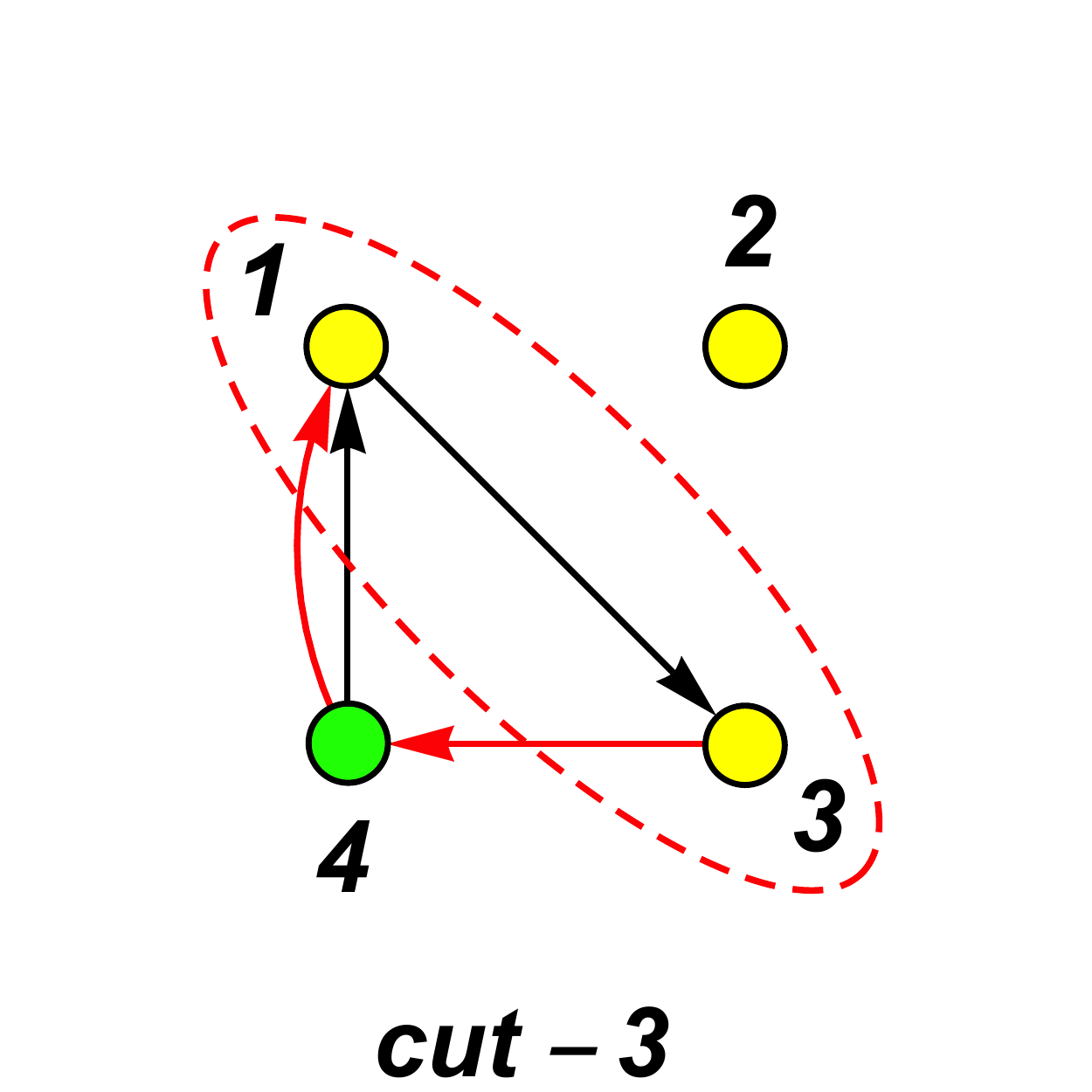}}    
   ~ .
   \qquad
\end{eqnarray}
}
\vskip-0.3cm\noindent
It is straightforward to see that the first contribution vanishes trivially,
\vspace{-0.1cm}
{\small
\begin{eqnarray}
&&
\hspace{-0.8cm}
 \parbox[c]{5.1em}{\includegraphics[scale=0.19]{Gal-cut1.pdf}} = \left[
 \hspace{-0.24cm}
  \parbox[c]{5.4em}{\includegraphics[scale=0.19]{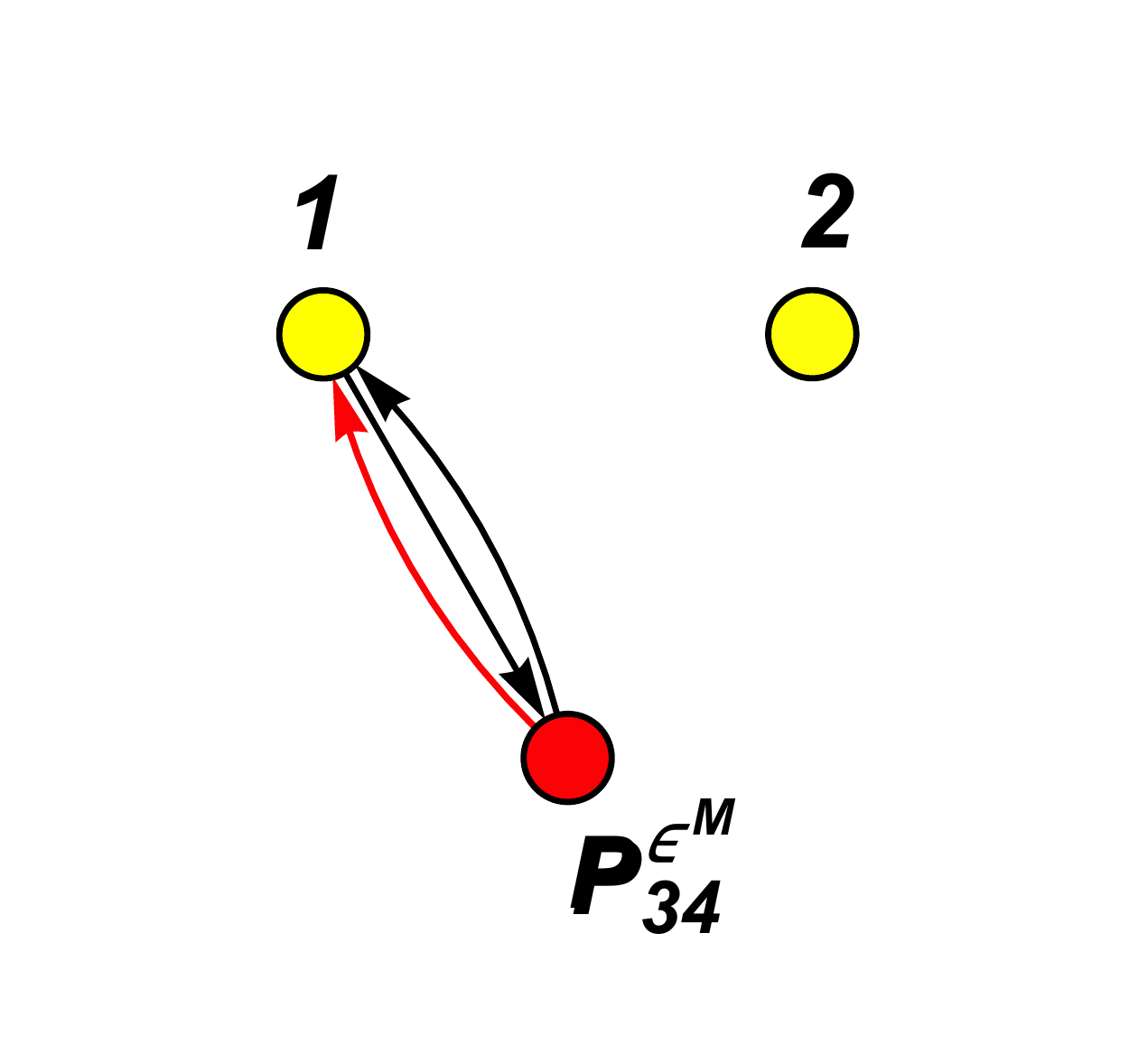}} \right] \times \left( \frac{1}{s_{34}} \right)  \times \left[
 \hspace{-0.24cm}  
   \parbox[c]{5.4em}{\includegraphics[scale=0.19]{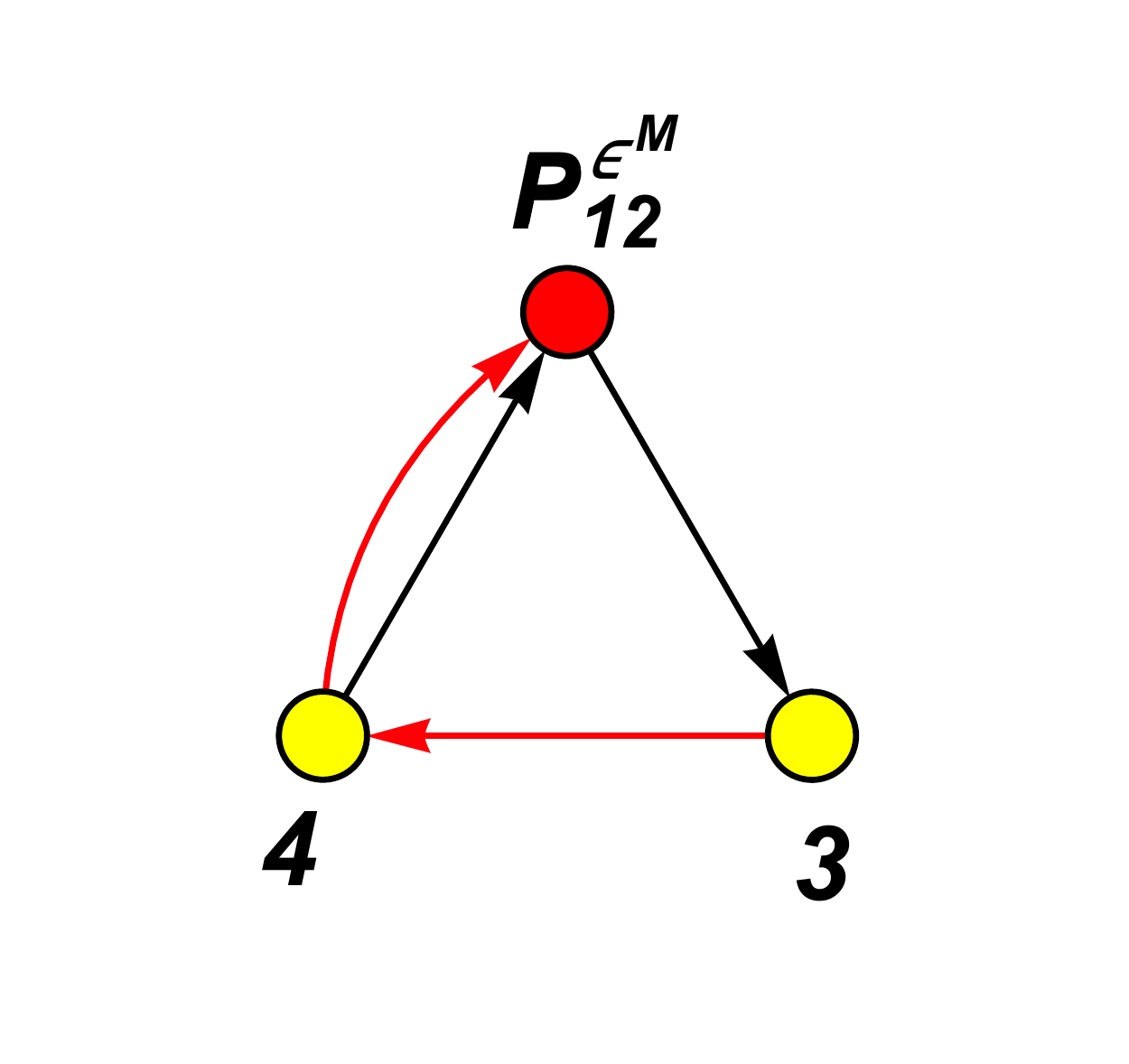}}  \right] = \sum_{M}\,  (\s_{12}\s_{2P_{34}} \s_{P_{34}1})^2 \times  \nonumber \\
   && {\rm PT}(1,P_{34})\,  {\rm det} \Big[(\mathsf{A}_3)^{1P_{34}}_{1P_{34}}\Big]\times 
   \frac{1}{\s_{P_{34} 1}}\,  {\rm det} \Big[(\mathsf{A}_3)^{P_{34}}_{1}\Big]\Big|_{P_{34} \rightarrow \frac{\eps^M_{34}}{\sqrt{2}}}
   \times \left( \frac{1}{s_{34}} \right)  \times \left[
 \hspace{-0.24cm}  
   \parbox[c]{5.4em}{\includegraphics[scale=0.19]{Gal-cut1-R2.pdf}}  \right] =0,
   \nonumber
   \qquad
\end{eqnarray}
}
\vskip-0.3cm\noindent
where we used the identity,  ${\rm det} \Big[(\mathsf{A}_3)^{1P_{34}}_{1P_{34}}\Big]=0$.  
The first and second reduced determinants correspond to the black and red arrows, respectively.
In the following, we associate the first reduced determinant with the black arrows, and the second reduced
determinant with the red arrows.
By a similar computation, the {\it cut-3} also vanishes, then,  the only non-zero contribution comes from the {\it cut-2}. 
\vspace{-0.1cm}
{\small
\begin{eqnarray}
&&
\hspace{-0.8cm}
 \parbox[c]{5.1em}{\includegraphics[scale=0.19]{Gal-cut2.pdf}} = \left[
 \hspace{-0.26cm}
  \parbox[c]{5.9em}{\includegraphics[scale=0.19]{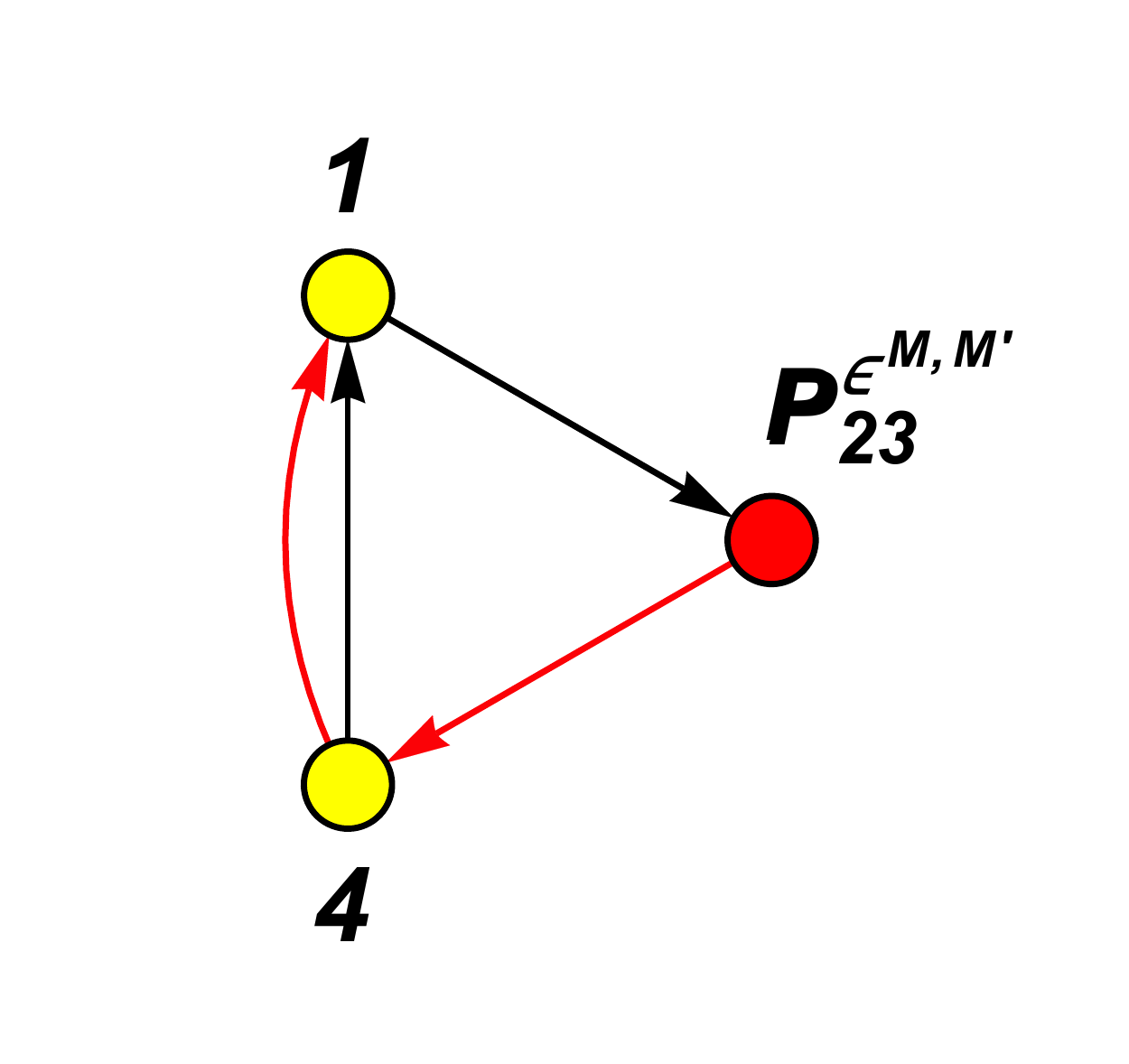}} \right] \times \left( \frac{1}{s_{14}} \right)  \times \left[
 \hspace{-0.24cm}  
   \parbox[c]{5.4em}{\includegraphics[scale=0.19]{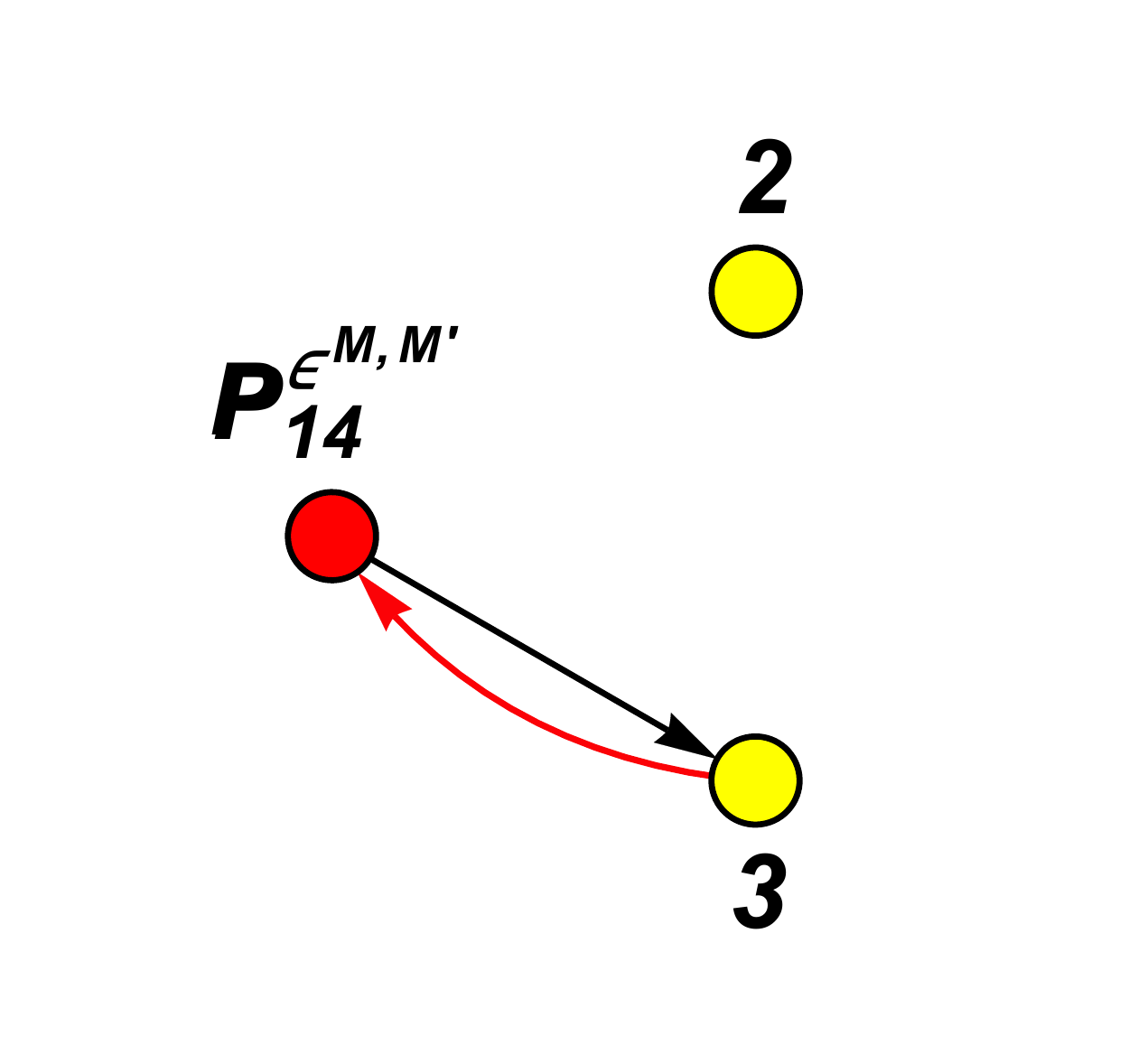}}  \right] = \sum_{M,M^\prime}\,  (\s_{41}\s_{1P_{23}} \s_{P_{23}4})^2 \times  \nonumber \\
   &&
   \left[
 \frac{1}{\s_{41} \s_{1P_{23}}}\,  {\rm det} \Big[(\mathsf{A}_3)^{41}_{1P_{23}}\Big]\Big|_{P_{23} \rightarrow \frac{\eps^M_{23}}{\sqrt{2}}}   
   \times 
    \frac{1}{\s_{P_{23} 4} \s_{41}}\,  {\rm det} \Big[(\mathsf{A}_3)^{P_{23}4}_{41}\Big]\Big|_{P_{23} \rightarrow \frac{\eps_{23}^{M^\prime}}{\sqrt{2}}} \right]
   \times \left( \frac{1}{s_{14}} \right)  \times
   \nonumber\\
   && 
(\s_{23}\s_{3P_{14}} \s_{P_{14}2})^2 \times \left[    
   \frac{1}{ \s_{P_{14}3 }}\,  {\rm det} \Big[(\mathsf{A}_3)_{3}^{P_{14}}\Big]\Big|_{P_{14} \rightarrow \frac{\eps^M_{14}}{\sqrt{2}}}   
   \times 
    \frac{1}{\s_{3P_{14} }}\,  {\rm det} \Big[(\mathsf{A}_3)_{P_{14}}^{3}\Big]\Big|_{P_{14} \rightarrow \frac{\eps_{14}^{M^\prime}}{\sqrt{2}}} \right]
   \qquad  \nonumber\\
   &&
 = -s_{12}\, s_{13} \, s_{14} \, \, ,
 \nonumber  
\end{eqnarray}
}
\vskip-0.3cm\noindent
where the completeness identities, $\sum_{M} \eps_{23}^{\mu\, M} \eps_{14}^{\nu\, M} = \eta^{\mu\nu}$ and $\sum_{ M^\prime} \eps_{23}^{\mu\, M^\prime} \eps_{14}^{\nu\, M^\prime} = \eta^{\mu\nu}$, have been used. Therefore, we obtain
\begin{equation}
A_4^{\rm sGal}(1,2,3,4)= - s_{12}\, s_{13} \, s_{14},
\end{equation}
which is the right answer.

Finally, it is straightforward to generalize this simple example to a higher number of points. Additionally, it would be interesting to understand the properties of the special Galileon theory similar to ones obtained for NLSM in 
\cref{section-Lcontribution,sec:boundaryterms,sec:recursion}.

\section{Conclusions}
\label{sec:conclusion}

The double-cover version of the CHY formalism is an intriguing extension that sheds new light on how scattering amplitudes can emerge as factorized
pieces. Focusing on the non-linear sigma model, we have illustrated how unphysical channels appear at intermediate steps, always
canceling in the end, and thus producing the right answer. The origin of factorizations is the appearance of one ``free" scattering
equation. 
This is the origin of the off-shell
channel through which the amplitudes factorize.

We have analyzed the factorizations obtained in the non-linear sigma model because they perfectly illustrate the mechanism, and the cancellations
that eventually render the full result free of unphysical poles. 
 For this theory, we have obtained three different factorization relationships, two of them emerged naturally from the double-cover framework (by using the $A_{2n}$ and $A^\prime_{2n}$ prescriptions), while the other one was obtained fortuitously by considering the longitudinal degrees of freedom of the 
cut-contributions from the new $A^\prime_{2n}$ prescription. 
By comparing to BCFW on-shell recursion relations we have found a perfect 
correspondence between the unphysical terms of the double-cover formalism and terms that arise from poles at infinity in the
BCFW formalism. In that sense, the double-cover version of CHY succeeds in evaluating what appears as poles at infinity in BCFW
recursion as simple CHY-type integrals of the double cover. It would be interesting if this correspondence could be made more explicit.
Certainly, it hints at the possibility that an alternative formulation of the problem of poles at infinity in BCFW recursion exists, without
recourse to the particular double-cover formalism.


Using the new prescription for the reduced determinant in the integrand, we found a
factorization relation where all the intermediate off-shell particles are spin-1 (gluons).
The corresponding momenta in the reduced determinants are replaced by polarization vectors.
We would like to investigate further the connection between this new object and the
integrand for generalized Yang-Mills-Scalar theory \cite{Cachazo:2014xea}.
At first sight, we thought that this new matrix could be related to the novel model proposed by  Cheung, Remmen, Shen,  and Wen in
\cite{Cheung:2016prv,Cheung:2017yef}, nevertheless, after comparing the numerators at the four-point computation, the relation among these two approaches is unclear.

On the other hand, when we replaced the off-shell gluons with only the longitudinal degrees of 
freedom, we were able to rewrite the factorized pieces in terms of lower-point NLSM
amplitudes in the new prescription, with up to three off-shell punctures. This is a very 
surprising result, and understanding the origin of this connection is left for future work.
The big advantage of being able to rewrite the factorized pieces is that we can iteratively
promote the lower-point NLSM amplitudes to the double cover, which would lead to further 
factorization. Thus, any NLSM amplitude can be factorized entirely in terms of 
off-shell three-point amplitudes. This is a novel off-shell recursion relation. The resulting 
expression is algebraic, and no scattering equation needs to be solved.  We have checked the 
validity of the recursion relation up to ten points (17 points for odd amplitudes).
We would like to find the connection between the recursion relation and Berends-Giele currents \cite{Berends:1987me,Chen:2014dfa,Chen:2013fya,Kampf:2013vha,Kampf:2012fn,Mizera:2018jbh}.

The novel recursion relation can also be used to investigate singular cuts and 
${\rm NLSM}\oplus \phi^3$ amplitudes through the soft limit. CCM showed how the 
soft limit of an NLSM amplitude can be expressed in terms of ${\rm NLSM}\oplus \phi^3$ amplitudes
\cite{Cachazo:2016njl}.
We calculated the soft limit of a six-point NLSM amplitude 
in two ways, using the CCM formula and using the novel recursion relation.
This gives a relation for a specific singular cut.
Further investigations into the nature of the soft limits might reveal insight into the singular
cuts in general.
Also, we were able to find a factorization relation for the ${\rm NLSM}\oplus \phi^3$ amplitudes.

Lastly, we showed how the special Galileon amplitudes can be calculated in a double 
cover language. One intriguing feature is that for some configurations, the off-shell
particle propagating between the lower-point pieces is spin-2 (graviton).
So, we have observed that for the NLSM, off-shell gluons appear, while for the 
special Galileon theory, both off-shell gluons and gravitons appear. This might
be connected to the fact that the NLSM originated as an effective theory of 
pion scattering, while the Galileon theories arise as 
effective field theories in the decoupling limit of massive gravity.
This also seems natural, as the special Galileon theory is the square of the 
NLSM, using the KLT relation.

It seems evident that there are numerous aspects of CHY on a double cover that need to be investigated.

\acknowledgments
We would like to thank N. E. J. Bjerrum-Bohr and P. H. Damgaard for valuable discussions.
We are grateful to C. Vergu for pointing out the Desnanot-Jacobi identity.
This work was supported in part by the Danish National Research Foundation (DNRF91).
The work of  H.G.  is supported by 
the Niels Bohr Institute - University of Copenhagen and the Santiago de Cali University (USC). 

\appendix

\section{Some Matrix Identities}\label{apx:matrixProof}

In this section, we are going to provide some useful properties of the determinant of the 
$\mathsf{A}_n$ matrix. Although we lack formal proofs for many of the relations, 
we have performed numerous checks, up to ten points.

\subsection{A New NLSM Prescription from CHY}\label{apx:matrixProof1}

In this appendix, we formulate two propositions which have been employed to redefine 
the $n$-point NLSM amplitude from the CHY framework.

{\bf Proposition 1:} Let $M$ be a  $2n\times 2n$ antisymmetric matrix. Then $M$ satisfy the identity
\begin{equation}\label{Prop1}
{\rm Pf}\Big[ (M)_{ik}^{ik}  \Big] \times {\rm Pf} \Big[ (M)_{kj}^{kj}  \Big]  =  {\rm det}\Big[ (M)^{ik}_{kj}  \Big],
\end{equation}
up to an overall sign.

{\bf Proof:}
We start with the Desnanot-Jacobi identity \cite{Jacobi}, given by
\begin{equation}\label{DJ-identity}
{\rm det}\left[ M  \right]\, {\rm det}\left[ (M)_{ij}^{ij}  \right] = 
{\rm det}\left[ (M)_{i}^{i}  \right] \, {\rm det}\left[ (M)_{j}^{j}  \right] - {\rm det}\left[ (M)^{i}_{j}  \right] \, {\rm det}\left[ (M)^{j}_{j}  \right].
\end{equation}
Now, let $M$ be a $2n\times 2n$ antisymmetric  matrix, therefore,  $(M)^k_k$ is a $(2n-1)\times (2n-1)$ antisymmetric matrix. Thus,  from the identity in eq.~\eqref{DJ-identity}, it is straightforward to see that
\begin{equation}
0= 
{\rm det}\left[ (M)_{ki}^{ki}  \right] \, {\rm det}\left[ (M)_{kj}^{kj}  \right] - {\rm det}\left[ (M)^{ki}_{kj}  \right] \, {\rm det}\left[ (M)_{ki}^{kj}  \right], 
\end{equation}
where we used the fact, ${\rm det}\left[ (M)_{k}^{k}  \right] ={\rm det}\left[ (M)_{kij}^{kij}  \right] =0$. Since, $[(M)^{kj}_{ki}]= [(M)^{ki}_{kj}]^{\rm t} = - [(M)^{ki}_{kj}] $, then
\begin{equation}
\left\{ {\rm Pf}\Big[ (M)_{ik}^{ik}  \Big] \times  {\rm Pf}  \Big[ (M)_{kj}^{kj}  \Big]  \right\}^2 =  \left\{ {\rm det} \Big[ (M)^{ik}_{kj}  \Big] \right\}^2,
\end{equation}
and {\bf proposition 1} has been proved.

{\bf Proposition 2:} Let $\mathsf{A}$ be the antisymmetric matrix defined in eq.~\eqref{eq:defA}. When its size is $(2n+1 )\times (2n+1)$, then
\begin{equation}\label{Prop2}
 {\rm det}\Big[ (  \mathsf{A}  )^{ik}_{kj}  \Big] =0.
\end{equation}

{\bf Proof:} Let us consider the $2n\times 2n$ antisymmetric matrix given by $( \mathsf{A} )^k_k$. Thus, from the Desnanot-Jacobi identity in eq.~\eqref{DJ-identity}, one has
\begin{equation}
{\rm det}\Big[ (\mathsf{A})^k_k  \Big]\times  {\rm det}\Big[ (\mathsf{A})_{kij}^{kij}  \Big] = 
- \left\{ {\rm det}\Big[ (\mathsf{A})^{ik}_{kj}  \Big] \right\}^2,
\end{equation}
where we used, ${\rm det}\Big[ (\mathsf{A})^{ki}_{ki}  \Big]={\rm det}\Big[ (\mathsf{A})^{kj}_{kj}  \Big]=0$. Under the support of the scattering equations, $S_a=0$, and the on-shell conditions, $k_a^2=0$,  it is simple to show that the $\mathsf{A} $ matrix has co-rank 2, therefore, ${\rm det}\Big[ (\mathsf{A})^k_k  \Big]=0$.  This implies that, 
${\rm det}\Big[ (\mathsf{A})^{ik}_{kj}  \Big] =0$, and the proof is completed.

\subsection{Off-shell Determinant Properties}\label{Pf-properties}

In this appendix we give some properties of the determinant when there is an off-shell particle. 
These properties involve the matrices, $\mathsf{A}_n$ and $\mathsf{A}_n\Big|_{P_i\rightarrow\frac{1}{\sqrt{2}} \eps_i }$.

This is very important to remark that those properties are supported on the solution of the scattering equations, and, although we do not have a formal proof, they have been checked up to ten points.

Let us consider $n$-particles with momenta, $(P_1,P_2,P_3,k_4,...,k_n)$, where the first three are off-shell,  {\ie}\, $P_i^2\neq 0$, and the momentum conservation condition is satisfied, $P_1+P_2+P_3+k_4\cdots + k_{n}=0$. Additionally, the three off-shell punctures are fixed, 
$\s_{P_1}=\text{c}_1, \,  \s_{P_2}=\text{c}_2, \,  \s_{P_3}=\text{c}_3,\, \text{c}_i \in \mathbb{C}$, where $\text{c}_1\neq \text{c}_2 \neq \text{c}_3$. Thus,  the ``$n-3$" scattering equations are given by 
\begin{eqnarray}
S_a=\frac{2\,k_a\cdot P_1}{\s_{aP_1}}+\frac{2\,k_a\cdot P_2}{\s_{aP_2}}+\frac{2\,k_a\cdot P_3}{\s_{aP_3}} + \sum_{b=4\atop a\neq b  }^{n} \frac{2\, k_a\cdot k_b}{\s_{ab}}=0, \quad  a=4, \ldots , n.
\end{eqnarray}

{\bf Properties:} \\
Under the support of the scattering equations and using the above setup, we have the following  properties
\begin{itemize}
\item[\bf I.] Let $n$ an odd number, $n=2m+1$, then
\begin{equation}\label{firstP}
{\rm det}\left[\left(\A_n\right)^{P_1}_{P_2}   \right] = (P_1^2-P_2^2-P^2_3) \times \frac{(-1) }{\s_{P_2\, P_3}} \, {\rm det}\left[\left(\A_n\right)^{P_1P_2}_{P_2P_3}   \right] .
\end{equation}
Notice that if  all particles are on-shell, $P_i^2=0$, the right hand side vanishes trivially by the overall factor, $(P_1^2-P_2^2-P^2_3)$.  

When the momentum $P^\mu_1$ is replaced by an off-shell polarization vector, $P^\mu_1\rightarrow \frac{1}{\sqrt{2}} \epsilon^\mu_{1}$, ($\eps_{1}\cdot P_1\neq 0$), the identity keeps the same form, namely
\begin{equation}
\left. {\rm det}\left[\left(\A_n\right)^{P_1}_{P_2}   \right]\right|_{P^\mu_1\rightarrow \frac{1}{\sqrt{2}} \epsilon^\mu_{1}} = (P_1^2-P_2^2-P^2_3) \times \frac{(-1) }{\s_{P_2 P_3}} \, \left. {\rm det}\left[\left(\A_n\right)^{P_1P_2}_{P_2P_3}   \right] \right|_{P^\mu_1\rightarrow \frac{1}{\sqrt{2}} \epsilon^\mu_{1}}.
\end{equation}
This identity is no longer satisfied if there are two off-shell polarization vectors.

\item[\bf II.] Let $n$ an even number, $n=2m$, then
\begin{eqnarray}\label{secP}
 \frac{(-1) }{\s_{P_1 P_2}} 
{\rm det}\left[\left(\A_n\right)^{P_1}_{P_2}   \right] &=& -(P_1^2+P_2^2+P^2_3) \times \frac{1 }{\s_{P_1 P_2}\, \s_{P_2 P_3} } \, {\rm det}\left[\left(\A_n\right)^{P_1P_2}_{P_2P_3}   \right] \nonumber  \\
&=& -(P_1^2+P_2^2+P^2_3) \times \frac{ (-1) }{\s_{P_1 P_2}\, \s_{P_2 P_1} } \, {\rm det}\left[\left(\A_n\right)^{P_1P_2}_{P_2P_1}   \right] .
\end{eqnarray}
If  all particles are on-shell, $P_i^2=0$, the right hand side vanishes trivially by the overall factor, $(P_1^2+P_2^2+P^2_3)$.  

When the momentum $P^\mu_1$ is replaced by an off-shell polarization vector, $P^\mu_1\rightarrow \frac{1}{\sqrt{2}} \epsilon^\mu_{1}$, ($\eps_{1}\cdot P_1\neq 0$), then, eq.~\eqref{secP}  is no longer an identity. Instead, we have a new identity given by
\begin{equation}
 \frac{(-1) }{\s_{P_1\, P_2}} 
\left. {\rm det}\left[\left(\A_n\right)^{P_1}_{P_2}   \right]\right|_{P^\mu_1\rightarrow \frac{1}{\sqrt{2}} \epsilon^\mu_{1}} =  \frac{1 }{\s_{P_1\, P_3}} 
\left. {\rm det}\left[\left(\A_n\right)^{P_1}_{P_3}   \right]\right|_{P^\mu_1\rightarrow \frac{1}{\sqrt{2}} \epsilon^\mu_{1}} .
\end{equation}
 If there are two off-shell polarization vectors, then, this  equality is no longer true.

\item[\bf III.] Let $n$ an odd number, $n=2m+1$, and  let us consider the particles $P_1$ and $P_2$ on-shell ($P_1^2=P_2^2=0$). Then, we have the following identities
\begin{equation}\label{thirdP}
\frac{ 1 }{\s_{P_1\, P_3}} \, {\rm det}\left[\left(\A_n\right)^{P_1}_{P_1}   \right] =  \frac{(-1) }{\s_{P_2\, P_3}} \, {\rm det}\left[\left(\A_n\right)^{P_1}_{P_2}   \right] ,
\end{equation}
\begin{equation}
{\rm det}\left[\left(\A_n\right)^{P_1}_{P_1}   \right] .
=\left[ P_1^2\times
\frac{ 1 }{\s_{P_2\, P_3}} \right]^2\, {\rm det}\left[\left(\A_n\right)^{P_1P_2P_3}_{P_1P_3P_3}   \right] .
\end{equation}

\end{itemize}

\section{Six-Point Computations}\label{6-pts-comp}

In this section we are going to explicitly calculate the six-point NLSM amplitudes 
$A_6^\prime(\mathbb{I}^{(123)})$, $A_6^\prime(\mathbb{I}^{(134)})$ and 
$A_6(\mathbb{I}^{(13)})$, where the two first are defined with the new integrand prescription,
while the third is defined with the standard integrand.
We will calculate some of the cut-contributions in detail, with the hope that the reader
becomes more familiar with the double cover formalism. 
The rest of the cut-contributions can be computed in a similar way.

\subsection{$A_{6}^{\prime}(\mathbb{I}^{(123)})$}
\label{sec:sixpointex1}

Let us consider the six-point NLSM amplitude, $\mathcal{A}_6 (1,2,3,4,5,6)$, 
with the gauge fixing, $(pqr|m)=(123|4)$, 
and the reduced matrix $[\mathsf{A}_n]^{12}_{23}$  (\ie\, $(ijk)=(123)$).  
Applying {\bf rule-I}, this amplitude 
has the following contributions
\vspace{-0.2cm}
{\small
\begin{equation}\label{cuts-6ptsAp}
	A_{6}^{\prime}(\mathbb{I}^{(123)}) =
\hspace{-0.15cm}
 \parbox[c]{6.1em}{\includegraphics[scale=0.19]{6pts-ex1-cut1.pdf}} +
 \hspace{-0.1cm}
  \parbox[c]{7.1em}{\includegraphics[scale=0.19]{6pts-ex1-cut2.pdf}}
 \hspace{-0.36cm}  
  +
   \parbox[c]{6.1em}{\includegraphics[scale=0.19]{6pts-ex1-cut3.pdf}} 
  +
  \hspace{-0.05cm}
   \parbox[c]{6.1em}{\includegraphics[scale=0.19]{6pts-ex1-cut4.pdf}}    
   ~ .
   \qquad
\end{equation}
}
\vskip-0.1cm\noindent

We will compute in detail the first contribution, which we call {\it cut-1}.
The other cuts can be evaluated using the same techniques. 

From the {\bf integration rules}, {\it cut-1} is evaluated as 
\vspace{-0.2cm}
{\small
\begin{equation}\label{cut1Ap}
\parbox[c]{6.2em}{\includegraphics[scale=0.19]{6pts-ex1-cut1.pdf}}=\sum_M
 \frac{ A^{\prime}_3(1,2, P^{M}_{3:6}) \times   A^{(P_{12}\,3)}_5(P^{M}_{12},3,4,5,6)}{s_{3:6}} . 
 \qquad~
\end{equation}
}
\vskip-0.1cm\noindent
The three-point amplitude was already computed in \cref{3pt-1}. We remind ourselves that the 
notation $P_{3:6}^{M}$ means that the off-shell momentum, $P^\mu_{3:6}$, 
must be replacement by the polarization vector, 
$P^\mu_{3:6} \rightarrow  \frac{1}{\sqrt{2}} \, \eps^{M\, \mu }_{3:6}$.  More precisely, the 
three-point amplitude becomes
\begin{equation}
A^{\prime}_3(1,2, P^{M}_{3:6}) = \sqrt{2}\, ( \epsilon_{3:6}^M\cdot k_1) ~ .
\end{equation}

Before computing the five-point amplitude in \cref{cut1Ap}, 
it is useful to use the identity, 
$ A^{(P_{12}\,3)}_5(P^{M}_{12},3,4,5,6)= P_{12}^2\times A^{\prime}_5( P^{M}_{12}, 3, 4,5,6)$. 
Thus, by applying the integration rules for $A^{\prime}_5( P^{M}_{12},3,4,5,6)$ one has
{\small
\begin{align}
&
A^{\prime}_5(P^{M}_{12},3,4,5,6) = 
\int d\mu_5^\L
\hspace{-0.3cm}
\parbox[c]{5.4em}{\includegraphics[scale=0.2]{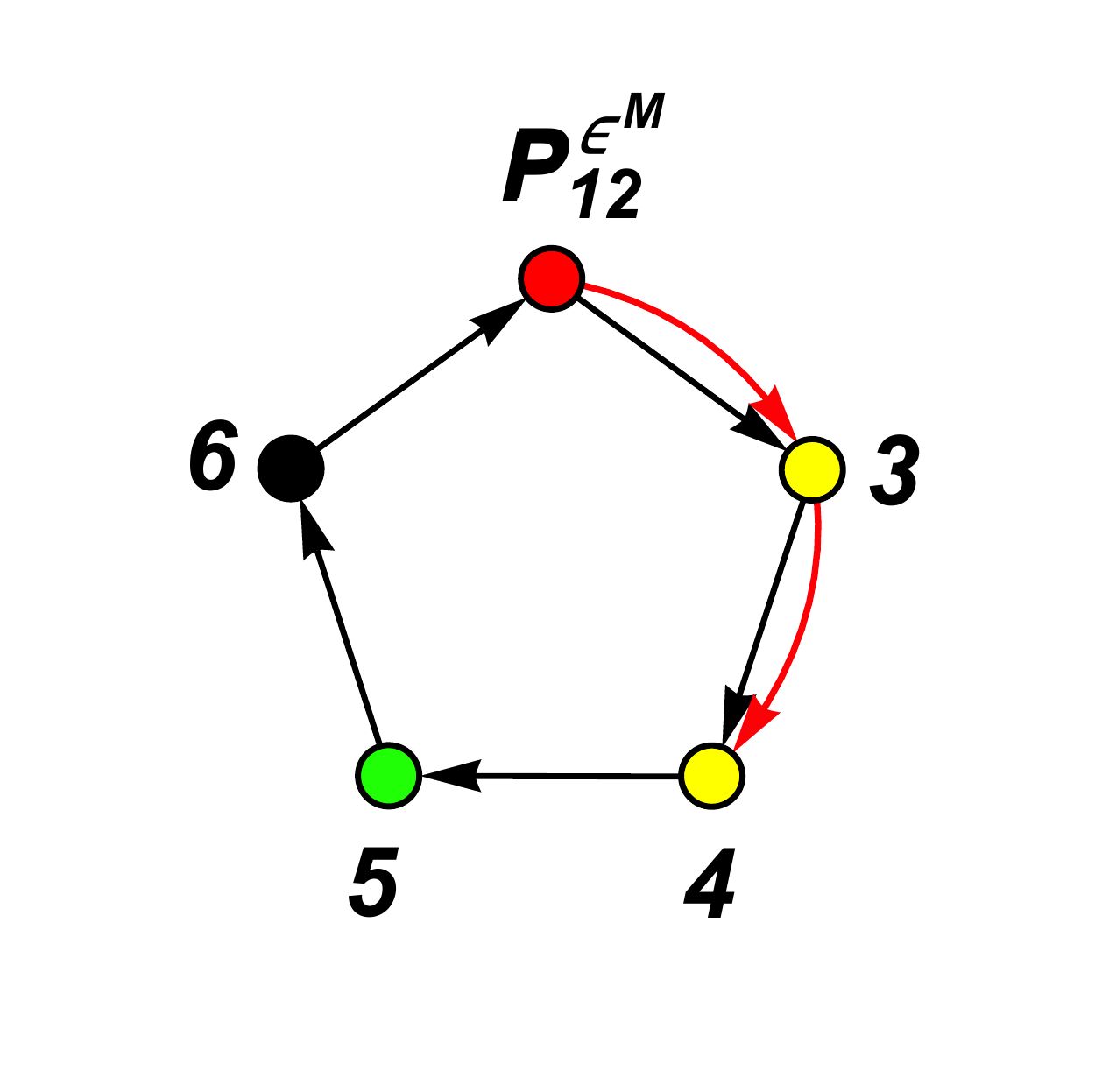}}
=\sum_N \left\{
 \frac{ A^{\prime}_3( P^{M}_{12},3, P^{N}_{4:6}) \times   A^{(P_{1:3}\,4)}_4(P^{N}_{1:3},4,5,6)}{s_{4:6}}  \,  + \right.  \nonumber\\
 &
 \frac{   
  A^{(P_{12}P_{34})}_4(P^{M}_{12},P^{N}_{34},5,6) 
 \times A^{\prime}_3( P^{N}_{5:2} ,3,4 ) } {s_{56P_{12}}}
  +
  \left.
 \frac{   
 A^{\prime}_4(P^{M}_{12}, 3 , P^{N}_{45},6) 
 \times  A^{( P_{6:3}\, 4 )}_3(  P^{N}_{6:3} , 4 , 5  ) } {s_{45} } \right\} \! ,
 ~\qquad~\label{A5-p12Ap}
\end{align}
}
\vskip-0.2cm\noindent
with $\sum_{N} \eps_{i}^{N\, \mu} \eps_{j}^{N\, \nu} = \eta^{\mu\nu}$.
From the building blocks in \cref{3pt-1,3point-BB}, 
the above three-point amplitudes are straightforward to compute. We find that
{\small
\begin{equation}\label{eq:a3-amplitudesAp}
 A^{\prime}_3( P^{M}_{12} ,3, P^{N}_{4:6})  
= \epsilon_{12}^M \cdot \epsilon_{4:6}^N , 
~~
 A^{\prime}_3( P^{N}_{5:2} ,3, 4 )  
= \sqrt{2}  \, \epsilon_{5:2}^N \cdot  k_4 ,
~~ 
 A^{( P_{6:3}\,4 )}_3(  P^{N}_{6:3} , 4, 5  )  
= \sqrt{2}  \, s_{45}\,  (\epsilon_{6:3}^N \cdot  k_5).
\end{equation}
}
\vskip-0.5cm\noindent
Next, using the same procedure as in eq.~\eqref{new-exp1},  
we evaluate the four-point graph,  
$  A^{\prime}_4(P^{M}_{12} , 3 ,P^{N}_{45},6)$, arriving at
{\small
\begin{equation}\label{eq:a4prime}
 A^{\prime}_4( P^{M}_{12} ,3, P^{N}_{45},6)  
= 2 (\eps^M_{12} \cdot k_6  ) \, (\eps^N_{45} \cdot k_6  ) \, \left( \frac{1 }{s_{6P_{45}}} + \frac{1 }{s_{6P_{12}}} \right) .
\end{equation}
}
\vskip-0.0cm\noindent
On the other hand, in order to avoid singular cuts when applying the integration rules over 
$A^{(P_{1:3}\,4)}_4(P^{N}_{1:3},4,5,6)$,  we  employ  the identity,  
$A^{(P_{1:3}\,4)}_4(P^{N}_{1:3},4,5,6)=A^{(P_{1:3}\,5)}_4(P^{N}_{1:3},4,5,6)$. 
Thus,  
{\small
	\begin{align}
A^{(P_{1:3}\,5)}_4(P^{N}_{1:3},4,5,6)
&=  
\int d\mu_4^\L
\hspace{-0.6cm}
\parbox[c]{5.4em}{\includegraphics[scale=0.21]{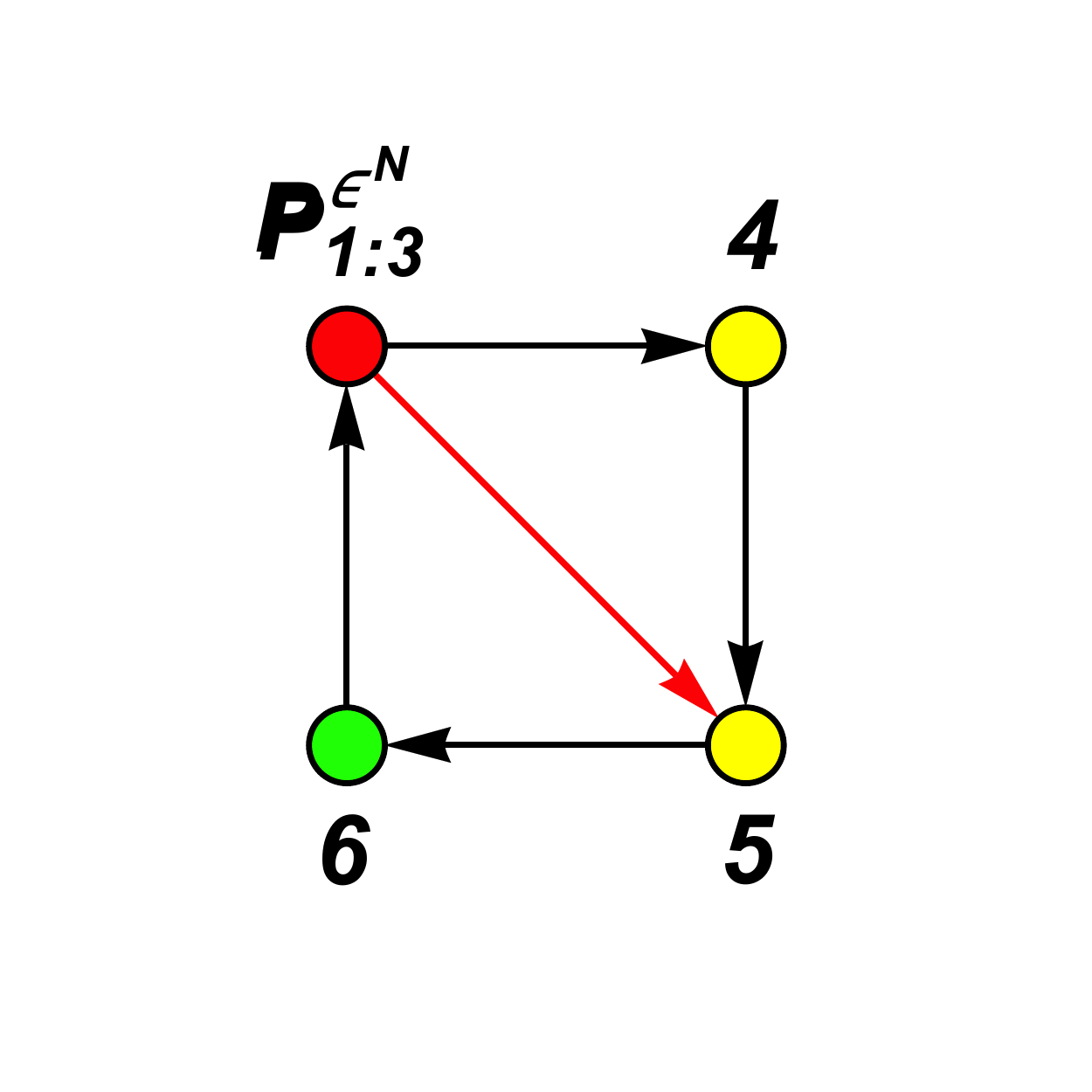}}
=
\hspace{-0.6cm}
\parbox[c]{5.4em}{\includegraphics[scale=0.21]{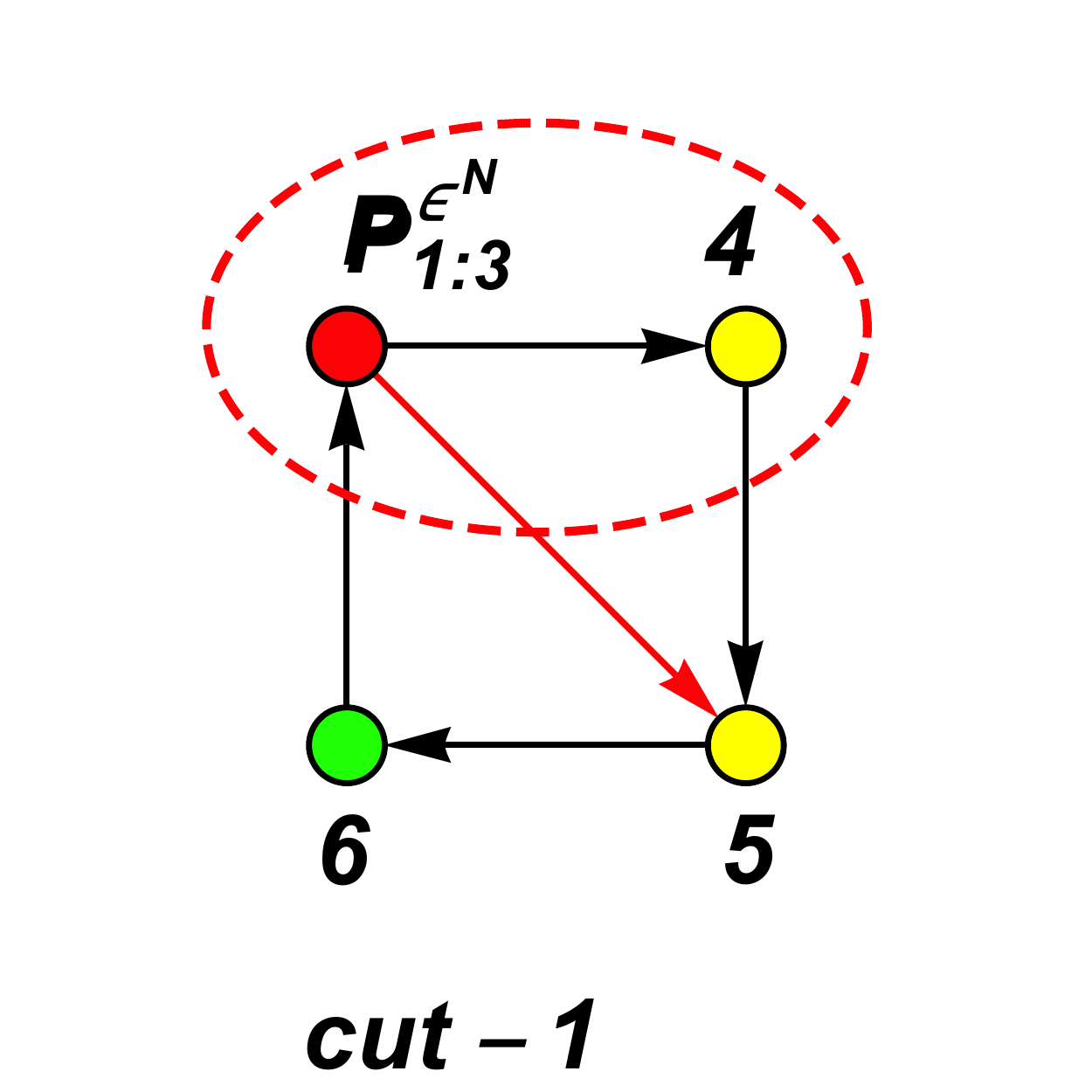}} +
\hspace{-0.6cm}
\parbox[c]{6.4em}{\includegraphics[scale=0.21]{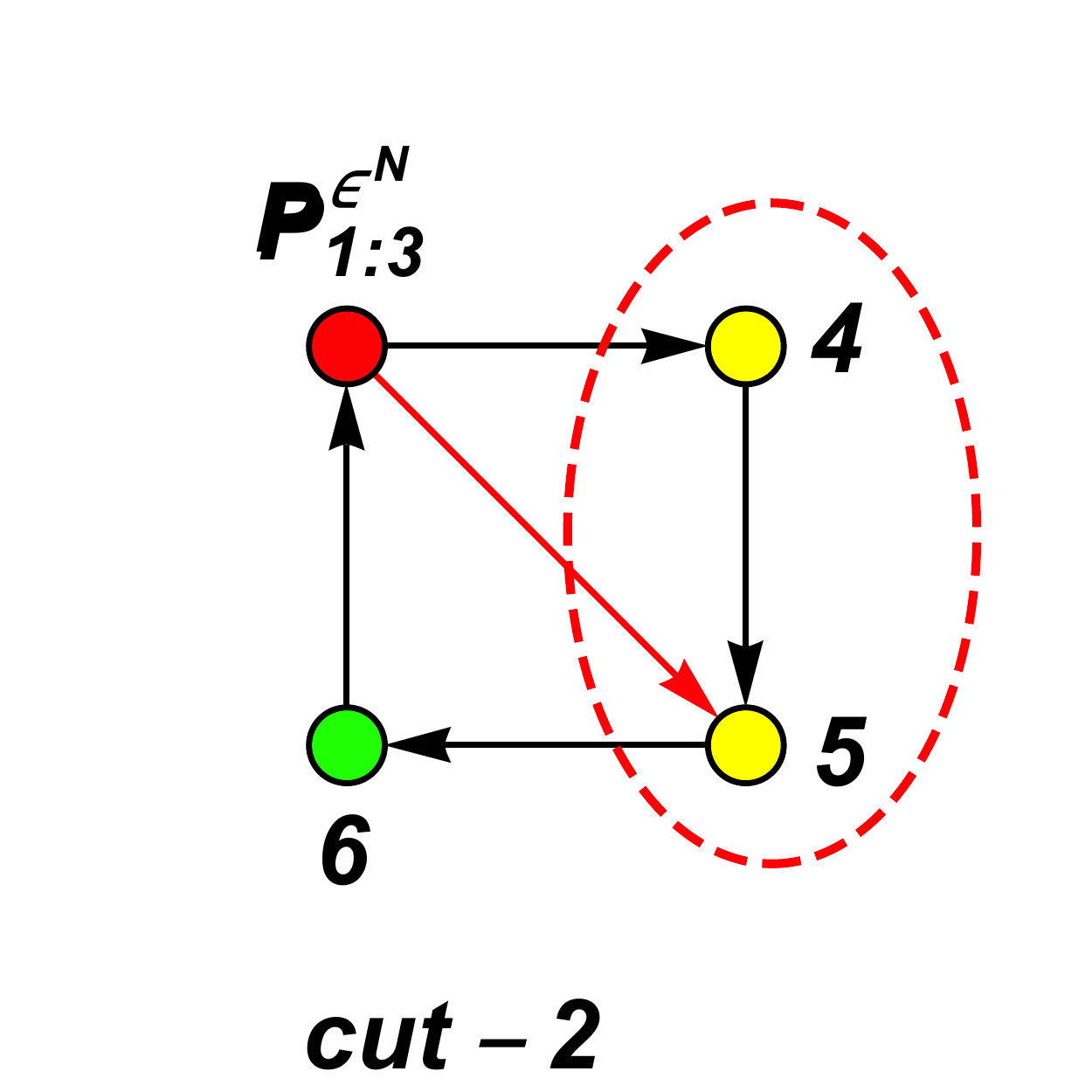}} +
\hspace{-0.45cm}
\parbox[c]{5.4em}{\includegraphics[scale=0.21]{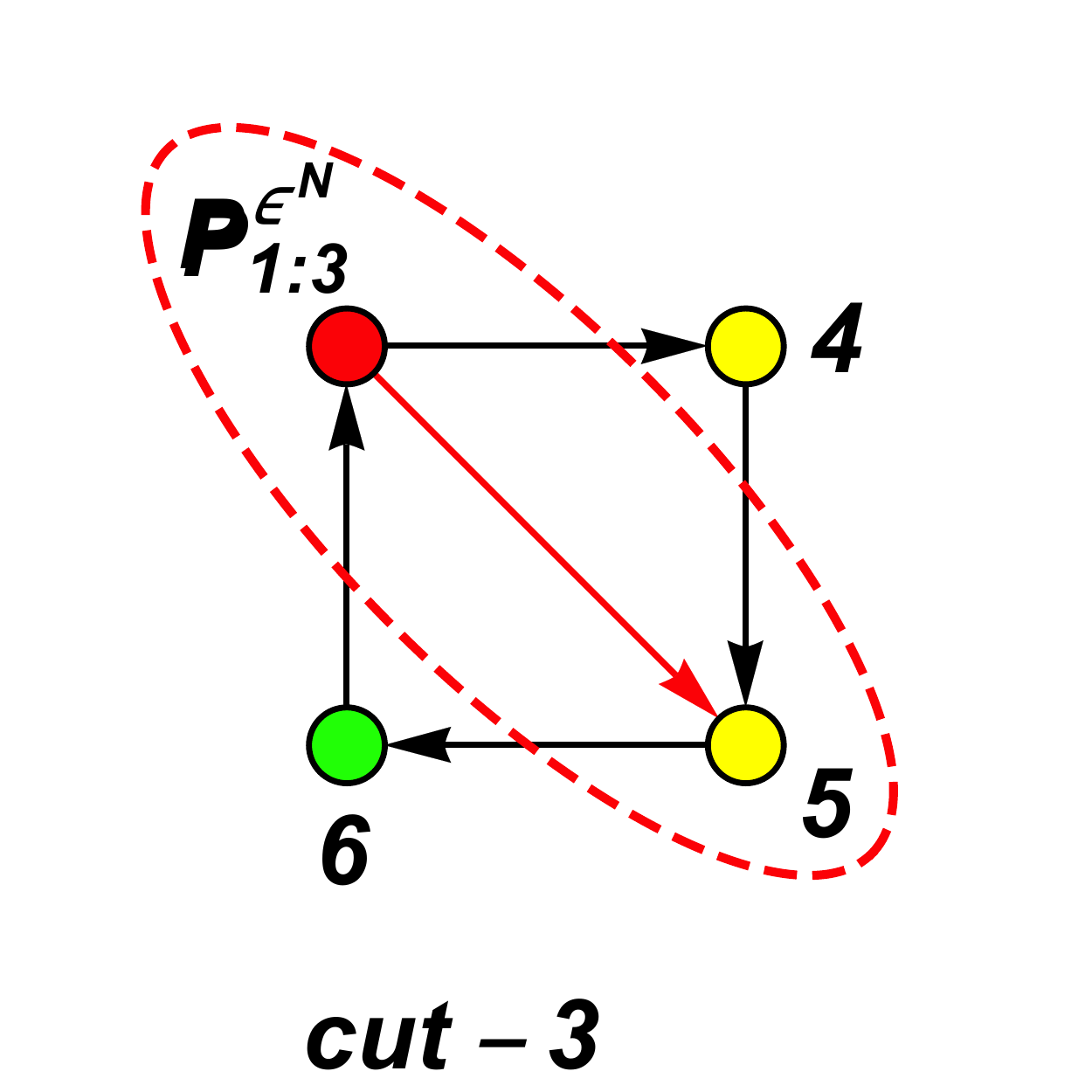}}
\nonumber\\\label{eq:a41red}
&= - \sqrt{2} \, s_{46} (\eps_{1:3}^N \cdot k_4) - \frac{ \sqrt{2} \, s_{46}\, s_{45}\, (\eps_{1:3}^N \cdot k_6)}{s_{6P_{1:3}}}  -  \sqrt{2} \, s_{46}\, (\eps_{1:3}^N \cdot k_5),
\end{align}
}
\vskip-0.0cm\noindent
where we again have used the three-point building blocks in \cref{3pt-1,3point-BB}. 
Lastly, since for the amplitude, 
$A^{(P_{12}\,P_{34})}_4(P^{M}_{12},P^{N}_{34},5,6)$, 
the above identity is no longer valid, namely\footnote{This is because there is more than 
one off-shell polarization vector.} 
$A^{(P_{12}\,P_{34})}_4(P^{M}_{12},P^{N}_{34},5,6) \neq  A^{(P_{12}\, 5 )}_4(P^{M}_{12},P^{N}_{34},5,6)$, 
we make use of the BCJ relation \cite{Bjerrum-Bohr:2016juj,Cardona:2016gon},     
$s_{65} \, {\rm PT}(5,6,P_{12}, P_{34}) + s_{6 P_{125}} \, {\rm PT} (5, P_{12},6, P_{34} ) = 0$.
From this we obtain the equality  
$A^{(P_{12}\,P_{34})}_4(P^{M}_{12},P^{N}_{34},5,6) = \Big( \frac{s_{6P_{34}}}{s_{56}} \Big) \times  A^{(P_{12}\, P_{34} )}_4(P^{M}_{12},6, P^{N}_{34},5)$. 
Now, applying the integration rules, one has
{\small
\begin{align}
&
\hspace{-0.2cm}
A^{(P_{12}\,P_{34})}_4(P^{M}_{12},P^{N}_{34},5,6)
=   \left( \frac{s_{6P_{34}}}{s_{56}} \right)\times
\int d\mu_4^\L
\hspace{-0.6cm}
\parbox[c]{5.4em}{\includegraphics[scale=0.21]{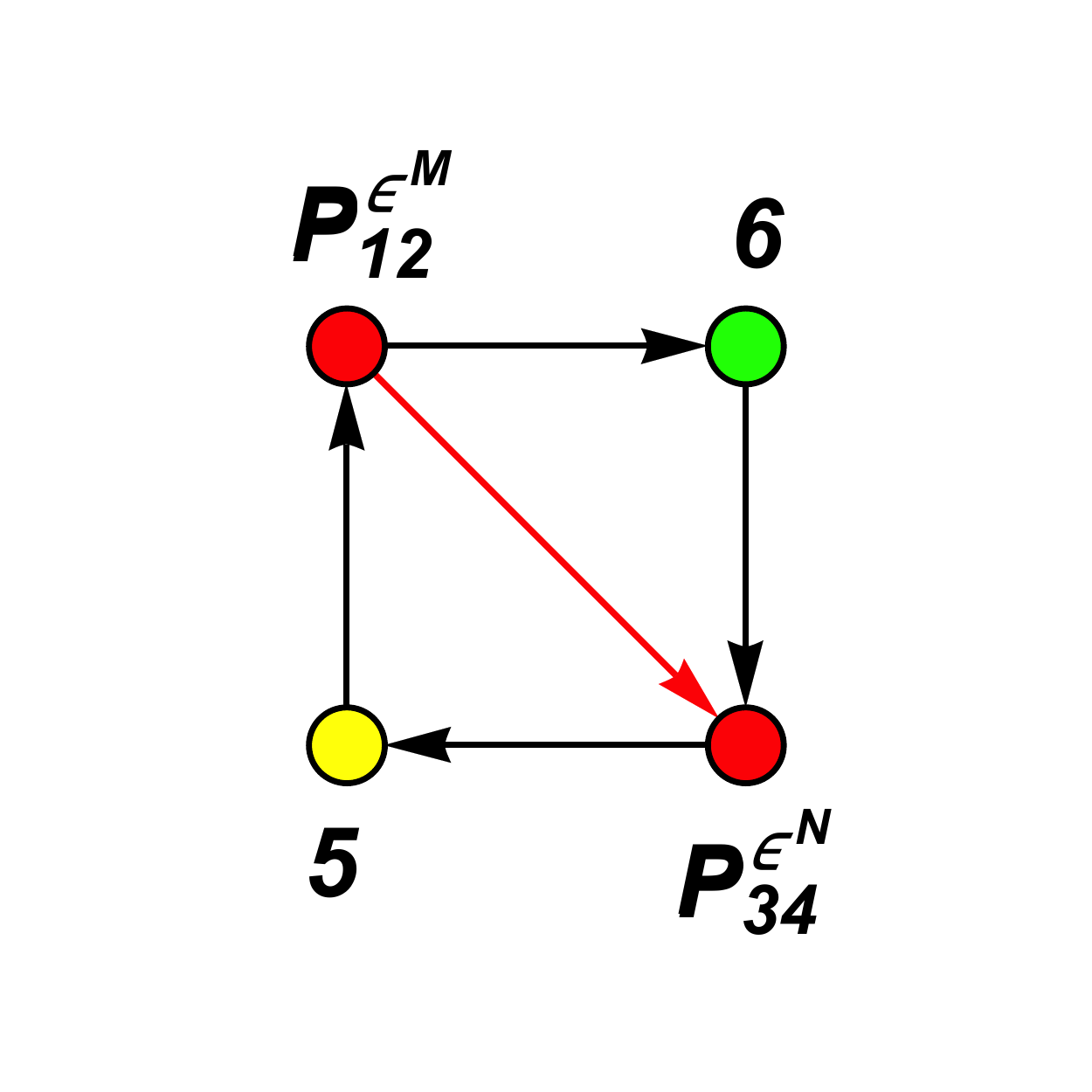}}
=
\hspace{-0.6cm}
\parbox[c]{5.4em}{\includegraphics[scale=0.21]{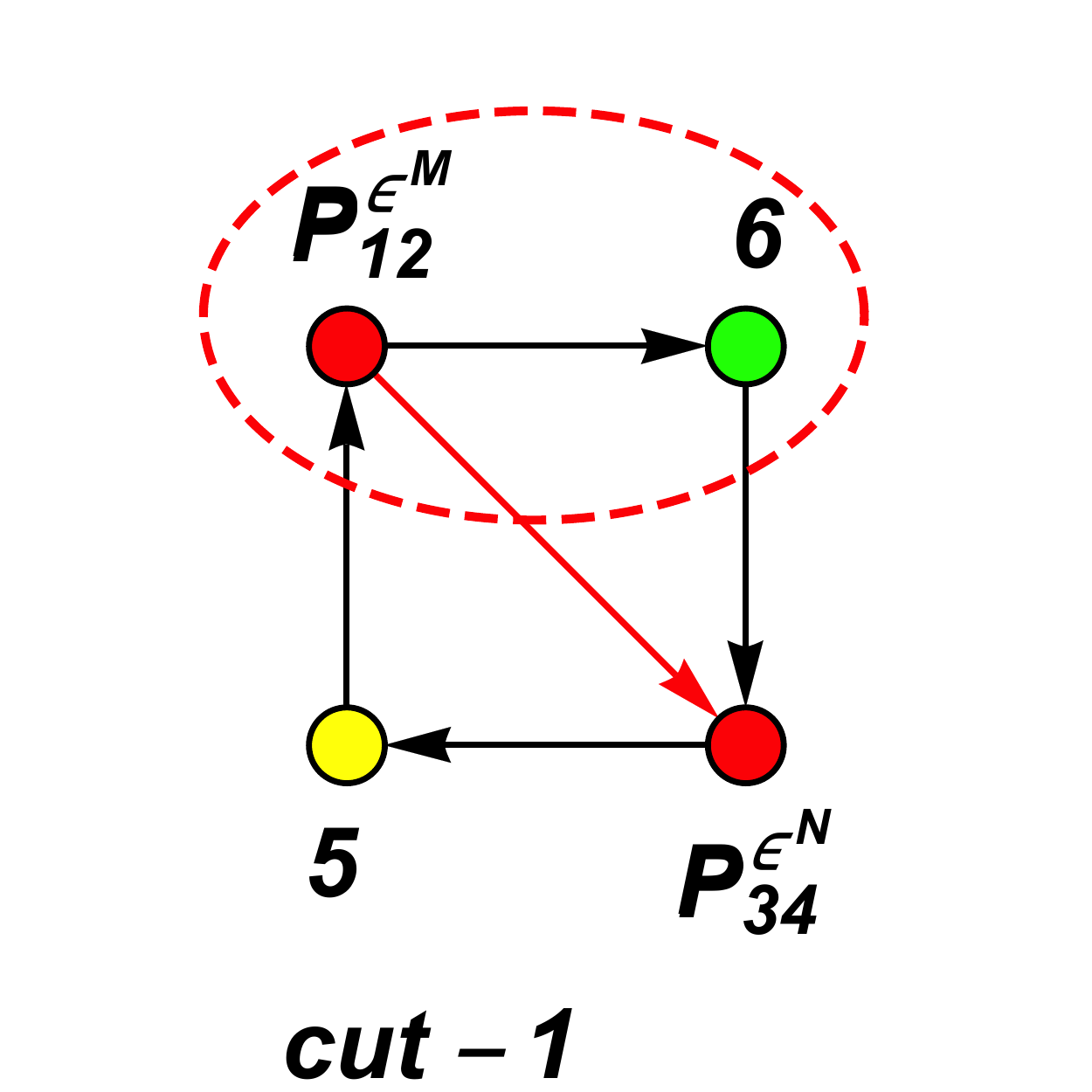}} +
\hspace{-0.6cm}
\parbox[c]{6.4em}{\includegraphics[scale=0.21]{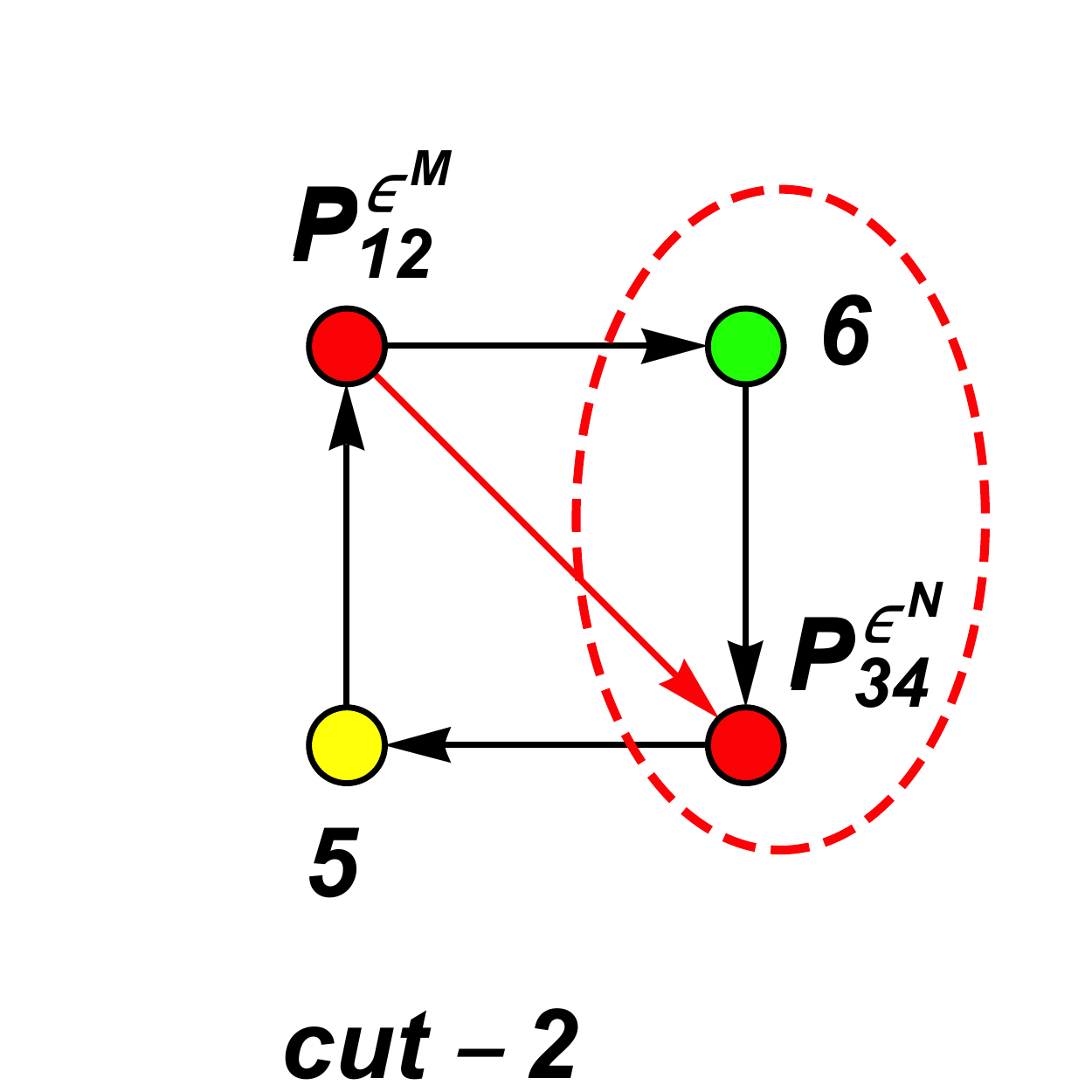}} +
\hspace{-0.45cm}
\parbox[c]{5.4em}{\includegraphics[scale=0.21]{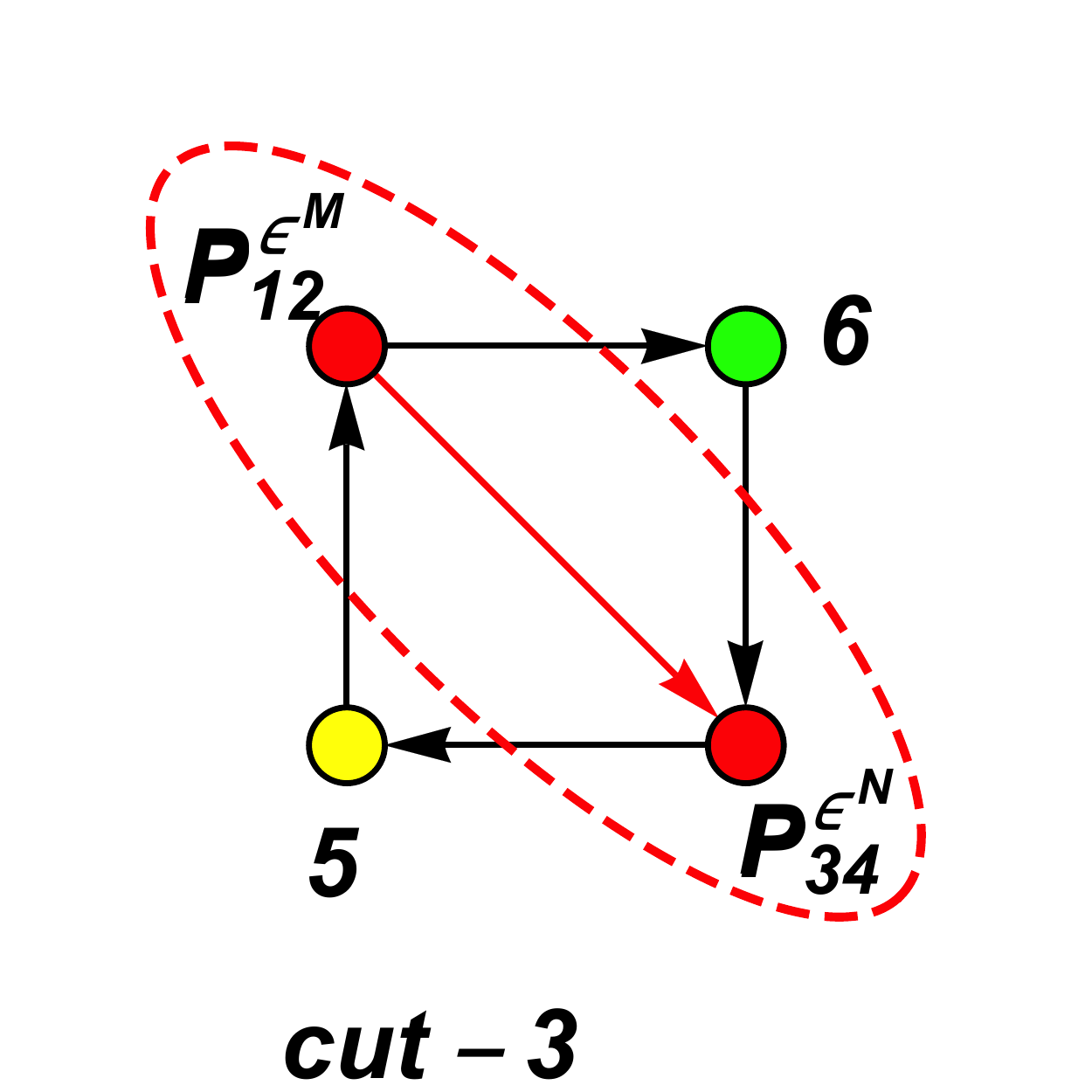}}
\nonumber\\
&\hspace{+1.6cm}
= - \frac{ 2\, s_{6P_{34}}\, (\eps_{12}^M \cdot k_6)\, (\eps_{34}^N \cdot k_5) }{s_{6P_{12}}}  -2 \, (\eps_{12}^M \cdot k_5)\, (\eps_{34}^N \cdot k_6)-  s_{6P_{34}} (\eps_{12}^M \cdot \eps_{34}^N ) . \label{a4-p1234}
\end{align}
}
\vskip-0.0cm\noindent

Utilizing the results obtained in \cref{eq:a3-amplitudesAp,eq:a4prime,eq:a41red,a4-p1234}, 
it is straightforward  to check the five-point amplitude,
$A^{(P_{12}\,3)}_5(P^{M}_{12},3,4,5,6)$, is given by
{\small
\begin{eqnarray}
&&
A^{(P_{12}\,3)}_5(P^{M}_{12},3,4,5,6)  = -  s_{12}\,  \sqrt{2} \left\{  \frac{ s_{46}}{s_{4:6}}\left[ (\eps_{12}^M\cdot k_4) + \frac{ s_{45}   (\eps_{12}^M\cdot k_6)   }{s_{6P_{1:3}} }  + (\eps_{12}^M\cdot k_5)  \right] \right. \nonumber\\
&& \left.
+ \frac{  s_{6P_{34}}  }{s_{56P_{12}}}\left[  \frac{ s_{45}\,  (\eps_{12}^M\cdot k_6)   }{s_{6P_{12}} } +  \frac{ s_{46}\,  (\eps_{12}^M\cdot k_5)   }{s_{6P_{34}} }  + (\eps_{12}^M\cdot k_4)  \right]
- s_{56}  \,(\eps_{12}^M\cdot k_6) \left[      \frac{1}{s_{6P_{45}} } +\frac{1}{s_{6P_{12}} }        \right] \right\},  ~~~\quad
\end{eqnarray}
}
\vskip-0.0cm\noindent
and therefore {\it cut-1} in eq.~\eqref{cut1Ap} is given by
{\small
\begin{eqnarray}\label{cut1ApR}
\parbox[c]{6.2em}{\includegraphics[scale=0.19]{6pts-ex1-cut1.pdf}} &= &
\sum_M
 \frac{A^{\prime}_3(1,2,P^{M}_{3:6}) \times A^{(P_{12}\,3)}_5(P^{M}_{12},3,4,5,6)}{s_{3:6}} =
  -   \left\{  \frac{ s_{46}}{s_{4:6}}\left[ s_{14} + \frac{ s_{45}   s_{16}   }{s_{6P_{1:3}} }  + s_{15}  \right] \right. \nonumber\\
&& \left.
+ \frac{  s_{6P_{34}}  }{s_{56P_{12}}}\left[  \frac{ s_{45}\,  s_{16}   }{s_{6P_{12}} } +  \frac{ s_{46}\,  s_{15}   }{s_{6P_{34}} }  + s_{14}  \right]
- s_{56}  \, s_{16} \left[      \frac{1}{s_{6P_{45}} } +\frac{1}{s_{6P_{12}} }        \right] \right\}.  ~~~\quad
\end{eqnarray}
}
\vskip-0.2cm\noindent

The other contributions, {\it cut-2,3,4}, are calculated in a similar fashion. 
We find that
\vspace{-0.2cm}
{\small
\begin{eqnarray}
\parbox[c]{6.2em}{\includegraphics[scale=0.19]{6pts-ex1-cut2.pdf}} &= &\sum_M
 \frac{A^{\prime}_5(1,2,P^{M}_{34},5,6) \times  A^{(P_{5:2}\,3)}_3(P^{M}_{5:2},3,4)}{s_{34}}  
 = 
 -   \left\{  \frac{ s_{15}}{s_{5:1}}\left[ s_{14} + \frac{ s_{45}   s_{16}   }{s_{5P_{2:4}} }  + s_{46}  \right] \right. \nonumber\\
&& \left.
+ \frac{  s_{5P_{12}}  }{s_{56P_{34}}}\left[  \frac{ s_{45}\,  s_{16}   }{s_{5P_{34}} } +  \frac{ s_{46}\,  s_{15}   }{s_{5P_{12}} }  + s_{14}  \right]
- s_{56}  \, s_{45} \left[      \frac{1}{s_{5P_{16}} } +\frac{1}{s_{5P_{34}} }        \right] \right\},  ~~~\quad
\\
\parbox[c]{6.2em}{\includegraphics[scale=0.19]{6pts-ex1-cut3.pdf}} &=&  \sum_M
 \frac{A^{\prime}_3(P^{M}_{4:1},2,3) \times  A^{(1\,P_{23})}_5(1,P^{M}_{23},4,5,6)}{s_{4:1}}  
 = 
 -   \left\{  \frac{ s_{46}}{s_{4:6}}\left[ s_{34} + \frac{ s_{45}   s_{36}   }{s_{6P_{1:3}} }  + s_{35}  \right] \right. \nonumber\\
&& \left.
- \frac{  s_{56}\, s_{36}  }{s_{6P_{45}}} 
+      \frac{ s_{6P_{2:4} } \, s_{34}   }{ s_{5:1} }  \right\},  ~~~\quad
\\
\parbox[c]{6.2em}{\includegraphics[scale=0.19]{6pts-ex1-cut4.pdf}} &=& \sum_M
 \frac{A^{\prime}_4(1,2,P^{M}_{3:5},6) \times  A^{(P_{6:2}\,3)}_4(P^{M}_{6:2},3,4,5)}{s_{3:5}}  
  = -\frac{s_{16}\, s_{35} }{s_{3:5}} \times  \nonumber \\
 && 
\left(   \frac{1}{s_{16}} + \frac{1}{ s_{6P_{3:5}} }        \right) 
 \times  \left( 
s_{36}  +  \frac{s_{34} \, s_{56} }{ s_{5P_{6:2}} } + s_{46}
 \right).
 \qquad~
\end{eqnarray}
}
\vskip-0.2cm\noindent

\subsection{$A_{6}^{\prime}(\mathbb{I}^{(134)})$}
\label{sec:sixpointex2}

In this section, we just write down the results found for the cut-contributions obtained  in eq.~\eqref{cuts-6pts-2}.  Using the same method presented above, it is straightforward to arrive
\vspace{-0.3cm}
{\small
\begin{eqnarray}
\parbox[c]{6.2em}{\includegraphics[scale=0.19]{6pts-ex2-cut1.pdf}} &= &
\sum_M
 \frac{A^{(1\,P_{3:6})}_3(1,2,P^{M}_{3:6}) \times  A^{\prime}_5(P^{M}_{12},3,4,5,6)}{s_{3:6}}   
 =
 \frac{ s_{46} }{s_{4:6}}\left[ s_{24} + \frac{ s_{45}   s_{26}   }{s_{6P_{1:3}} }  + s_{25}  \right]  \nonumber\\
&& 
+ \frac{  s_{6P_{34}}  }{s_{56P_{12}}}\left[  \frac{ s_{45}\,  s_{26}   }{s_{6P_{12}} } +  \frac{ s_{46}\,  s_{25}   }{s_{6P_{34}} }  + s_{24}  \right]
- s_{56}  \, s_{26} \left[      \frac{1}{s_{6P_{45}} } +\frac{1}{s_{6P_{12}} }        \right]  ,  ~~~\quad
\\
\parbox[c]{6.2em}{\includegraphics[scale=0.19]{6pts-ex2-cut2.pdf}} &= &\sum_M
 \frac{A^{(1\,P_{34})}_5(1,2,{ P^{M}_{34}},5,6) \times  A^{\prime}_3(P^{M}_{5:2},3,4)}{s_{34}}  
 =
-  \frac{  s_{26} \, s_{56} \, s_{45}  }{s_{5P_{34}} \, s_{6P_{3:5}}  } 
+        \frac{   s_{24}  \, s_{6P_{2:4} }    }{s_{5:1} } 
 \nonumber\\
&& 
+
 \frac{ 1 }{s_{ P_{34}56 }}  \left[ s_{25} \, s_{46} + \frac{ s_{26}  \, s_{6P_{34}} \, s_{45}   }{s_{6P_{12}} }  + s_{24}\, s_{6P_{34}}  \right] , 
\quad
\\
\parbox[c]{6.2em}{\includegraphics[scale=0.19]{6pts-ex3-cut3.pdf}} &=& 
  \frac{  A_3(3 ,P_{4:1},2) \times  A^{\prime}_5(1,P_{23},4,5,6)   } {s_{4:1}}
 = 0
 ,  ~~~\quad
\\
\parbox[c]{6.2em}{\includegraphics[scale=0.19]{6pts-ex4-cut4.pdf}} &=& \sum_M
  \frac{A^{(1\,P_{3:5})}_4(1,2,{P^{M}_{3:5}},6) \times  A^{\prime}_4(P^{M}_{6:2},3,4,5)}{s_{3:5}}  
  = -\frac{s_{26}\, s_{45} }{s_{3:5}} \times  \nonumber \\
 && 
\left(   \frac{1}{s_{45}} + \frac{1}{ s_{5P_{6:2}} }        \right) 
 \times  \left( 
s_{15}  +  \frac{s_{12} \, s_{56} }{ s_{6P_{3:5}} } + s_{25}
 \right).
 \qquad~
\end{eqnarray}
}
\vskip-0.2cm\noindent

\subsection{$A_{6}(\mathbb{I}^{(13)})$}
\label{sec:sixpointex3}

Now, we focus on applying the {\bf integration rules} for $A_{6}( \mathbb{I}^{(13)})$.
We recall that this notation means that the reduced Pfaffian is given by 
$-{\rm PT}^T(1,3)\times {\rm det}[(\mathsf{A}_6^\L)^{13}_{13}]$. 
In addition, such as in the previous examples, we fix the gauge by $(pqr|m)=(123|4)$. 
Thus, from the eq.~\eqref{cuts-6pts-1}, we have that
\vspace{-0.3cm}
{\small
\begin{eqnarray}\label{cuts-6pts-Ap}
	A_{6}(\mathbb{I}^{(13)}) =
\int d\mu_6^\L
\hspace{-0.25cm}
 \parbox[c]{6.1em}{\includegraphics[scale=0.21]{6pts-gauge1.pdf}}
=
\hspace{-0.3cm}
 \parbox[c]{6.1em}{\includegraphics[scale=0.21]{gauge1-cut1.pdf}} +
 \hspace{-0.3cm}
  \parbox[c]{6.1em}{\includegraphics[scale=0.21]{gauge1-cut2.pdf}} +
  \hspace{-0.25cm}
   \parbox[c]{6.1em}{\includegraphics[scale=0.21]{gauge1-cut3.pdf}} ~ .
   \qquad\nonumber
\end{eqnarray}
}
\vskip-0.15cm\noindent
Applying the {\bf integration rules}, {\it cut-1} is split into
\vspace{-0.3cm}
{\small
\begin{eqnarray}\label{G1-cut1-RG}
\parbox[c]{5.9em}{\includegraphics[scale=0.21]{gauge1-cut1.pdf}}&=&
\int d\mu_{4}^{\rm CHY}
\hspace{-0.51cm}
\parbox[c]{5.3em}{\includegraphics[scale=0.21]{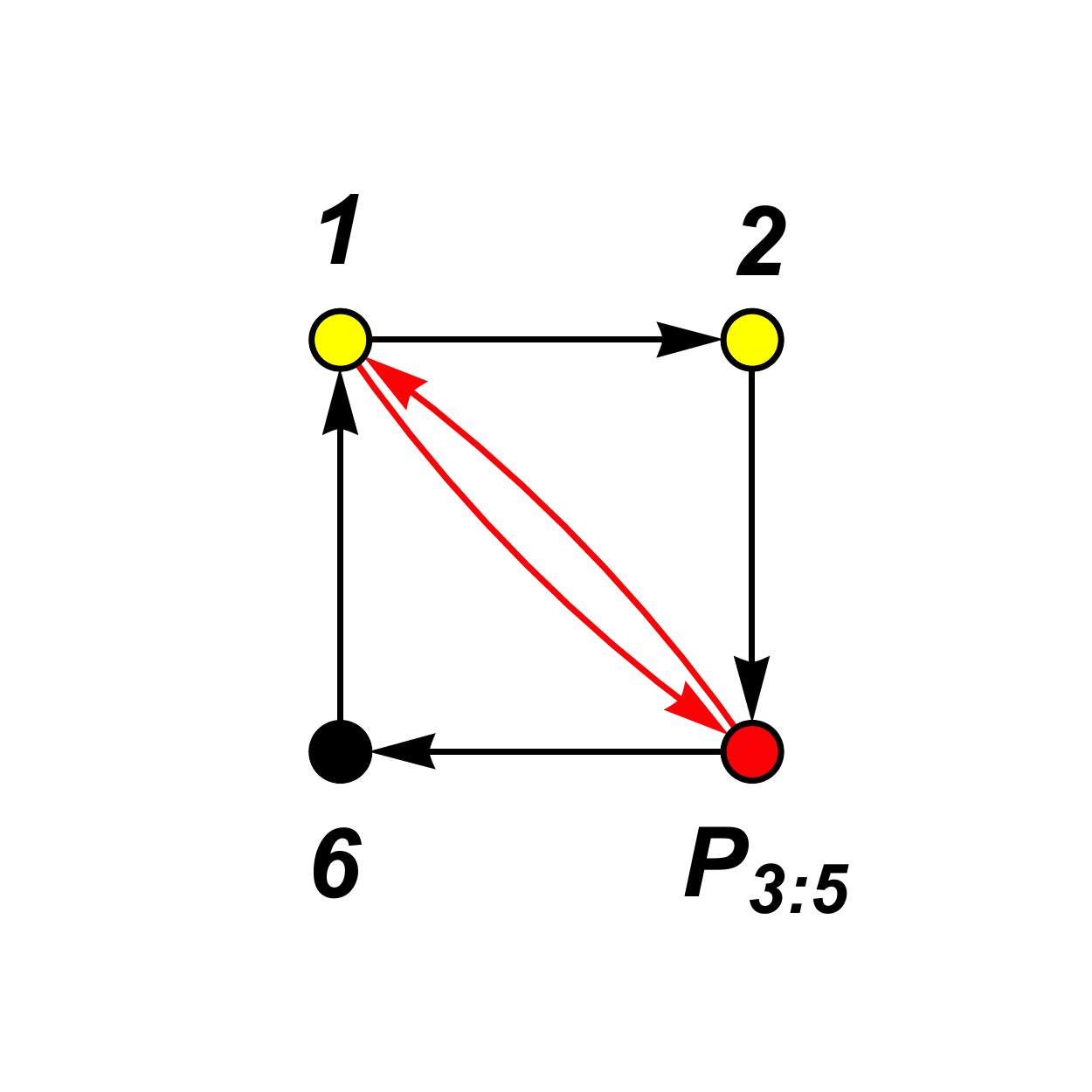}}
\times \left(  \frac{1}{ s_{345}}  \right) \times 
\int d\mu_{4}^{\rm CHY}
\hspace{-0.51cm}
\parbox[c]{10em}{\includegraphics[scale=0.21]{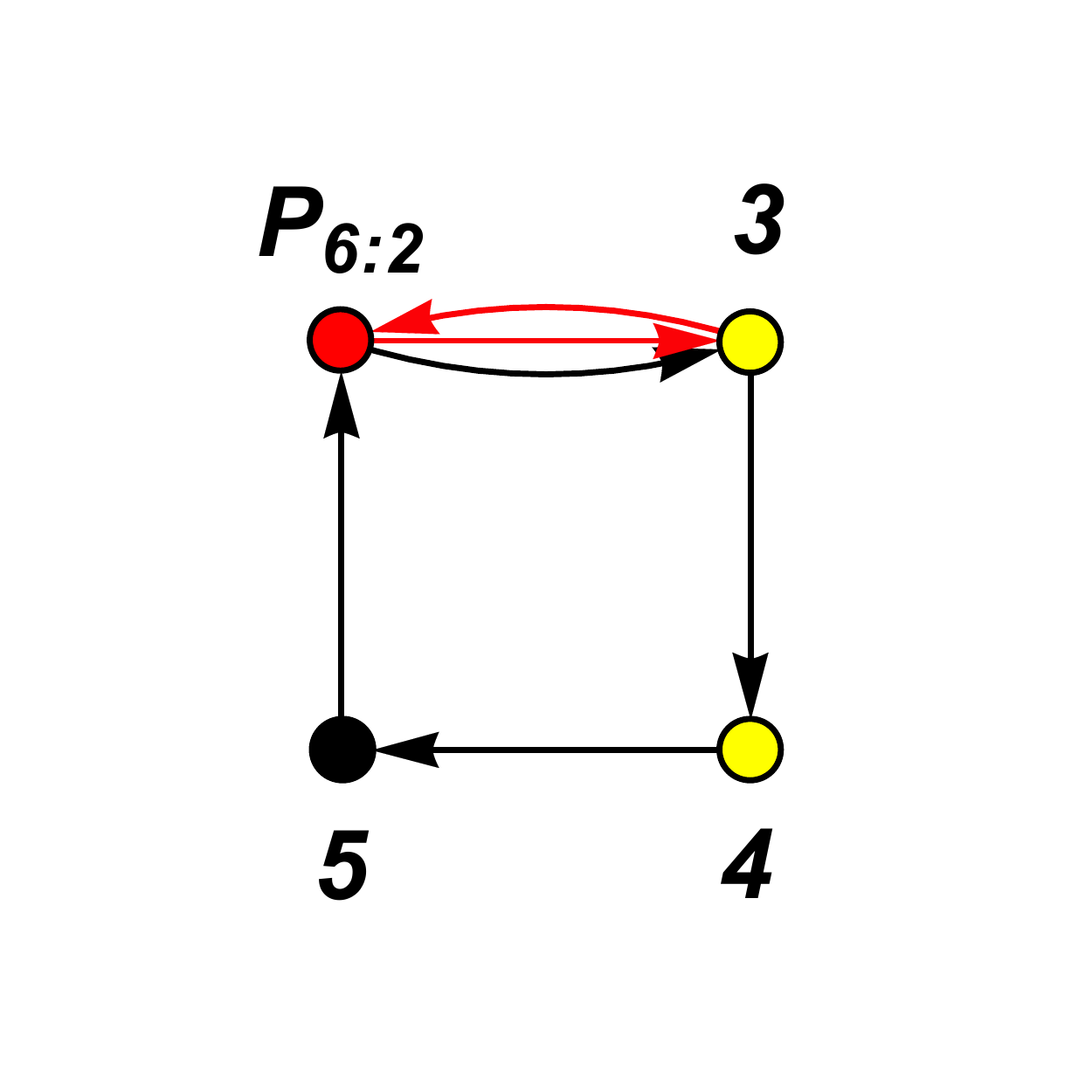}} \nonumber \\
& =&
 \frac{  A_4(\bm{1},2,\bm{P_{3:5}},6) \times  A_4(\bm{P_{6:2}}, \bm{3} ,4,5) }
{s_{3:5}}  \,\, 
 =\frac{ s_{26} \, s_{35}}{ s_{3:5}} \, . 
\end{eqnarray}
}
\vskip-0.25cm\noindent
On the last equality we used the identity, 
$A_4(\bm{P_{6:2}}, \bm{3} ,4,5) =A_4(\bm{P_{6:2}},3, \bm{4} ,5)$ 
(in order to avoid singular cuts), and the same procedure as in eq.~\eqref{4pts-Gg-cuts}. 
This identity is supported over the off-shell Pfaffian properties given in  appendix \ref{Pf-properties}.

The following contribution is the {\it cut-2} ({\it strange-cut}), which, by the {\bf integration rules},  is broken as  
\vspace{-0.3cm}
{\small
\begin{eqnarray}
\parbox[c]{6.2em}{\includegraphics[scale=0.21]{gauge1-cut2.pdf}}=
\int d\mu_5^{\rm CHY}
 \hspace{-0.25cm}
  \parbox[c]{5.8em}{\includegraphics[scale=0.21]{G1-C2-RG1.pdf}}
\times \left(  \frac{1}{s_{4:6,2}}  \right) \times 
\hspace{-0.65cm}
   \parbox[c]{6.0em}{\includegraphics[scale=0.21]{G1-C2-RG2.pdf}} .
\end{eqnarray}
}
\vskip-0.2cm\noindent
Notice that on the first graph the our method can not be employed. 
Nevertheless, similar to Yang-Mills theory~\cite{Gomez:2018cqg}, 
this strange-cut can be rewritten in the following way  
\vspace{-0.3cm}
{\small
\begin{eqnarray}
\int d\mu_5^{\rm CHY}
 \hspace{-0.25cm} 
\parbox[c]{5.8em}{\includegraphics[scale=0.21]{G1-C2-RG1.pdf}}
\times 
\hspace{-0.65cm}
   \parbox[c]{5.8em}{\includegraphics[scale=0.21]{G1-C2-RG2.pdf}} &=&\,(-1)\,
\int d\mu_5^{\rm CHY}
 \hspace{-0.25cm}
  \parbox[c]{5.8em}{\includegraphics[scale=0.21]{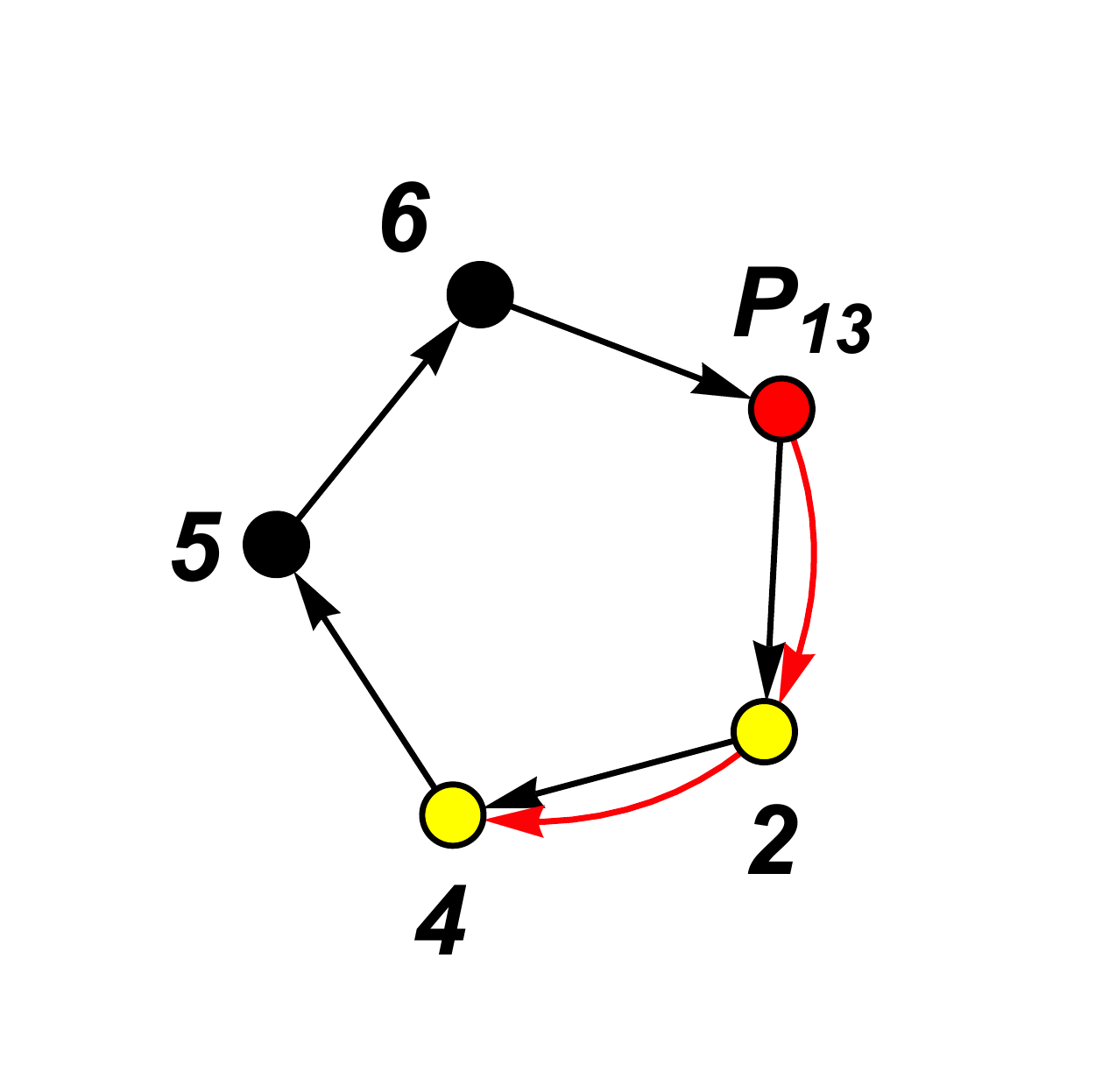}}
\times 
\hspace{-0.65cm}
   \parbox[c]{6.0em}{\includegraphics[scale=0.21]{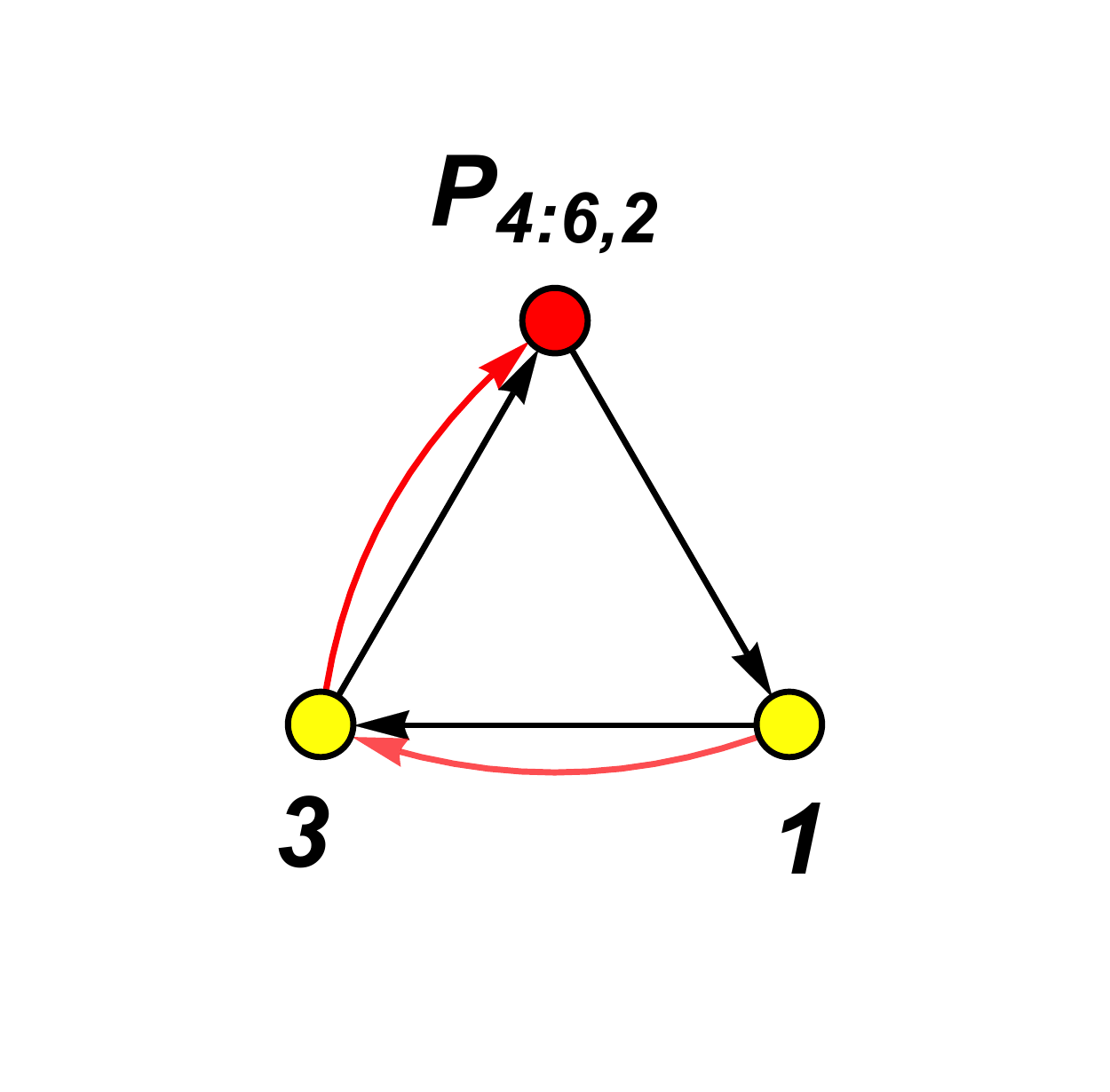}}   \nonumber
 \\
& =& (-1)\,
A^{\prime}_5(P_{13},2,4,5,6) \times  A^{\prime}_3(1, 3,P_{4:6,2})  .\,\, ~~     
\end{eqnarray}
}
\vskip-0.2cm\noindent
where we used the 
identities formulated in appendix \ref{Pf-properties}. Therefore, this cut turns into
\vspace{-0.3cm}
{\small
\begin{eqnarray}
\hspace{-0.3cm}
\parbox[c]{6.2em}{\includegraphics[scale=0.21]{gauge1-cut2.pdf}}= (-1)\,
 \frac{A^{\prime}_5(P_{13},2,4,5,6) \times  A^{\prime}_3(1,3,P_{4:6,2})}{s_{4:6,2}}  
 =
 s_{13} \left[
 \frac{ s_{46}}{s_{456}}+ \frac{ s_{26}+ s_{46}}{s_{56P_{13}}}
 \right]
 , \qquad
\end{eqnarray}
}
\vskip-0.2cm\noindent
The five-point amplitude, $A^{\prime}_5(P_{13},2,4,5,6)$, was already calculated in \cref{A5-p12Ap} 
and the three-point function is given in \cref{3pt-1}.

Lastly, the strange {\it cut-3} is
\vspace{-0.3cm}
{\small
\begin{eqnarray}
\hspace{-0.3cm}
\parbox[c]{6.2em}{\includegraphics[scale=0.21]{gauge1-cut3.pdf}}=(-1)\,
 \frac{A^{\prime}_3(P_{5:1,3},2,4) \times  A^{\prime}_5(1,3,P_{24},5,6)}{s_{24}}  
 = 
 s_{24} \left[
 \frac{ s_{26}+s_{46}}{s_{56P_{24}}}+ \frac{ s_{26}+s_{36} +s_{46}  }{s_{561}}
 \right].\nonumber \\
  \qquad~
\end{eqnarray}
}
\vskip-0.2cm\noindent

\subsection{Longitudinal Contributions}\label{appendix-Lcontribution}

In this section,  we consider just the longitudinal degrees of freedom of all 
cut-contributions obtained from 
$A_{6}^{\prime}(\mathbb{I}^{(123)})$ and 
$A_{6}^{\prime}(\mathbb{I}^{(134)})$.  
Those results are used in section \ref{section-Lcontribution}. 

First, we begin with the cut-structure given in eq.~\eqref{cuts-6ptsAp} for 
$A_{6}^{\prime}(\mathbb{I}^{(123)})$.
We replace $\epsilon^M \rightarrow \epsilon^L$, and use \cref{eq:epsL}.
The longitudinal contributions become
{\small
\begin{eqnarray}\label{cut1ApL}
&&
\sum_L
 \frac{ A^{\prime}_3(1,2,P^{L}_{3:6}) \times  A^{(P_{12}3)}_5(P^{L}_{12},3,4,5,6)}{s_{3:6}} = \frac{ s_{1P_{3:6}} } {2\, s_{12}}
  \times   \left\{  \frac{ s_{46}}{s_{4:6}}\left[ s_{P_{12}P_{45}} + \frac{ s_{45}   s_{P_{12}6}   }{s_{6P_{1:3}} }   \right] \right. \nonumber\\
&& \left.
+ \frac{  s_{6P_{34}}  }{s_{56P_{12}}}\left[  \frac{ s_{45}\,  s_{P_{12}6}   }{s_{6P_{12}} } +  \frac{ s_{46}\,  s_{P_{12}5}   }{s_{6P_{34}} }  + s_{P_{12}4}  \right]
- s_{56}  \, s_{P_{12}6} \left[      \frac{1}{s_{6P_{45}} } +\frac{1}{s_{6P_{12}} }        \right] \right\}.  ~~~\quad
\end{eqnarray}
}
\vskip-0.2cm\noindent
\vspace{-0.2cm}
{\small
\begin{eqnarray}
&&\sum_L
 \frac{A^{\prime}_5(1,2,P^{L}_{34},5,6) \times   A^{(P_{5:2}3)}_3(P^{L}_{5:2},3,4)}{s_{34}}  
 = 
 \frac{ s_{4P_{5:2}} } {2\, s_{34}}  \times 
  \left\{  \frac{ s_{15}}{s_{5:1}}\left[   s_{P_{34} P_{16} } + \frac{ s_{P_{34}5}   s_{16}   }{s_{5P_{2:4}} }    \right] \right. \nonumber\\
&& \left.
+ \frac{  s_{5P_{12}}  }{s_{56P_{34}}}\left[  \frac{ s_{P_{34}5}\,  s_{16}   }{s_{5P_{34}} } +  \frac{ s_{P_{34}6}\,  s_{15}   }{s_{5P_{12}} }  + s_{1P_{34}}  \right]
- s_{56}  \, s_{5P_{34}} \left[      \frac{1}{s_{5P_{16}} } +\frac{1}{s_{5P_{34}} }        \right] \right\},  ~~~\quad
\end{eqnarray}
}
\vskip-0.2cm\noindent
\vspace{-0.3cm}
{\small
\begin{eqnarray}
&&  \sum_L
 \frac{A^{\prime}_3(P^{L}_{4:1},2,3) \times  A^{(1P_{23})}_5(1,P^{L}_{23},4,5,6)}{s_{4:1}}  
 = 
  \frac{ s_{3P_{4:1}} } {2\, s_{23}}  \times 
 \left\{  \frac{ s_{46}}{s_{4:6}}\left[ s_{P_{23}P_{45}} + \frac{ s_{45}   s_{P_{23}6}   }{s_{6P_{1:3}} }   \right] \right. \nonumber\\
&& \left.
- \frac{  s_{56}\, s_{P_{23}6}  }{s_{6P_{45}}} 
+      \frac{ s_{6P_{2:4} } \, s_{P_{23}4}   }{ s_{5:1} }  \right\},  ~~~\quad
\end{eqnarray}
}
\vskip-0.2cm\noindent
\vspace{-0.3cm}
{\small
\begin{eqnarray}
&& \sum_L
 \frac{A^{\prime}_4(1,2,P^{L}_{3:5},6) \times  A^{(P_{6:2}3)}_4(P^{L}_{6:2},3,4,5)}{s_{3:5}}  
  = \frac{s_{16}\, s_{6P_{3:5}} }{2\, s_{3:5} } \times  \left(   \frac{1}{s_{16}} + \frac{1}{ s_{6P_{3:5}} }        \right)  \times  \nonumber \\
 && 
\frac{s_{35}}{ s_{3:5}}  \times \left( 
s_{P_{6:2}P_{34} }  +  \frac{s_{34} \, s_{5P_{6:2} } }{ s_{5P_{6:2}} }
 \right).
 \qquad~
\end{eqnarray}
}
\vskip-0.2cm\noindent

To end, we carry out the longitudinal contributions for all cut-contributions of 
$ A_{6}^{\prime}(\mathbb{I}^{(134)})$, 
{\small
\begin{eqnarray}
&&
\sum_L
\frac{ A^{(1P_{3:6})}_3(1,2,P^{L}_{3:6}) \times  A^{\prime}_5(P^{L}_{12},3,4,5,6)}{s_{3:6}}   
 =-
  \frac{ s_{2P_{3:6}} } {2\, s_{12}}
  \times   \left\{  \frac{ s_{46}}{s_{4:6}}\left[ s_{P_{12}P_{45}} + \frac{ s_{45}   s_{P_{12}6}   }{s_{6P_{1:3}} }   \right] \right. \nonumber\\
&& \left.
+ \frac{  s_{6P_{34}}  }{s_{56P_{12}}}\left[  \frac{ s_{45}\,  s_{P_{12}6}   }{s_{6P_{12}} } +  \frac{ s_{46}\,  s_{P_{12}5}   }{s_{6P_{34}} }  + s_{P_{12}4}  \right]
- s_{56}  \, s_{P_{12}6} \left[      \frac{1}{s_{6P_{45}} } +\frac{1}{s_{6P_{12}} }        \right] \right\}.  ~~~\quad
\end{eqnarray}
}
\vskip-0.2cm\noindent
\vspace{-0.2cm}
{\small
\begin{eqnarray}
 & &\sum_L
 \frac{A^{(1P_{34})}_5(1,2,{ P^{L}_{34}},5,6) \times A^{\prime}_3(P^{L}_{5:2},3,4)}{s_{34}}  
 =
-  
  \frac{ s_{4P_{5:2}} } {2\, s_{34}}
  \times
\left\{
\frac{  s_{26} \, s_{56} \, s_{P_{34}5}  }{s_{5P_{34}} \, s_{6P_{3:5}}  } 
+        \frac{   s_{2P_{34}}  \, s_{6P_{2:4} }    }{s_{5:1} } \right.
 \nonumber\\
&& \left.
+
 \frac{ 1 }{s_{ P_{34}56 }}  \left[ s_{25} \, s_{P_{34}6} + \frac{ s_{26}  \, s_{6P_{34}} \, s_{5P_{34}}   }{s_{6P_{12}} }  + s_{2P_{34}}\, s_{6P_{34}}  \right]\right\} , 
\quad
\end{eqnarray}
}
\vskip-0.2cm\noindent
\vspace{-0.3cm}
{\small
\begin{eqnarray}
 && \sum_L
  \frac{A^{(1P_{3:5})}_4(1,2,{P^{L}_{3:5}},6) \times  A^{\prime}_4(P^{L}_{6:2},3,4,5)}{s_{3:5}}  
  = \frac{s_{5P_{6:2}}\, s_{45} }{s_{3:5}} \times \left(   \frac{1}{s_{45}} + \frac{1}{ s_{5P_{6:2}} }        \right)  \times  \nonumber \\
 && 
 \frac{s_{26} }{s_{3:5}} 
 \times  \left( 
s_{1P_{3:5}}  +  \frac{s_{12} \, s_{P_{3:5}6} }{ s_{6P_{3:5}} } + s_{2P_{3:5}}
 \right).
 \qquad~
\end{eqnarray}
}
\vskip-0.2cm\noindent
\vspace{-0.3cm}
{\small
\begin{eqnarray}
  \frac{  A_3(3 ,P_{4:1},2) \times  A^{\prime}_5(1,P_{23},4,5,6)   } {s_{4:1}}
 = 0
 ,  ~~~\quad
\end{eqnarray}
}
\vskip-0.2cm\noindent


\bibliographystyle{JHEP}
\bibliography{mybib}


\end{document}